\documentclass[usenatbib]{mn2e}

\usepackage{epsfig}
\usepackage{graphics}
\usepackage{afterpage}
\usepackage[draft]{hyperref}

\newlength{\colwidth}
\newcommand{\cm}{{\rm cm}}
\newcommand{\g}{{\rm g}}
\newcommand{\s}{{\rm s}}
\newcommand{\kms}{{\rm km}\,{\rm s}^{-1}}
\newcommand{\K}{{\rm K}}
\newcommand{\pc}{{\rm pc}}

\newcommand{\kpc}{{\rm kpc}}

\newcommand{\hcMpc}{h^{-1}\,{\rm cMpc}}
\newcommand{\erg}{{\rm erg}}

\newcommand{\Msun}{{{\rm M}_\odot}}

\newcommand{\Zsun}{{\rm Z}_\odot}
\newcommand{\nHtc}{n_{{\rm H},t_{\rm c}}}

\newcommand{\yr}{{\rm yr}}
\newcommand{\Gyr}{{\rm Gyr}}

\newcommand{\Msolyrkpcsq}{{\rm M}_\odot\,{\rm yr}^{-1}\,{\rm kpc}^{-2}}
\newcommand{\ion}[2]{\hbox{#1\,{\sc #2}}}
\newcommand{\ionsubscript}[2]{\hbox{\scriptsize #1\,{\tiny #2}}}

\newcommand{\HI}{\ion{H}{i}}

\newcommand{\HeII}{\ion{He}{ii}}

\newcommand{\CIV}{\ion{C}{iv}}

\newcommand{\OVI}{\ion{O}{vi}}

\newcommand{\lya}{Ly$\alpha$}

\newcommand{\eagle}{EAGLE}
\newcommand{\owls}{OWLS}
\newcommand{\cosmo}{cosmo-OWLS}
\newcommand{\gimic}{GIMIC}
\newcommand{\ICgen}{{\sc ic\_2lpt\_gen}}
\newcommand{\skirt}{{\sc skirt}}

\title[The \eagle\ simulation project]{The \eagle\ project: Simulating the evolution and assembly of galaxies and their environments} 
\author[J. Schaye et al.]{%
Joop~Schaye,$^1$\thanks{E-mail: schaye@strw.leidenuniv.nl}  
Robert A. Crain,$^1$ 
Richard G. Bower,$^2$ 
Michelle Furlong,$^2$ \newauthor 
Matthieu Schaller,$^2$ 
Tom Theuns,$^{2,3}$ 
Claudio~Dalla~Vecchia,$^{4,5}$  
Carlos S. Frenk,$^2$ \newauthor 
I.~G. McCarthy,$^6$ 
John C. Helly,$^2$ 
Adrian Jenkins,$^2$ 
Y.~M. Rosas-Guevara,$^2$ \newauthor 
Simon D. M. White,$^7$ 
Maarten Baes,$^8$ 
C.~M.~Booth,$^{1,9}$ 
Peter Camps,$^8$ \newauthor 
Julio F. Navarro,$^{10}$ 
Yan Qu,$^2$ 
Alireza Rahmati,$^7$ 
Till Sawala,$^2$ 
Peter A. Thomas,$^{11}$ \newauthor 
James Trayford$^2$
\\
$^1$ Leiden Observatory, Leiden University, P.O. Box 9513, 2300 RA Leiden, the Netherlands\\
$^2$ Institute for Computational Cosmology, Department of Physics, University of Durham, South Road, Durham, DH1 3LE, UK\\
$^3$ Department of Physics, University of Antwerp, Groenenborgerlaan 171, B-2020 Antwerpen, Belgium\\
$^4$ Instituto de Astrof\'isica de Canarias, C/ V\'ia L\'actea s/n,38205 La Laguna, Tenerife, Spain\\
$^5$ Departamento de Astrof\'sica, Universidad de La Laguna, Av.~del Astrof\'isico Franciso S\'anchez s/n, 38206 La Laguna, Tenerife, Spain \\
$^6$ Astrophysics Research Institute, Liverpool John Moores University, 146 Brownlow Hill, Liverpool L3 5RF, UK\\ 
$^7$ Max-Planck-Institut fur Astrophysik, Karl-Schwarzschild-Str. 1, D-85748 Garching, Germany\\
$^8$ Sterrenkundig Observatorium, Universiteit Gent, Krijgslaan 281-S9, B-9000 Gent, Belgium\\
$^9$ Department of Astronomy \& Astrophysics, The University of Chicago, Chicago, IL 60637, USA\\
$^{10}$ Department of Physics and Astronomy, University of Victoria, Victoria, BC V8P 5C2, Canada\\
$^{11}$ Astronomy Centre, University of Sussex, Falmer, Brighton BN1 9QH, UK
}

\begin{document}

\maketitle

\begin{abstract}
We introduce the Virgo Consortium's \eagle\ project, a suite of hydrodynamical simulations that follow the formation of galaxies and supermassive black holes in cosmologically representative volumes of a standard $\Lambda$CDM universe. We discuss the limitations of such simulations in light of their finite resolution and poorly constrained subgrid physics, and how these affect their predictive power. One major improvement is our treatment of feedback from massive stars and AGN in which thermal energy is injected into the gas without the need to turn off cooling or decouple hydrodynamical forces, allowing winds to develop without predetermined speed or mass loading factors. Because the feedback efficiencies cannot be predicted from first principles, we calibrate them to the present-day galaxy stellar mass function and the amplitude of the galaxy-central black hole mass relation, also taking galaxy sizes into account. The observed galaxy stellar mass function is reproduced to $\la 0.2$~dex over the full resolved mass range, $10^8 < M_\ast/\Msun \la 10^{11}$, a level of agreement close to that attained by semi-analytic models, and unprecedented for hydrodynamical simulations. We compare our results to a representative set of low-redshift observables not considered in the calibration, and find good agreement with the observed galaxy specific star formation rates, passive fractions, Tully-Fisher relation, total stellar luminosities of galaxy clusters, and column density distributions of intergalactic \CIV\ and \OVI. While the mass-metallicity relations for gas and stars are consistent with observations for $M_\ast \ga 10^9\,\Msun$ ($M_\ast \ga 10^{10}\,\Msun$ at intermediate resolution), they are insufficiently steep at lower masses. For the reference model the gas fractions and temperatures are too high for clusters of galaxies, but for galaxy groups these discrepancies can be resolved by adopting a higher heating temperature in the subgrid prescription for AGN feedback.  The \eagle\ simulation suite, which also includes physics variations and higher-resolution zoomed-in volumes described elsewhere, constitutes a valuable new resource for studies of galaxy formation. 
\end{abstract}


\section{Introduction}

Cosmological simulations have greatly improved our understanding of the physics of galaxy formation and are widely used to guide the interpretation of observations and the design of new observational campaigns and instruments. Simulations enable astronomers to ``turn the knobs'' much as experimental physicists are able to in the laboratory. While such numerical experiments can be valuable even if the simulations fail to reproduce observations, in general our confidence in the conclusions drawn from simulations, and the number of applications they can be used for, increases with the level of agreement between the best-fit model and the observations.

For many years the overall agreement between hydrodynamical simulations and observations of galaxies was poor. Most simulations produced galaxy mass functions with the wrong shape and normalisation, the galaxies were too massive and too compact, and the stars formed too early. Star formation in high-mass galaxies was not quenched and the models could not simultaneously reproduce the stellar masses and the thermodynamic properties of the gas in groups and clusters (e.g.\ \citealt{Scannapieco2012Aquila} and references therein).

Driven in part by the failure of hydrodynamical simulations to reproduce key observations, semi-analytic and halo-based models have become the tools of choice for detailed comparisons between galaxy surveys and theory (see \citealt{Baugh2006SAMReview} and \citealt{Cooray2002HaloModelReview} for reviews). Thanks to their flexibility and relatively modest computational expense, these approaches have proven valuable for many purposes. Examples include the interpretation of observations of galaxies within the context of the cold dark matter framework, relating galaxy populations at different redshifts, the creation of mock galaxy catalogues to investigate selection effects or to translate measurements of galaxy clustering into information concerning the occupation of dark matter haloes by galaxies. 

However, hydrodynamical simulations have a number of important advantages over these other approaches. The risk that a poor or invalid approximation may lead to over-confidence in an extrapolation, interpretation or application of the model is potentially smaller, because they do not need to make as many simplifying assumptions. Although the subgrid models employed by current hydrodynamical simulations often resemble the ingredients of semi-analytic models, there are important parts of the problem for which subgrid models are no longer required. Since hydrodynamical simulations evolve the dark matter and baryonic components self-consistently, they automatically include the back-reaction of the baryons on the collisionless matter, both inside and outside of haloes. The higher resolution description of the baryonic component provided by hydrodynamical simulations also enables one to ask more detailed questions and to compare with many more observables. Cosmological hydrodynamical simulations can be used to model galaxies and the intergalactic medium (IGM) simultaneously, including the interface between the two, which may well be critical to understanding the fuelling and feedback cycles of galaxies.

The agreement between hydrodynamical simulations of galaxy formation and observations has improved significantly in recent years. Simulations of the diffuse IGM already broadly reproduced quasar absorption line observations of the \lya\ forest two decades ago \citep[e.g.][]{Cen1994LyaForest,Zhang1995LyaForest,Hernquist1996LyaForest,Theuns1998LyaForest,Dave1999LyaForest}. The agreement is sufficiently good that comparisons between theory and observation can be used to measure cosmological and physical parameters \citep[e.g.][]{Croft1998LyaPowerSpectrum,Schaye2000IGMTemp,Viel2004LyaPowerSpectrum,McDonald2005LyaPowerSpectrum}. More recently, simulations that have been re-processed using radiative transfer of ionizing radiation have succeeded in matching key properties of the high-column density \HI\ absorbers \citep[e.g.][]{Pontzen2008DLAs,Altay2011HICDDF,McQuinn2011LLS,Rahmati2013HICDDF}.

Reproducing observations of galaxies and the gas in clusters of galaxies has proven to be more difficult than matching observations of the low-density IGM, but several groups have now independently succeeded in producing disc galaxies with more realistic sizes and masses
(e.g.\ \citealt{Governato2004Disk,Governato2010Disk,Okamoto2005Disks,Agertz2011Disk,Guedes2011Eris,McCarthy2012RotSize,Brook2012MagicDisks,Stinson2013MAGICC,Munshi2013StellarHaloMass,Aumer2013Disks,Hopkins2013FIRE,Vogelsberger2013IllustrisModel,Vogelsberger2014Illustris,Marinacci2014Disks}).
For the thermodynamic properties of groups and clusters of galaxies the progress has also been rapid \citep[e.g.][]{Puchwein2008AGN,McCarthy2010AGN,Fabjan2010AGN,LeBrun2014CosmoOWLS}. The improvement in the realism of the simulated galaxies has been accompanied by better agreement between simulations and observations of the metals in circumgalactic and intergalactic gas \citep[e.g.][]{Stinson2012CGM,Oppenheimer2012IGMMetals}, which suggests that a more appropriate description of galactic winds may have been responsible for much of the progress. 

Indeed, the key to the increase in the realism of the simulated galaxies has been the use of subgrid models for feedback from star formation that are more effective in generating galactic winds and, at the high-mass end, the inclusion of subgrid models for feedback from active galactic nuclei (AGN). The improvement in the resolution afforded by increases in computing power and code efficiency has also been important, but perhaps mostly because higher resolution has helped to make the implemented feedback more efficient by reducing spurious, numerical radiative losses. Improvements in the numerical techniques to solve the hydrodynamics have also been made \citep[e.g.][]{Price:2008kx,Springel2010Arepo,Read:2010uq,Saitoh:2013uq,Hopkins:2013lr} and may even be critical for particular applications \citep[e.g.][]{Agertz:2007fk,Bauer2012SubsonicTurbulence}, but overall their effect appears to be small compared to reasonable variations in subgrid models for feedback processes \citep[][]{Scannapieco2012Aquila}. 

Here we present the \eagle\ project\footnote{\eagle\ is a project of the Virgo consortium for cosmological supercomputer simulations.}, which stands for Evolution and Assembly of GaLaxies and their Environments. \eagle\ consists of a suite of cosmological, hydrodynamical simulations of a standard $\Lambda$CDM universe.
The main models were run in volumes of 25 to 100 comoving Mpc (cMpc) on a side and employ a resolution that is sufficient to marginally resolve the Jeans scales in the warm ($T\sim 10^4\,\K$) interstellar medium (ISM). The simulations use state-of-the-art numerical techniques and subgrid models for radiative cooling, star formation, stellar mass loss and metal enrichment, energy feedback from star formation, gas accretion onto, and mergers of, supermassive black holes (BHs), and AGN feedback. The efficiency of the stellar feedback and the BH accretion were calibrated to broadly match the observed $z\sim 0$ galaxy stellar mass function (GSMF) subject to the constraint that the galaxy sizes must also be reasonable, while the efficiency of the AGN feedback was calibrated to the observed relation between stellar mass and BH mass. The goal was to reproduce these observables using, in our opinion, simpler and more natural prescriptions for feedback than used in previous work with similar objectives.

By ``simpler'' and ``more natural'', which are obviously subjective terms, we mean the following. Apart from stellar mass loss, we employ only one type of stellar feedback, which captures the collective effects of processes such as stellar winds, radiation pressure on dust grains, and supernovae. These and other feedback mechanisms are often implemented individually, but we believe they cannot be properly distinguished at the resolution of $10^2$--$10^3\,\pc$ that is currently typical for simulations that sample a representative volume of the universe. Similarly, we employ only one type of AGN feedback (as opposed to e.g.\ both a ``radio'' and ``quasar'' mode). Contrary to most previous work, stellar (and AGN) feedback is injected in thermal form without turning off radiative cooling and without turning off hydrodynamical forces. Hence, galactic winds are generated without specifying a wind direction, velocity, mass loading factor, or metal mass loading factor. We also do not need to boost the BH Bondi-Hoyle accretion rates by an ad-hoc factor. Finally, the amount of feedback energy (and momentum) that is injected per unit stellar mass depends on local gas properties rather than on non-local or non-baryonic properties such as the dark matter velocity dispersion or halo mass.

The \eagle\ suite includes many simulations that will be presented elsewhere. It includes higher-resolution simulations that zoom into individual galaxies or galaxy groups \citep[e.g.][]{Sawala2014EagleZooms}. It also includes variations in the numerical techniques \citep{Schaller2014EagleSPH} and in the subgrid models \citep{Crain2014EagleModels} that can be used to test the robustness of the predictions and to isolate the effects of individual processes. 

This paper is organised as follows. We begin in \S\ref{sec:general} with a discussion of the use and pitfalls of cosmological hydrodynamical simulations in light of the critical role played by subgrid processes. We focus in particular on the implications for the interpretation and the predictive power of the simulations, and the role of numerical convergence. In \S\ref{sec:simulations} we describe the simulations and our definition of a galaxy. This section also briefly discusses the numerical techniques and subgrid physics. The subgrid models are discussed in depth in \S\ref{sec:subgrid}; readers not interested in the details may wish to skip this section. In \S\ref{sec:cal_obs} we show the results for observables that were considered in the calibration of the subgrid models, namely the $z\sim 0$ GSMF, the related relation between stellar mass and halo mass, galaxy sizes, and the relations between BH mass and stellar mass. We also consider the importance of the choice of aperture used to measure stellar masses and investigate both weak and strong convergence (terms that are defined in \S\ref{sec:general}). In \S\ref{sec:otherobs} we present a diverse and representative set of predictions that were not used for the calibration, including specific star formation rates and passive fractions, the Tully-Fisher relation, the mass-metallicity relations, various properties of the intracluster medium, and the column density distributions of intergalactic metals. All results presented here are for $z\sim 0$. We defer an investigation of the evolution to \citet{Furlong2014EagleEvolution} and other future papers. We summarize and discuss our conclusions in \S\ref{sec:summary}. Finally, our implementation of the hydrodynamics and our method for generating the initial conditions are summarized in Appendices~\ref{app:hydro} and \ref{app:ics}, respectively.

\section{Implications of the critical role of subgrid models for feedback}
\label{sec:general}

In this section we discuss what, in our view, the consequences of our reliance on subgrid models for feedback are for the predictive power of the simulations (\S\ref{sec:need}) and for the role of numerical convergence (\S\ref{sec:conv_discussion}).

\subsection{The need for calibration}
\label{sec:need}

Because the recent improvement in the match between simulated and observed galaxies can, for the most part, be attributed to the implementation of more effective subgrid models for feedback, the success of the hydrodynamical simulations is subject to two important caveats that are more commonly associated with semi-analytic models. 

First, while it is clear that effective feedback is required, the simulations can only provide limited insight into the nature and source of the feedback processes. For example, suppose that the implemented subgrid model for supernovae is too inefficient because, for numerical reasons, too much of the energy is radiated away, too much of the momentum cancels out, or the energy/momentum are coupled to the gas at the wrong scale. If we were unaware of such numerical problems, then we might erroneously conclude that additional feedback processes such as radiation pressure are required. The converse is, of course, also possible: the implemented feedback can also be too efficient, for example because the subgrid model underestimates the actual radiative losses. The risk of misinterpretation is real, because it can be shown that many simulations underestimate the effectiveness of feedback due to excessive radiative losses \citep[e.g.][]{DallaVecchia2012Winds}, which themselves are caused by a lack of resolution and insufficiently realistic modelling of the ISM.

Second, the \emph{ab initio} predictive power of the simulations is currently limited when it comes to the properties of galaxies. If the efficiency of the feedback processes depends on subgrid prescriptions that may not be good approximations to the outcome of unresolved processes, or if the outcome depends on resolution, then the true efficiencies cannot be predicted from first principles. Note that the use of subgrid models does not in itself remove predictive power. If the physical processes that operate below the resolution limit and their connection with the physical conditions on larger scales are fully understood and can be modelled or observed, then it may be possible to create a subgrid model that is sufficiently realistic to retain full predictive power. However, this is currently not the case for feedback from star formation and AGN. As we shall explain below, this implies that simulations that appeal to a subgrid prescription for the generation of outflows are unable to predict the stellar masses of galaxies. Similarly, for galaxies whose evolution is controlled by AGN feedback, such simulations cannot predict the masses of their central BHs. 

To illustrate this, it is helpful to consider a simple model. Let us assume that galaxy evolution is self-regulated, in the sense that galaxies tend to evolve towards a quasi-equilibrium state in which the gas outflow rate balances the difference between the gas inflow rate and the rate at which gas is locked up in stars and BHs. The mean rate of inflow (e.g.\ in the form of cold streams) evolves with redshift and tracks the accretion rate of dark matter onto haloes, which is determined by the cosmological initial conditions. For simplicity, let us further assume that the outflow rate is large compared to the rate at which the gas is locked up. Although our conclusions do not depend on the validity of this last assumption, it simplifies the arguments because it implies that the outflow rate balances the inflow rate, when averaged 
over appropriate length and time scales. Note that the observed low 
efficiency of galaxy formation (see Fig.~\ref{fig:eta} in \S\ref{sec:eta}) suggests that this may actually be a reasonable approximation, particularly for low-mass galaxies. 

This toy model is obviously incorrect in detail. For example, it ignores the re-accretion of matter ejected by winds, the recycling of stellar mass loss, and the interaction of outflows and inflows. However, recent numerical experiments and analytic models provide some support for the general idea \citep[e.g.][]{Finlator2008MZ,Schaye2010OWLS,Booth2010DMHaloesBHs,Dave2012EquilModel,Haas2013OwlsI,Haas2013OwlsII,Feldmann2013SFLaw,Dekel2013ToyModel,Altay2013DLAs,Lilly2013EquilModel,Sanchez2014Review}. This idea in itself is certainly not new and follows from the existence of a feedback loop \citep[e.g.][]{White1991GF}, as can be seen as follows. If the inflow rate exceeds the outflow rate, then the gas fraction will increase and this will in turn increase the star formation rate (and/or, on a smaller scale, the BH accretion rate) and hence also the outflow rate. If, on the other hand, the outflow rate exceeds the inflow rate, then the gas fraction will decrease and this will in turn decrease the star formation rate (and/or the BH accretion rate) and hence also the outflow rate.

In this self-regulated picture of galaxy evolution the outflow rate is determined by the inflow rate. Hence, the outflow rate is \emph{not} determined by the efficiency of the implemented feedback. Therefore, if the outflow is driven by feedback from star formation, then the star formation rate will adjust until the outflow rate balances the inflow rate, irrespective of the (nonzero) feedback efficiency. However, the star formation rate for which this balance is achieved, and hence also ultimately the stellar mass, do depend on the efficiency of the implemented feedback. If the true feedback efficiency cannot be predicted, then neither can the stellar mass. Similarly, if the outflow rate is driven by AGN feedback, then the BH accretion rate will adjust until the outflow rate balances the inflow rate (again averaged over appropriate length and time scales). The BH accretion rate, and hence the BH mass, for which this balance is achieved depend on the efficiency of the implemented feedback, which has to be assumed. According to this toy model, which appears to be a reasonable description of the evolution of simulated galaxies, the stellar and BH masses are thus determined by the efficiencies of the (subgrid) implementations for stellar and AGN feedback, respectively. 

The simulations therefore need to be calibrated to produce the correct stellar and BH masses. Moreover, if the true efficiency varies systematically with the physical conditions on a scale resolved by the simulations, then the implemented subgrid efficiency would also have to be a function of the local physical conditions in order to produce the correct mass functions of galaxies and BHs. 

A similar story applies to the gas fractions of galaxies or, more precisely, for the amount of gas above the assumed star formation threshold, even if the simulations have been calibrated to produce the correct GSMF. We can see this as follows. If the outflow rate is determined by the inflow rate, then it is \emph{not} determined by the assumed subgrid star formation law. Hence, if we modify the star formation law,\footnote{The argument breaks down if the gas consumption time scale becomes longer than the Hubble time.} then the mean outflow rate should remain unchanged. And if the outflow rate remains unchanged, then so must the star formation rate because for a fixed feedback efficiency the star formation rate will adjust to the rate required for outflows to balance inflows. If the star formation rate is independent of the star formation law, then the galaxies must adjust the amount of star-forming gas that they contain when the star formation law is changed. 

Hence, to predict the correct amount of star-forming gas, we need to calibrate the subgrid model for star formation to the observed star formation law.
Fortunately, the star formation law is relatively well characterised observationally on the $\sim 10^2 - 10^3\,\pc$ scales resolved by large-volume simulations, although there are important unanswered questions, e.g.\ regarding the dependence on metallicity. Ultimately the star formation law must be predicted by simulations and will probably depend on the true efficiency of feedback processes within the ISM, but resolving such processes is not yet possible in simulations of cosmological volumes.

It is not obvious how the efficiency of feedback from star formation should be calibrated. We could choose to calibrate to observations of outflow rates relative to star formation rates. However, those outflow rates are highly uncertain and may be affected by AGN feedback. It is also unclear on what scale the outflow rate should be calibrated. In addition, the outflow velocity and the wind mass loading may be individually important. Moreover, unless the interaction of the wind with the circumgalactic medium is modelled correctly and resolved, then obtaining a correct outflow rate on the scale used for the calibration does not necessarily imply that it is also correct for the other scales that matter. 

We choose to calibrate the feedback efficiency using the observed present-day GSMF, as is also common practice for semi-analytic models. We do this mostly because it is relatively well constrained observationally and because obtaining the correct stellar mass - halo mass relation, and hence the correct GSMF if the cosmological initial conditions are known, is a pre-condition for many applications of cosmological simulations. For example, the physical properties of the circumgalactic medium (CGM) are likely sensitive to the halo mass, but because halo mass is difficult to measure, observations and simulations of the CGM are typically compared for galaxies of the same stellar mass. 

One may wonder what the point of hydrodynamical simulations (or, indeed, semi-analytic models) is if they cannot predict stellar masses or BH masses. This is a valid question for which there are several answers. One is that the simulations can still make predictions for observables that were not used for the calibration, and we will present such predictions in \S\ref{sec:otherobs} and in subsequent papers. However, which observables are unrelated is not always unambiguous. One way to proceed, and an excellent way to learn about the physics of galaxy formation, is to run multiple simulations with varying subgrid models. It is particularly useful to have multiple prescriptions calibrated to the same observables. \eagle\ comprises many variations, including several that reproduce the $z\sim 0$ GSMF through different means \citep{Crain2014EagleModels}.

A second answer is that making good use of simulations of galaxy formation does not necessarily mean making quantitative predictions for observables of the galaxy population. We can use the simulations to gain insight into physical processes, to explore possible scenarios, and to make qualitative predictions. How does gas get into galaxies? What factors control the size of galaxies? What is the origin of scatter in galaxy scaling relations? What is the potential effect of outflows on cosmology using weak gravitational lensing or the \lya\ forest? The list of interesting questions is nearly endless. 

A third answer is that cosmological, hydrodynamical simulations can make robust, quantitative predictions for more diffuse components, such as the low-density IGM and perhaps the outer parts of clusters of galaxies.

A fourth answer is that calibrated simulations can be useful to guide the interpretation and planning of observations, as the use of semi-analytic and halo models has clearly demonstrated. In this respect hydrodynamical simulations can provide more detailed information on both the galaxies and their gaseous environments. 

\subsection{Numerical convergence}
\label{sec:conv_discussion}

The need to calibrate the efficiency of the feedback and the associated limits on the predictive power of the simulations call the role of numerical convergence into question. The conventional point of view is that subgrid models should be designed to yield numerically converged predictions. Convergence is clearly a necessary condition for predictive power. However, we have just concluded that current simulations cannot, in any case, make \emph{ab initio} predictions for some of the most fundamental observables of the galaxy population. 

While it is obvious that we should demand convergence for predictions that are relatively robust to the choice of subgrid model, e.g.\ the statistics of the \lya\ forest, it is less obvious that the same is required for observables that depend strongly and directly on the efficiency of the subgrid feedback. One could argue that, instead, we only need convergence after recalibration of the subgrid model. We will call this ``weak convergence'', as opposed to the ``strong convergence'' that is obtained if the results do not change with resolution when the model is held fixed.

If only weak convergence is required, then the demands placed on the subgrid model are much reduced, which has two advantages:

First, we can take better advantage of increases in resolution. The subgrid scale can now move along with the resolution limit, so we can potentially model the physics more faithfully if we adopt higher resolution.

A second advantage of demanding only weak convergence is that we do not have to make the sacrifices that are required to improve the strong convergence and that might have undesirable consequences. We will provide three examples of compromises that are commonly made.

Simulations that sample a representative volume currently lack the resolution and the physics to predict the radiative losses to which outflows are subject within the ISM. Strong convergence can nevertheless be achieved if these losses are somehow removed altogether, for example, by temporarily turning off radiative cooling and calibrating the criterion for switching it back on \citep[e.g.][]{Gerritsen1997PhD,Stinson2006Winds}. However, it is then unclear for which gas the cooling should be switched off. Only the gas elements into which the subgrid feedback was directly injected? Or also the surrounding gas that is subsequently shock-heated? 

Other ways to circumvent radiative losses in the ISM are to generate the outflow outside the galaxy or to turn off the hydrodynamic interaction between the wind and the ISM \citep[e.g.][]{Springel2003Multiphase,Oppenheimer2006Wpot,Oppenheimer2010GSMF,Puchwein2013GSMF,Vogelsberger2013IllustrisModel,Vogelsberger2014Illustris}. This is a valid choice, but one that eliminates the possibility of capturing any aspect of the feedback other than mass loss, such as puffing up of discs, blowing holes, driving turbulence, collimating outflows, ejecting gas clouds, generating small-scale galactic fountains, etc. Furthermore, it necessarily introduces new parameters that control where the outflow is generated and when the hydrodynamics is turned back on. These parameters may directly affect results of interest, including the state of gas around galaxies, and may also re-introduce resolution effects. A potential solution to this problem is to never re-couple and hence to evaluate all wind interactions using a subgrid model, even outside the galaxies, as is done in semi-analytic models.

However, bypassing radiative losses in the ISM is not by itself sufficient to achieve strong convergence. In addition, the feedback must not depend on physical conditions in the ISM since those are unlikely to be converged. Instead, one can make the feedback depend on properties defined by the dark matter, such as its local velocity dispersion or halo mass \citep[e.g.][]{Oppenheimer2006Wpot,Okamoto2010Sats,Oppenheimer2010GSMF,Puchwein2013GSMF,Vogelsberger2013IllustrisModel,Vogelsberger2014Illustris}, which are generally better converged than the properties of the gas. As was the case for turning off cooling or hydrodynamic forces, this choice makes the simulations less ``hydrodynamical'', moving them in the direction of more phenomenological approaches, and it also introduces new problems. How do we treat satellite galaxies given that their subhalo mass and dark matter velocity dispersion are affected by the host halo? Or worse, what about star clusters or tidal dwarf galaxies that are not hosted by dark matter haloes?

In practice, however, the distinction between weak and strong convergence is often unclear. One may surmise that keeping the physical model fixed is equivalent to keeping the code and subgrid parameters fixed (apart from the numerical parameters controlling the resolution), but this is not necessarily the case because of the reliance on subgrid prescriptions and the inability to resolve the first generations of stars and BHs. For typical subgrid prescriptions, the energy, the mass, and the momentum involved in individual feedback events, and the number or intermittency of feedback events do not all remain fixed when the resolution is changed. Any such changes could affect the efficiency of the feedback. Consider, for example, a star-forming region and assume that feedback energy from young stars is distributed locally at every time step. If the resolution is increased, then the time step and the particle mass will become smaller. If the total star formation rate remains the same, then the feedback energy that is injected per time step will be smaller because of the decrease in the time step. If the gas mass also remains the same, then the temperature increase per time step will be smaller. A lower post-feedback temperature often leads to larger thermal losses. If, instead, the subgrid model specifies the temperature jump (or wind velocity), then the post-feedback temperature will remain the same when the resolution is increased, but the number of heating events will increase because the same amount of feedback energy has to be distributed over lower-mass particles. There is no guarantee that more frequent, lower-energy events drive the same outflows as less frequent, higher-energy events. 
 
Moreover, for cosmological initial conditions, higher resolution implies resolving smaller haloes, and hence tracing the progenitors of present-day galaxies to higher redshifts. If these progenitors drive winds, then this may impact the subsequent evolution. 

In \S\ref{sec:gsmf} we investigate both the weak and strong convergence of our simulations, focusing on the GSMF. 
We test the weak convergence for a wide variety of predictions in sections~\ref{sec:cal_obs} and \ref{sec:otherobs}.

\section{Simulations}
\label{sec:simulations}

\begin{table} 
\caption{The cosmological parameters used for the \eagle\ simulations: $\Omega_{\rm m}$,
$\Omega_\Lambda$, and $\Omega_{\rm b}$ are the average densities
of matter, dark energy and baryonic matter
in units of the critical density at redshift zero; $H_0$ is the Hubble
parameter, $\sigma_8$ is the square root of the linear variance of the
matter distribution when smoothed with a top-hat filter of radius
$8~\hcMpc$, $n_{\rm s}$ is the scalar power-law index of the power
spectrum of primordial adiabatic perturbations, and $Y$ is the primordial abundance of helium.}
\begin{center}
\begin{tabular}{|l|r|}
\hline
      Cosmological parameter  &            Value \\
\hline
   $\Omega_{\rm m}$             &    0.307\phantom{00} \\
   $\Omega_\Lambda$             &    0.693\phantom{00} \\
   $\Omega_{\rm b}$             &    0.04825\phantom{} \\ 
   $h \equiv H_0$/(100 km\,s$^{-1}$\,Mpc$^{-1})$ &  0.6777\phantom{0} \\
   $\sigma_8$                  &    0.8288\phantom{0} \\
   $ n_{\rm s}$                 &    0.9611\phantom{0} \\
   $Y$                         &    0.248\phantom{00}\\
\hline
\end{tabular}
\end{center}
\label{tbl:cosmo_params}
\end{table}

\eagle\ was run using a modified version of the $N$-Body Tree-PM smoothed particle hydrodynamics (SPH) code \textsc{gadget}~3, which was last described in \citet{Springel2005Gadget2}. The main modifications are the formulation of SPH, the time stepping and, most importantly, the subgrid physics.

The subgrid physics used in \eagle\ is based on that developed for \owls\ \citep{Schaye2010OWLS}, and used also in \gimic\ \citep{Crain2009GIMIC} and \cosmo\ \citep{LeBrun2014CosmoOWLS}. We include element-by-element radiative cooling for 11 elements, star formation, stellar mass loss, energy feedback from star formation, gas accretion onto and mergers of supermassive black holes (BHs), and AGN feedback. As we will detail in \S\ref{sec:subgrid}, we made a number of changes with respect to \owls. The most important changes concern the implementations of energy feedback from star formation (which is now thermal rather than kinetic), the accretion of gas onto BHs (which now accounts for angular momentum), and the star formation law (which now depends on metallicity). 

In the simulations presented here the amount of feedback energy that is injected per unit stellar mass decreases with the metallicity and increases with the gas density. It is bounded between one third and three times the energy provided by supernovae and, on average, it is about equal to that amount. The metallicity dependence is motivated by the fact that we expect greater (unresolved) thermal losses when the metallicity exceeds $\sim 10^{-1}\,{\rm Z}_\odot$, the value for which metal-line cooling becomes important. The density dependence compensates for spurious, numerical radiative losses which, as expected, are still present at our resolution even though they are greatly reduced by the use of the stochastic prescription of \citet{DallaVecchia2012Winds}. The simulations were calibrated against observational data by running a series of high-resolution 12.5~cMpc and intermediate resolution 25~cMpc test runs with somewhat different dependencies on metallicity and particularly density. From the models that predicted reasonable physical sizes for disc galaxies, we selected the one that best fit the $z\sim 0$ GSMF. For more details on the subgrid model for energy feedback from star formation we refer the reader to \S\ref{sec:snii}. 

As described in more detail in Appendix~\ref{app:hydro}, we make use of the conservative pressure-entropy formulation of SPH derived by \citet{Hopkins:2013lr}, the artificial viscosity switch from \citet{Cullen:2010qy}, an artificial conduction switch similar to that of \citet{Price:2008kx}, the $C^2$ \citet{Wendland:1995} kernel and the time step limiters of \citet{Durier:2012fj}. We will refer to these numerical methods collectively as ``Anarchy''. Anarchy will be described in more detail by Dalla Vecchia (in preparation), who also demonstrates its good performance on standard hydrodynamical tests (see \citealt{Hu2014SPHGal} for tests of a similar set of methods). In \citet{Schaller2014EagleSPH} we will show the relevance of the new hydrodynamical techniques and time stepping scheme for the results of the \eagle\ simulations. Although the Anarchy implementation yields dramatic improvements in the performance on some standard hydrodynamical tests as compared to the original implementation of the hydrodynamics in \textsc{gadget}~3, we generally find that the impact on the results of the cosmological simulations is small compared to those resulting from reasonable variations in the subgrid physics (see also \citealt{Scannapieco2012Aquila}).

\begin{table*} 
\begin{center}
\caption{Box sizes and resolutions of the main \eagle\ simulations.  From
  left-to-right the columns show: simulation name suffix; comoving box size;
  number of dark matter particles (there is initially an equal number of baryonic particles); initial baryonic particle mass; dark matter
  particle mass; comoving, Plummer-equivalent gravitational
  softening length; maximum proper softening length.} 
\label{tbl:sims}
\begin{tabular}{lrrrrrrr}
\hline
Name & $L$ & $N$ & $m_{\rm g}$ & $m_{\rm dm}$ &
$\epsilon_{\rm com}$ & $\epsilon_{\rm prop}$ \\  
& (comoving Mpc) & & ($\Msun$) & ($\Msun$) & (comoving kpc) & (proper kpc)\\
\hline 
L025N0376 &  25 & $376^3$ & $1.81\times 10^6$ & $9.70\times 10^6$ & 2.66 &0.70\\
L025N0752 &  25 & $752^3$ & $2.26\times 10^5$ & $1.21\times 10^6$ & 1.33 & 0.35\\
L050N0752 &  50 & $752^3$ & $1.81\times 10^6$ & $9.70\times 10^6$ & 2.66 &0.70\\
L100N1504 & 100 & $1504^3$ & $1.81\times 10^6$ & $9.70\times 10^6$ & 2.66 &0.70\\
\hline
\end{tabular}
\end{center}
\end{table*}

The values of the cosmological parameters used for the \eagle\ simulations are taken from the most recent Planck results \citep[][Table 9]{PlanckI} and are listed in Table~\ref{tbl:cosmo_params}. A transfer function with these
parameters was generated using CAMB \citep[version Jan\_12]{CAMB}.
The linear matter power spectrum was generated by multiplying a
power-law primordial power spectrum with an index of $n_{\rm s} = 0.9611$ by the
square of the dark matter transfer function evaluated at redshift
zero\footnote{The CAMB input parameter file and the linear power spectrum are available at \url{http://eagle.strw.leidenuniv.nl/}.}. Particles arranged in a glass-like initial configuration were displaced according to 2nd-order Lagrangian perturbation theory using the method of \citet{Jenkins20102lpt} and the public Gaussian white noise field \emph{Panphasia} \citep{Jenkins2013ICs,Jenkins2013Panphasia}. The methods used to generate the initial conditions are described in detail in Appendix~\ref{app:ics}.

\begin{figure*} 
\resizebox{\textwidth}{!}{\includegraphics{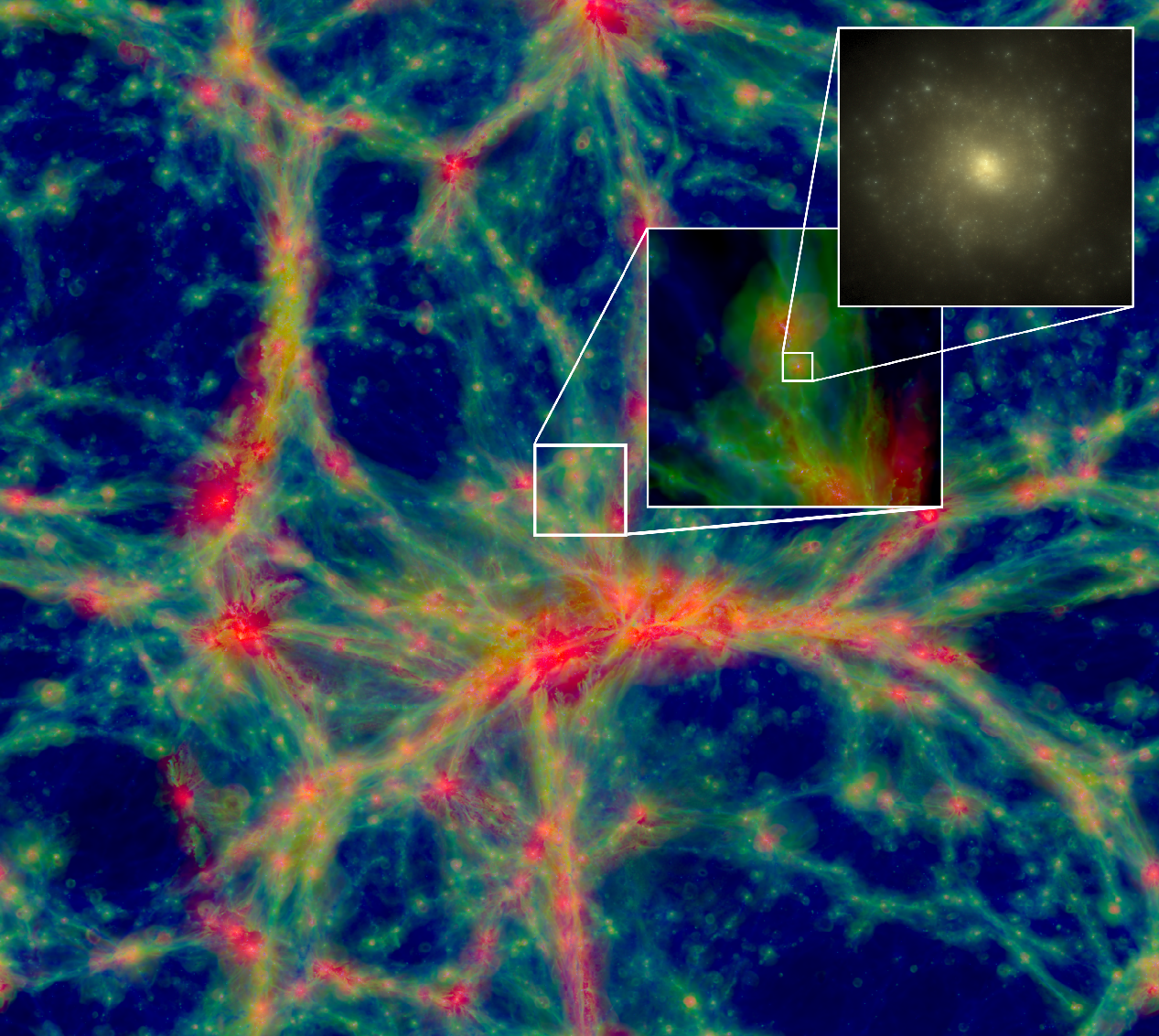}} 
\caption{A $100\times 100\times 20$ cMpc slice through the Ref-L100N1504 simulation at $z=0$. The intensity shows the gas density while the colour encodes the gas temperature using different colour channels for gas with $T<10^{4.5}\,\K$ (blue), $10^{4.5}\,\K<T<10^{5.5}\,\K$ (green), and $T>10^{5.5}\,\K$ (red). The insets show regions of 10 cMpc and 60 ckpc on a side and zoom into an individual galaxy with a stellar mass of $3\times 10^{10}\,\Msun$. The 60 ckpc image shows the stellar light based on monochromatic u, g and r band SDSS filter means and accounting for dust extinction. It was created using the radiative transfer code \skirt\ \citep{Baes2011SKIRT}.}
\label{fig:zoom} 
\end{figure*}

\begin{figure*} 
\resizebox{\textwidth}{!}{\includegraphics{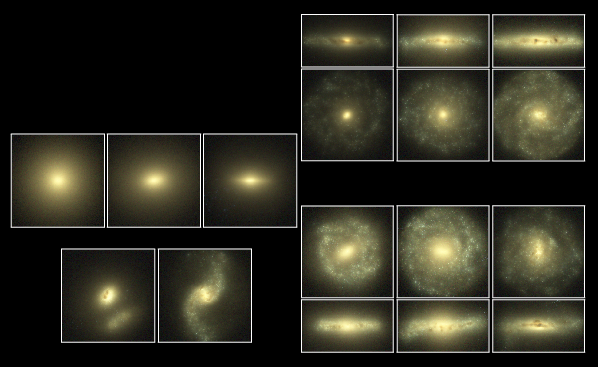}} 
\caption{Examples of galaxies taken from simulation Ref-L100N1504 illustrating the $z=0$ Hubble sequence of galaxy morphologies. The images were created with the radiative transfer code \skirt\ \citep{Baes2011SKIRT}. They show the stellar light based on monochromatic u, g and r band SDSS filter means and accounting for dust extinction. Each image is 60 ckpc on a side. For disc galaxies both face-on and edge-on projections are shown. Except for the 3rd elliptical from the left, which has a stellar mass of $1\times 10^{11}\,\Msun$, and the merger in the bottom-left, which has a total stellar mass of $8\times 10^{10}\,\Msun$, all galaxies shown have stellar masses of 5--$6\times 10^{10}\,\Msun$.} 
\label{fig:morphology} 
\end{figure*}

Table~\ref{tbl:sims} lists box sizes and resolutions of the main \eagle\ simulations. All simulations were run to redshift $z=0$. Note that contrary to convention, box sizes, particles masses and gravitational softening lengths are \emph{not} quoted in units of $h^{-1}$. The gravitational softening was kept fixed in comoving units down to $z=2.8$ and in proper units thereafter. We will refer to simulations with the same mass and spatial resolution as L100N1504 as intermediate resolution runs and to simulations with the same resolution as L025N0752 as high-resolution runs. 

Particle properties were recorded for 29 snapshots between redshifts 20 and 0. In addition, we saved a reduced set of particle properties (``snipshots'') at 400 redshifts between 20 and 0. The largest simulation, L100N1504, took about 4.5~M CPU hours to reach $z=0$ on a machine with 32~TB of memory, with the \eagle\ subgrid physics typically taking less than 25 per cent of the CPU time. 

The resolution of \eagle\ suffices to marginally resolve the Jeans scales in the warm ISM. The Jeans mass and length for a cloud with gas fraction, $f_{\rm g}$, are, respectively, $M_{\rm J} \approx 1 \times 10^7\,\Msun\, f_{\rm g}^{3/2} (n_{\rm H} /10^{-1}\,\cm^{-3})^{-1/2} (T / 10^4\,\K)^{3/2}$ and $L_{\rm J} \approx 2~\kpc ~f_{\rm g}^{1/2} (n_{\rm H} / 10^{-1}\,\cm^{-3})^{-1/2} (T / 10^4\,\K)^{1/2}$, where $n_{\rm H}$ and $T$ are the total hydrogen number density and the temperature, respectively. These Jeans scales can be compared to the gas particle masses and maximum proper gravitational softening lengths listed in columns 4 and 7 of Table~\ref{tbl:sims}.

Simulations with the same subgrid physics and numerical techniques as used for L100N1504 were carried out for all box sizes (12.5 -- 100 cMpc) and particles numbers ($188^3$ -- $1504^3$). We will refer to this physical model as the reference model and will indicate the corresponding simulations with the prefix ``Ref-'' (e.g.\ Ref-L100N1504). As detailed in \S\ref{sec:subgrid}, we re-ran the high-resolution simulations with recalibrated parameter values for the subgrid stellar and AGN feedback to improve the match to the observed $z\sim 0$ GSMF. We will use the prefix ``Recal-'' when referring to the simulations with this alternative set of subgrid parameters (e.g.\ Recal-L025N0752). Note that in terms of weak convergence, Ref-L100N1504 is more similar to model Recal-L025N0752 than to model Ref-L025N0752 (see \S\ref{sec:conv_discussion} for a discussion of weak and strong convergence). In addition, we repeated the L050N0752 run with adjusted AGN parameters in order to further improve the agreement with observations for high-mass galaxies. We will refer to this model with the prefix ``AGNdT9''. Table~\ref{tbl:subgridpars} summarizes the values of the four subgrid parameters that vary between the models presented here. \citet{Crain2014EagleModels} and \citet{Schaller2014EagleSPH} will present the remaining \eagle\ simulations, which concern variations in the subgrid physics and the numerical techniques, respectively. Finally, \citet{Sawala2014EagleZooms} present very high-resolution zoomed simulations of Local Group like systems run with the \eagle\ code and a physical model that is nearly identical to the one used for the Ref-L100N1504 model described here.

\begin{table} 
\begin{center}
\caption{Values of the subgrid parameters that vary between the models presented here. The parameters $n_{\rm H,0}$ and $n_n$ control, respectively, the characteristic density and the power-law slope of the density dependence of the energy feedback from star formation (see equation \ref{eq:f(Z,n)} in \S\ref{sec:calibration}). The parameter $C_{\rm visc}$ controls the sensitivity of the BH accretion rate to the angular momentum of the gas (see equation \ref{eq:mdotaccr} in \S\ref{sec:bh_accretion}) and $\Delta T_{\rm AGN}$ is the temperature increase of the gas during AGN feedback (see \S\ref{sec:AGNfeedback}). } 
\label{tbl:subgridpars}
\begin{tabular}{lrrrl}
\hline
Prefix & $n_{\rm H,0}$ & $n_n$ & $C_{\rm visc}$ & $\Delta T_{\rm AGN}$ \\  
& ($\cm^{-3}$) & & & (K) \\
\hline 
Ref    & 0.67 & $2/\ln10$ & $2\pi$ & $10^{8.5}$ \\
Recal  & 0.25 & $1/\ln10$ & $2\pi \times 10^3$ & $10^9$ \\
AGNdT9 & 0.67 & $2/\ln10$ & $2\pi \times 10^2$ & $10^9$ \\
\hline
\end{tabular}
\end{center}
\end{table}

Figure~\ref{fig:zoom} illustrates the large dynamic range of \eagle. It shows the large-scale gas distribution in a thick slice through the $z=0$ output of the Ref-L100N1504 run, colour-coded by the gas temperature. The insets zoom in on an individual galaxy. The first zoom shows the gas, but the last zoom shows the stellar light after accounting for dust extinction. This image was created using three monochromatic radiative transfer simulations with the code \skirt\ \citep{Baes2011SKIRT} at the effective wavelengths of the Sloan Digital Sky Survey (SDSS) u, g \& r filters. Dust extinction is implemented using the metal distribution predicted by the simulations and assuming that 30 per cent of the metal mass is locked up in dust grains. Only material within a spherical aperture with a radius of 30 pkpc is included in the radiative transfer calculation. More examples of \skirt\ images of galaxies are shown in Figure~\ref{fig:morphology}, in the form of a Hubble sequence. This figure illustrates the wide range of morphologies present in \eagle. Note that \citet{Vogelsberger2014IllustrisNature} showed a similar figure for their Illustris simulation. In future work we will investigate how morphology correlates with other galaxy properties. More images, as well as videos, can be found on the \eagle\ web sites at Leiden, \url{http://eagle.strw.leidenuniv.nl/}, and Durham, \url{http://icc.dur.ac.uk/Eagle/}.

We define galaxies as gravitationally bound subhaloes identified by the \textsc{subfind} algorithm \citep{Springel2001Subfind,Dolag2009Substructure}. The procedure consists of three main steps. First we find haloes by running the Friends-of-Friends (FoF; \citealt{Davis1985FoF}) algorithm on the dark matter particles with linking length 0.2 times the mean interparticle separation. Gas and star particles are assigned to the same, if any, FoF halo as their nearest dark matter particles. Second, \textsc{subfind} defines substructure candidates by identifying overdense regions within the FoF halo that are bounded by saddle points in the density distribution. Note that whereas FoF considers only dark matter particles, \textsc{subfind} uses all particle types within the FoF halo. Third, particles that are not gravitationally bound to the substructure are removed and the resulting substructures are referred to as subhaloes. Finally, we merged subhaloes separated by less than the minimum of 3~pkpc and the stellar half-mass radius. This last step removes a very small number of very low-mass subhaloes whose mass is dominated by a single particle such as a supermassive BH.

For each FoF halo we define the subhalo that contains the particle with the lowest value of the gravitational potential to be the central galaxy while any remaining subhaloes are classified as satellite galaxies. The position of each galaxy is defined to be the location of the particle belonging to the subhalo for which the gravitational potential is minimum.  

The stellar mass of a galaxy is defined to be the sum of the masses of all star particles that belong to the corresponding subhalo and that are within a 3-D aperture with radius 30~pkpc. Unless stated otherwise, other galaxy properties, such as the star formation rate, metallicity, and half-mass radius, are also computed using only particles within the 3-D aperture.  
In \S\ref{sec:aperture} we show that this aperture gives a nearly identical GSMF as the 2-D Petrosian apertures that are frequently used in observational studies.

We find the effect of the aperture to be negligible for $M_\ast < 10^{11}\,\Msun$ for all galaxy properties that we consider. However, for more massive galaxies the aperture reduces the stellar masses somewhat by cutting out intracluster light. For example, at a stellar mass $M_\ast = 10^{11}\,\Msun$ as measured using a 30 pkpc aperture, the median subhalo stellar mass is 0.1~dex higher (see \S\ref{sec:aperture} for the effect on the GSMF). Without the aperture, metallicities are slightly lower and half-mass radii are slightly larger for $M_\ast > 10^{11}\,\Msun$, but the effect on the star formation rate is negligible. 

\section{Subgrid physics}
\label{sec:subgrid}

In this section we provide a thorough description and motivation for the subgrid physics implemented in \eagle: radiative cooling (\S\ref{sec:cooling}), reionisation (\S\ref{sec:reionisation}), star formation (\S\ref{sec:sf}), stellar mass loss and metal enrichment (\S\ref{sec:chemo}), energy feedback from star formation (\S\ref{sec:snii}), and supermassive black holes and AGN feedback (\S\ref{sec:BHs}). These subsections can be read separately. Readers who are mainly interested in the results may skip this section. 

\subsection{Radiative cooling}
\label{sec:cooling}

Radiative cooling and photoheating are implemented element-by-element following \citet{Wiersma2009Cooling}, including all 11 elements that they found to be important: H, He, C, N, O, Ne, Mg, Si, S, Ca, and Fe. \citet{Wiersma2009Cooling} used \textsc{cloudy} version\footnote{Note that \owls\ used tables based on version 05.07.} 07.02 \citep{Ferland1998Cloudy} to tabulate the rates as a function of density, temperature, and redshift assuming the gas to be in ionisation equilibrium and exposed to the cosmic microwave background (CMB) and the \citet{Haardt2001UVB} model for the evolving UV/X-ray background from galaxies and quasars. By computing the rates element-by-element, we account not only for variations in the metallicity, but also for variations in the relative abundances of the elements. 

We caution that our assumption of ionisation equilibrium and the neglect of local sources of ionizing radiation may cause us to overestimate the cooling rate in certain situations, e.g.\ in gas that is cooling rapidly \cite[e.g.][]{Oppenheimer2013Nonequil} or that has recently been exposed to radiation from a local AGN \citep{Oppenheimer2013AGNFossils}. 

We have also chosen to ignore self-shielding, which may cause us to underestimate the cooling rates in dense gas. While we could have accounted for this effect, e.g.\ using the fitting formula of \citet{Rahmati2013HICDDF}, we opted against doing so because there are other complicating factors. Self-shielding is only expected to play a role for $n_{\rm H} > 10^{-2}\,\cm^{-3}$ and $T\la 10^4\,\K$ \citep[e.g.][]{Rahmati2013HICDDF}, but at such high densities the radiation from local stellar sources, which we neglect here, is expected to be at least as important as the background radiation \citep[e.g.][]{Schaye2001MaxHI,Rahmati2013LocalSources}. 

\subsection{Reionization}
\label{sec:reionisation}

Hydrogen reionization is implemented by turning on the time-dependent,  spatially-uniform ionizing background from \citet{Haardt2001UVB}. This is done at redshift $z=11.5$, consistent with the optical depth measurements from \citet{PlanckI}. At higher redshifts we use net cooling rates for gas exposed to the CMB and the photo-dissociating background obtained by cutting the $z = 9$ \citet{Haardt2001UVB} spectrum above 1~Ryd. 

To account for the boost in the photoheating rates during reionization relative to the optically thin rates assumed here, we inject 2~eV per proton mass. This ensures that the photoionised gas is quickly heated to $\sim 10^4\,\K$. For H this is done instantaneously, but for \HeII\ the extra heat is distributed in redshift with a Gaussian centred on $z=3.5$ of width $\sigma(z)=0.5$. \citet{Wiersma2009Chemo} showed that this choice results in broad agreement with the thermal history of the intergalactic gas as measured by \citet{Schaye2000IGMTemp}. 

\subsection{Star formation}
\label{sec:sf}

Star formation is implemented following \citet{Schaye2008SF}, but with the metallicity-dependent density threshold of \citet{Schaye2004SF} and a different temperature threshold, as detailed below. Contrary to standard practice, we take the star formation rate to depend on pressure rather than density. As demonstrated by \citet{Schaye2008SF}, this has two important advantages. First, under the assumption that the gas is self-gravitating, we can rewrite the observed Kennicutt-Schmidt star formation law \citep{Kennicutt1998Law}, $\dot{\Sigma}_\ast = A (\Sigma_{\rm g}/1~\Msun\,\pc^{-2})^n$, as a pressure law: 
\begin{equation}
\dot{m}_\ast = m_{\rm g} A \left (1~\Msun\,\pc^{-2}\right
)^{-n} \left ({\gamma \over G} f_{\rm g} P\right )^{(n-1)/2},
\label{eq:sflaw}
\end{equation}
where $m_{\rm g}$ is the gas particle mass, $\gamma=5/3$ is the
ratio of specific heats, $G$ is the gravitational constant, $f_{\rm g}$ is the mass fraction in gas (assumed to be unity), and $P$ is the total pressure. Hence, the free parameters $A$ and $n$ are determined by observations of the gas and star formation rate surface densities of galaxies and no tuning is necessary. Second, if we impose an equation of state, $P=P_{\rm eos}(\rho)$, then the observed Kennicutt-Schmidt star formation law will still be reproduced without having to change the star formation parameters. In contrast, if star formation is implemented using a volume density rather than a pressure law, then the predicted Kennicutt-Schmidt law will depend on the thickness of the disc and thus on the equation of state of the star forming gas. Hence, in that case the star formation law not only has to be calibrated, it has to be recalibrated if the imposed equation of state is changed. In practice, this is rarely done.

Equation (\ref{eq:sflaw}) is implemented stochastically. The probability that a gas particle is converted into a collisionless star particle during a time step $\Delta t$ is $\min(\dot{m}_\ast \Delta t/m_{\rm g},1)$.

We use $A=1.515\times10^{-4}~\Msolyrkpcsq$ and $n=1.4$, where we have decreased the amplitude by a factor 1.65 relative to the value used by \citet{Kennicutt1998Law} because we use a Chabrier rather than a Salpeter stellar initial mass function (IMF). We increase $n$ to 2 for $n_{\rm H} > 10^3\,\cm^{-3}$, because there is some evidence for a steepening at high densities \citep[e.g.][]{Liu2011KSLaw,Genzel2010SFLaw}, but this does not have a significant effect on the results since only $\sim 1$\% of the stars form at such high densities in our simulations. 

Star formation is observed to occur in cold ($T\ll 10^4\,\K$), molecular gas. Because simulations of large cosmological volumes, such as ours, lack the resolution and the physics to model the cold, interstellar gas phase, it is appropriate to impose a star formation threshold at the density above which a cold phase is expected to form. In \owls\ we used a constant threshold of $n_{\rm H}^* = 10^{-1}\,\cm^{-3}$, which was motivated by theoretical considerations and yields a critical gas surface density $\sim 10~\Msun\,\pc^{-2}$ \citep{Schaye2004SF,Schaye2008SF}. The critical volume density, $n_{\rm H} = 0.1~\cm^{-3}$, is also similar to the value used in other work of comparable resolution \citep[e.g.][]{Springel2003Multiphase,Vogelsberger2013IllustrisModel}. Here we instead use the metallicity-dependent density threshold of \citet{Schaye2004SF} as implemented in \owls\ model ``SFTHRESZ'' (eq.\ 4 of \citealt{Schaye2010OWLS}; equations 19 and 24 of
\citealt{Schaye2004SF}), 
\begin{equation} 
\label{eq:sfthresz} 
n_{\rm H}^*(Z)=10^{-1}\,\cm^{-3} \left ({Z \over 0.002}\right )^{-0.64}, 
\end{equation} 
where $Z$ is the gas metallicity (i.e.\ the fraction of the gas mass in elements heavier than helium). In the code the threshold is evaluated as a mass density rather than a total hydrogen number density. To prevent an additional dependence on the hydrogen mass fraction (beyond that implied by equation \ref{eq:sfthresz}), we convert $n_{\rm H}$ into a mass density assuming the initial hydrogen mass fraction, $X=0.752$. Because the \citet{Schaye2004SF} relation diverges at low metallicities, we impose an upper limit of $n_{\rm H}^*=10~\cm^{-3}$. To prevent star formation in low overdensity gas at very high redshift, we also require the gas density to exceed 57.7 times the cosmic mean, but the results are insensitive to this value. 

The metallicity dependence accounts for the fact that the transition from a warm, neutral to a cold, molecular phase occurs at lower densities and pressures if the metallicity, and hence also the dust-to-gas ratio, is higher. The phase transition shifts to lower pressures if the metallicity is increased due to the higher formation rate of molecular hydrogen, the increased cooling due to metals and the increased shielding by dust \citep[e.g.][]{Schaye2001MaxHI,Schaye2004SF,Pelupessy2006H2,Krumholz2008HIH2,Gnedin2009H2form,Richings2014Shielding}. Our metallicity-dependent density threshold causes the critical gas surface density below which the Kennicutt-Schmidt law steepens to decrease with increasing metallicity. 

Because our simulations do not model the cold gas phase, we impose a temperature floor, $T_{\rm eos}(\rho_{\rm g})$, corresponding to the equation of state $P_{\rm eos}\propto \rho_{\rm g}^{4/3}$, normalised to\footnote{For the purpose of imposing temperature floors, $T_{\rm eos}(\rho_{\rm g})$ is converted into an entropy assuming a fixed mean molecular weight of 1.2285, which corresponds to an atomic, primordial gas. Other conversions in the code use the actual mean molecular weight and hydrogen abundance, but we keep them fixed here to prevent particles with different abundances from following different effective equations of state.} $T_{\rm eos} = 8\times 10^3\,\K$ at $n_{\rm H} = 10^{-1}\,\cm^{-3}$, a temperature that is typical for the warm ISM \citep[e.g.][]{Richings2014Shielding}. The slope of $4/3$ guarantees that the Jeans mass, and the ratio of the Jeans length to the SPH kernel, are independent of the density, which prevents spurious fragmentation due to the finite resolution \citep{Schaye2008SF,Robertson2008SFlaw}. Following \citet{DallaVecchia2012Winds}, gas is eligible to form stars if $\log_{10} T < \log_{10}T_{\rm eos} + 0.5$ and $n_{\rm H} > n_{\rm H}^*$, where $n_{\rm H}^*$ depends on metallicity as specified above.

Because of the existence of a temperature floor, the temperature of star forming (i.e.\ interstellar) gas in the simulation merely reflects the effective pressure imposed on the unresolved, multiphase ISM, which may in reality be dominated by turbulent rather than thermal pressure. If the temperature of this gas needs to be specified, e.g.\ when computing neutral hydrogen fractions in post-processing, then one should assume a value based on physical considerations rather than use the formal simulation temperatures at face value. 

In addition to the minimum pressure corresponding to the equation of state with slope $4/3$, we impose a temperature floor of 8000~K for densities $n_{\rm H}>10^{-5}\,\cm^{-3}$ in order to prevent very metal-rich particles from cooling to temperatures characteristic of cold, interstellar gas. This constant temperature floor was not used in \owls\ and is unimportant for our results. We impose it because we do not wish to include a cold interstellar phase since we do not model all the  physical processes that are needed to describe it. We only impose this limit for densities $n_{\rm H}>10^{-5}\,\cm^{-3}$, because we should not prevent the existence of cold, adiabatically cooled, intergalactic gas, which our algorithms can model accurately. 

\subsection{Stellar mass loss and type Ia supernovae}
\label{sec:chemo}

Star particles are treated as simple stellar populations (SSPs) with a \citet{Chabrier2003IMF} IMF in the range $0.1-100~\Msun$. The implementation of stellar mass loss is based on  \citet{Wiersma2009Chemo}. At each time step\footnote{To reduce the computational cost associated with neighbour finding for stars, we implement the enrichment every 10 gravitational time steps for star particles older than 0.1 Gyr; for the high-resolution run, Recal-L025N0752, this is further reduced to once every 100 time steps for star particles older than 1 Gyr. We have verified that our results are unaffected by this reduction in the sampling of stellar mass loss from older SSPs.} and for each stellar particle, we compute which stellar masses reach the end of the main sequence phase using the metallicity-dependent lifetimes of \citet{Portinari1998Chemo}. The fraction of the initial particle mass reaching this evolutionary stage is used, together with the initial elemental abundances, to compute the mass of each element that is lost through winds from AGB stars, winds from massive stars, and core collapse supernovae using the nucleosynthetic yields from
\citet{Marigo2001AGByields} and \citet{Portinari1998Chemo}. The elements H, He, C, N, O, Ne, Mg, Si, and Fe are tracked individually, while for Ca and S we assume fixed mass ratios relative to Si of 0.094 and 0.605, respectively \citep{Wiersma2009Chemo}. In addition, we compute the mass and energy lost through supernovae of type Ia.

The mass lost by star particles is distributed among the neighbouring SPH particles using the SPH kernel, but setting the mass of the gas particles equal to the constant initial value, $m_{\rm g}$. Each SPH neighbour $k$ that is separated by a distance $r_k$ from a star particle with smoothing length $h$ then receives a fraction $\frac{m_{\rm g}}{\rho_k} W(r_k,h)/\Sigma_i \frac{m_{\rm g}}{\rho_i}W(r_i,h)$ of the mass lost during the time step, where $W$ is the SPH kernel and the sum is over all SPH neighbours. To speed up the calculation, we use only 48 neighbours for stellar mass loss rather than the 58 neighbours used for the SPH.

In \citet{Wiersma2009Chemo} and \owls\ we used the current gas particle masses rather than the constant, initial gas particle mass when computing the weights. The problem with that approach is that gas particles that are more massive than their neighbours, due to having received more mass lost by stars, carry more weight and therefore become even more massive relative to their neighbours. We found that this runaway process can cause a very small fraction of particles to end up with masses that far exceed the initial particle mass. The fraction of very massive particles is always small, because massive particles are typically also metal rich and relatively quickly converted into star particles. Nevertheless, it is still undesirable to preferentially direct the lost mass to relatively massive gas particles. We therefore removed this bias by using the fixed initial particle mass rather than the current particle mass, effectively taking the dependence on gas particle mass out of the equation for the distribution of stellar mass loss. 

We also account for the transfer of momentum and energy associated with the transfer of mass from star to gas particles. We refer here to the momentum and energy related to the difference in velocity between the star particle and the receiving gas particles, in addition to that associated with the mass loss process itself (e.g.\ winds or supernovae). We assume that winds from AGB stars have a velocity of $10~\kms$ \citep{Bergeat2005AGBMassLoss}. After adjusting the velocities of the receiving gas particles to conserve momentum, energy conservation is achieved by adjusting their entropies. Momentum and energy transfer may, for example, play a role if the differential velocity between the stellar and gas components is similar to or greater than the sound speed of the gas, although we should keep in mind that the change in the mass of a gas particle during a cooling time is typically small.  

As in \citet{Wiersma2009Chemo}, the abundances used to evaluate the radiative cooling rates are computed as the ratio of the mass density of an element to the total gas density, where both are calculated using the SPH formalism. Star particles inherit their parent gas particles' kernel-smoothed abundances\footnote{Note that this implies that metal mass is only approximately conserved. However, \citet{Wiersma2009Chemo} demonstrated that the error in the total metal mass is negligible even for simulations that are much smaller than \eagle.} and we use those to compute their lifetimes and yields. The use of SPH-smoothed abundances, rather than the mass fractions of the elements stored in each particle, is consistent with the SPH formalism. It helps to alleviate the symptoms of the lack of metal mixing that occurs when metals are fixed to particles. However, as discussed in \citet{Wiersma2009Chemo}, it does not solve the problem that SPH may underestimate metal mixing. The implementation of diffusion can be used to increase the mixing \citep[e.g.][]{Greif2009Mixing,Shen2010IGMEnrichment}, but we have opted not to do this because the effective diffusion coefficients that are appropriate for the ISM and IGM remain unknown. 

\begin{figure} 
\resizebox{\colwidth}{!}{\includegraphics{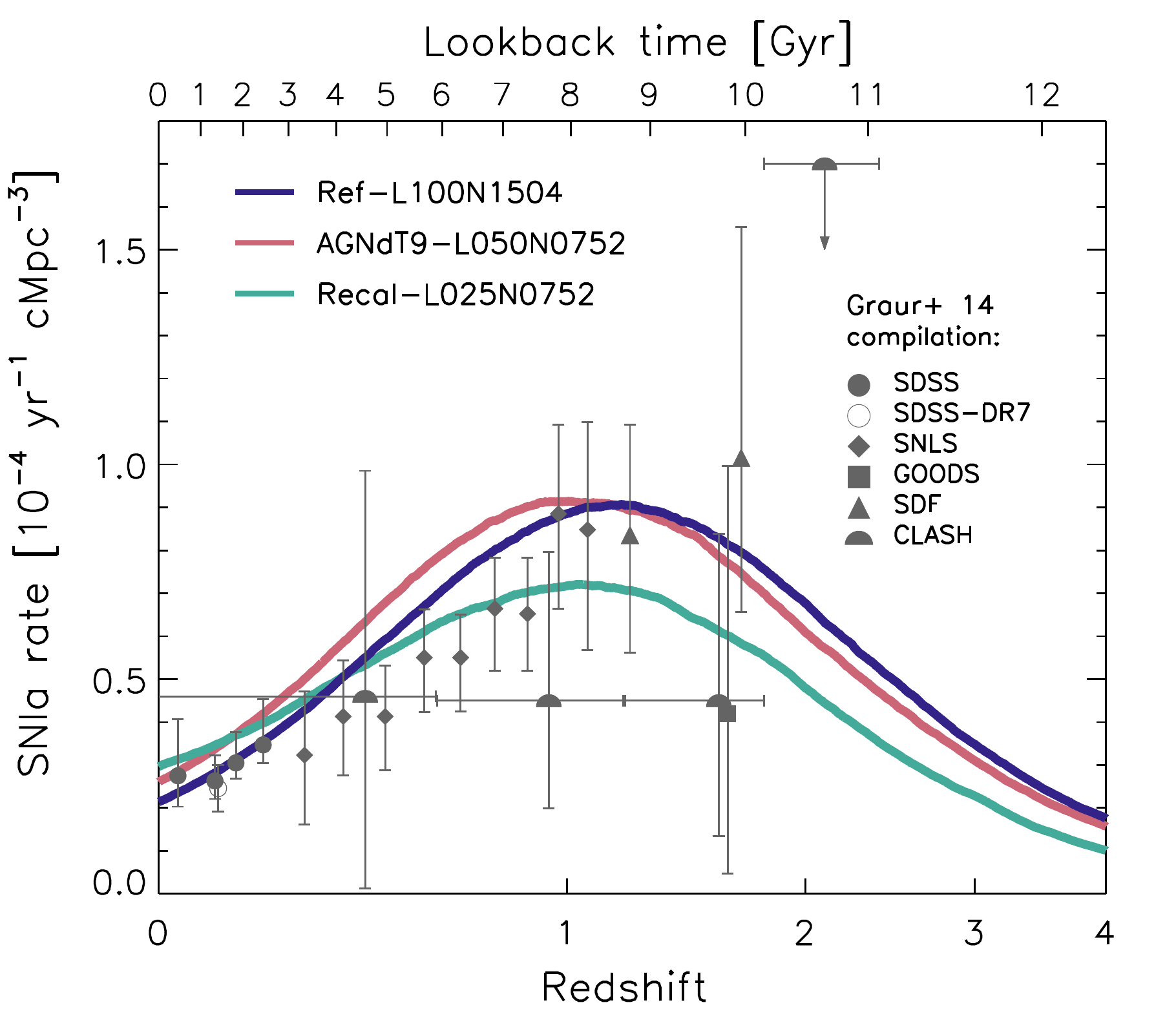}} 
\caption{The evolution of the supernova Ia rate density. Data points show observations from SDSS Stripe 82 \citep{Dilday2010SNIa}, SDSS-DR7 \citep{Graur2013SNIa}, SNLS \citep{Perrett2012SNIa}, GOODS \citep{Dahlen2008SNIa}, SDF \citep{Graur2011SNIa}, and CLASH \citep{Graur2014SNIa}, as compiled by \citet{Graur2014SNIa}. Only data classified by \citealt{Graur2014SNIa} as the ``most accurate and precise measurements'' are shown. The $1\sigma$ error bars account for both statistical and systematic uncertainties. The simulations assume that the rate is a convolution of the star formation rate density with an exponential delay time distribution (eq.~\ref{eq:snia}) with e-folding time $\tau = 2~$Gyr, normalised to yield $\nu = 2\times 10^{-3}\,\Msun^{-1}$ supernovae Ia per unit stellar mass when integrated over all time.}
\label{fig:snia} 
\end{figure}

The rate of supernovae of type Ia (SNIa) per unit initial stellar mass is given by, 
\begin{equation}
\dot{N}_{\rm SNIa} = \nu \frac{e^{-t/\tau}}{\tau},
\label{eq:snia}
\end{equation}
where $\nu$ is the total number of SNIa per unit initial stellar mass and $\exp(-t/\tau)/\tau$ is a normalised, empirical delay time distribution function. 
We set $\tau = 2$~Gyr and $\nu = 2\times 10^{-3}\,\Msun^{-1}$. Figure~\ref{fig:snia} shows that these choices yield broad agreement with the observed evolution of the SNIa rate density for the intermediate resolution simulations, although the AGNdT9-L050N0752 may overestimate the rate by $\sim 30$ per cent for lookback times of 4--7 Gyr. The high-resolution model, Recal-L025N0752, is consistent with the observations at all times. 

At each time step for which the mass loss is evaluated, star particles transfer the mass and energy associated with SNIa ejecta to their neighbours. We use the SNIa yields of the W7 model of \citet{Thielemann2003SNIayields}. Energy feedback from SNIa is implemented identically as for prompt stellar feedback using the stochastic thermal feedback model of \citet{DallaVecchia2012Winds} summarized in \S\ref{sec:snii}, using $\Delta T = 10^{7.5}\,\K$ and $10^{51}\,\erg$ per SNIa. 

\subsection{Energy feedback from star formation}
\label{sec:snii}

Stars can inject energy and momentum into the ISM through stellar winds, radiation, and supernovae. These processes are particularly important for massive and hence short-lived stars. If star formation is sufficiently vigorous, the associated feedback can drive large-scale galactic outflows \citep[e.g.][]{Veilleux2005WindsReview}. 

Cosmological, hydrodynamical simulations have traditionally struggled to make stellar feedback as efficient as is required to match observed galaxy masses, sizes, outflow rates and other data. If the energy is injected thermally, it tends to be quickly radiated away rather than to drive a wind \citep[e.g.][]{Katz1996TreeSPH}. This ``overcooling'' problem is typically attributed to a lack of numerical resolution. If the simulation does not contain dense, cold clouds, then the star formation is not sufficiently clumpy and the feedback energy is distributed too smoothly. Moreover, since in reality cold clouds contain a large fraction of the mass of the ISM, in simulations without a cold interstellar phase the density of the warm, diffuse phase, and hence its cooling rate, is overestimated. 

While these factors may well contribute to the problem, \citet[][see also \citealt{DallaVecchia2008Winds},  \citealt{Creasey2011Overcooling} and \citealt{Keller2014Winds}]{DallaVecchia2012Winds} argued that the fact that the energy is distributed over too much mass may be a more fundamental issue. 
For a standard IMF there is $\sim 1$ supernova per 100~$\Msun$ of SSP mass and, in reality, all the associated mechanical energy is initially deposited in a few solar masses of ejecta, leading to very high initial temperatures (e.g.\ $\sim 2\times 10^8\,\K$ if $10^{51}\,\erg$ is deposited in $10~\Msun$ of gas). In contrast, in SPH simulations that distribute the energy produced by a star particle over its SPH neighbours, the ratio of the heated mass to the mass of the SSP will be much greater than unity. The mismatch in the mass ratio implies that the maximum temperature of the directly heated gas is far lower than in reality, and hence that its radiative cooling time is much too short. Because the mass ratio of SPH to star particles is independent of resolution, to first order this problem is independent of resolution. At second order, higher resolution does help, because the thermal feedback can be effective in generating an outflow if the cooling time is large compared with the sound crossing time across a resolution element, and the latter decreases with increasing resolution (but only as $m_{\rm g}^{1/3}$).  

Thus, subgrid models are needed to generate galactic winds in large-volume cosmological simulations. Three types of prescriptions are widely used: injecting energy in kinetic form \citep[e.g.][]{Navarro1993KineticFeedback,Springel2003Multiphase,DallaVecchia2008Winds,Dubois2008Winds} often in combination with temporarily disabling hydrodynamical forces acting on wind particles \citep[e.g.][]{Springel2003Multiphase,Okamoto2005Disks,Oppenheimer2006Wpot}, temporarily turning off radiative cooling \citep[e.g.][]{Gerritsen1997PhD,Stinson2006Winds}, and explicitly decoupling different thermal phases (also within single particles) \citep[e.g.][]{Marri2003Multiphase,Scannapieco2006Multiphase,Murante2010Multiphase,Keller2014Winds}. Here we follow \citet[][see also \citealt{Kay2003XrayGroups}]{DallaVecchia2012Winds} and opt for a different type of solution: stochastic thermal feedback. By making the feedback stochastic, we can control the amount of energy per feedback event even if we fix the mean energy injected per unit mass of stars formed. We specify the temperature jump of gas particles receiving feedback energy, $\Delta T$, and use the fraction of the total amount of energy from core collapse supernovae per unit stellar mass that is injected on average, $f_{\rm th}$, to set the probability that an SPH neighbour of a young star particle is heated. We perform this operation only once, when the stellar particle has reached the age $3\times 10^7\,\yr$, which corresponds to the maximum lifetime of stars that explode as core collapse supernovae. 

The value $f_{\rm th}=1$ corresponds to an expectation value for the injected energy of $8.73\times 10^{15}\,\erg\,\g^{-1}$ of stellar mass formed, which corresponds to the energy available from core collapse supernovae for a Chabrier IMF if we assume $10^{51}\,\erg$ per supernova and that stars with mass $6-100~\Msun$ explode ($6-8~\Msun$ stars explode as electron capture  supernovae in models with convective overshoot; e.g.\ \citealt{Chiosi1992Overshoot}). 

If $\Delta T$ is sufficiently high, then the initial (spurious, numerical) thermal losses will be small and we can control the overall efficiency of the feedback using $f_{\rm th}$. This freedom is justified, because there will be \emph{physical} radiative losses in reality that we cannot predict accurately for the ISM. Moreover, because the true radiative losses likely depend on the physical conditions, we may choose to vary $f_{\rm th}$ with the relevant, local properties of the gas. 

By considering the ratio of the cooling time to the sound crossing time across a resolution element, \citet{DallaVecchia2012Winds} derive the maximum density for which the thermal feedback can be efficient (their equation 18),
\begin{equation}
\nHtc \sim 10~\cm^{-3} \left (\frac{T}{10^{7.5}\,\K}\right )^{3/2} \left (\frac{m_{\rm g}}{10^6\,\Msun}\right )^{-1/2}, 
\label{eq:nhtc}
\end{equation}
where $T> \Delta T$ is the temperature after the energy injection and we use $\Delta T = 10^{7.5}\,\K$.
This expression assumes that the radiative cooling rate is dominated by free-free emission and will thus significantly overestimate the
value of $\nHtc$ when line cooling dominates, i.e.\ for $T \ll 10^7\,\K$. In our simulations some stars do, in fact, form in gas that far exceeds the critical value $\nHtc$, particularly in massive galaxies. Although the density of the gas in which the stars inject their energy will generally be lower than that of the gas from which the star particle formed, since the star particles move relative to the gas during the $3\times 10^7\,\yr$ delay between star formation and feedback, this does mean that for stars forming at high gas densities
the radiative losses may well exceed those that would occur in a simulation that has the resolution and the physics required to resolve the small-scale structure of the ISM. As we calibrate the total amount of energy that is injected per unit stellar mass to achieve a good match to the observed GSMF, this implies that we may overestimate the required amount of feedback energy. At the high-mass end AGN feedback controls the efficiency of galaxy formation in our simulations. If the radiative losses from stellar feedback are overestimated, then this could potentially cause us to overestimate the required efficiency of AGN feedback. 

The critical density, $\nHtc$, increases with the numerical resolution, but also with the temperature jump, $\Delta T$. We could therefore reduce the initial thermal losses by increasing $\Delta T$. However, for a fixed amount of energy per unit stellar mass, i.e.\ for a fixed value of $f_{\rm th}$, the probability that a particular star particle generates feedback is inversely proportional to $\Delta T$. \citet{DallaVecchia2012Winds} show that, for the case of equal mass particles, the expectation value for the number of heated gas particles per star particle is (their equation 8)
\begin{equation}
\left < N_{\rm heat}\right > \approx 1.3 f_{\rm th} \left (\frac{\Delta T}{10^{7.5}\,\K}\right )^{-1}
\end{equation}
for our Chabrier IMF and only accounting for supernova energy (assuming that supernovae associated with stars in the range 6-100 $\Msun$ each yield $10^{51}\,\erg$). Hence, using $\Delta T \gg 10^{7.5}\,\K$ or $f_{\rm th} \ll 1$ would imply that most star particles do not inject any energy from core collapse supernovae into their surroundings, which may lead to poor sampling of the feedback cycle. We therefore keep the temperature jump set to  $\Delta T = 10^{7.5}\,\K$. Although the stochastic implementation enables efficient thermal feedback without the need to turn off cooling, the thermal losses are unlikely to be converged with numerical resolution for simulations such as \eagle. Hence, recalibration of $f_{\rm th}$ may be necessary when the resolution is changed.

\subsubsection{Dependence on local gas properties}
\label{sec:calibration}

We expect the true thermal losses in the ISM to increase when the metallicity becomes sufficiently high for metal-line cooling to become important. For temperatures of $10^5\,\K < T < 10^7\K$ this happens when $Z \ga 10^{-1}\,\Zsun$ \citep[e.g.][]{Wiersma2009Cooling}. Although the exact dependence on metallicity cannot be predicted without full knowledge of the physical conditions in the ISM, we can capture the expected, qualitative transition from cooling losses dominated by H and He to losses dominated by metals by making $f_{\rm th}$ a function of metallicity,
\begin{equation}
f_{\rm th} = f_{\rm th,min} + \frac{f_{\rm th,max} - f_{\rm th,min}}
{1 + \left (\frac{Z}{0.1\Zsun}\right )^{n_Z}},
\label{eq:f(Z)}
\end{equation}
where $\Zsun = 0.0127$ is the solar metallicity and $n_Z>0$. Note that $f_{\rm th}$ asymptotes to $f_{\rm th,max}$ and $f_{\rm th,min}$ for $Z \ll 0.1\Zsun$ and $Z\gg 0.1\Zsun$, respectively.

Since metallicity decreases with redshift at fixed stellar mass, this physically motivated metallicity dependence tends to make feedback relatively more efficient at high redshift.
As we show in \citet{Crain2014EagleModels}, this leads to good agreement with the observed, present-day GSMF. In fact, \citet{Crain2014EagleModels} show that using a constant $f_{\rm th} =1$ appears to yield even better agreement with the low-redshift mass function, but we keep the metallicity dependence because it is physically motivated: we do expect larger radiative losses for $Z \gg 0.1Z_\odot$ than for $Z \ll 0.1Z_\odot$. If we were only interested in the GSMF, then equation (\ref{eq:f(Z)}) (or $f_{\rm th} =1$) would suffice. However, we find that pure metallicity dependence results in galaxies that are too compact, which indicates that the feedback is too inefficient at high gas densities. As discussed above, this is not unexpected given the resolution of our simulations. Indeed, we found that increasing the resolution reduces the problem.

We therefore found it desirable to compensate for the excessive initial, thermal losses at high densities by adding a density dependence to $f_{\rm th}$:
\begin{equation}
f_{\rm th} = f_{\rm th,min} + \frac{f_{\rm th,max} - f_{\rm th,min}}
{1 + \left (\frac{Z}{0.1\Zsun}\right )^{n_Z} \left (\frac{n_{\rm H,birth}}{n_{{\rm H},0}}\right )^{-n_n}},
\label{eq:f(Z,n)}
\end{equation}
where $n_{\rm H,birth}$ is the density inherited by the star particle, i.e.\ the density of its parent gas particle at the time it was converted into a stellar particle. Hence, $f_{\rm th}$ increases with density at fixed metallicity, while still respecting the original asymptotic values. We use $n_Z = n_n = 2/\ln10$. The seemingly unnatural value $2/\ln10 \approx 0.87$ of the exponent is a leftover from an equivalent, but more complicated expression that was originally used in the code. Using the round number 1 instead of 0.87 would have worked equally well. We use $n_{{\rm H},0} = 0.67~\cm^{-3}$, a value that was chosen after comparing a few test simulations to the observed present-day GSMF and galaxy sizes. The higher resolution simulation Recal-L025N0752 instead uses $n_{{\rm H},0} = 0.25~\cm^{-3}$ and a power-law exponent for the density term of $-1/\ln10$ rather than $-2/\ln10$ (see Table~\ref{tbl:subgridpars}), which we found gives better agreement with the GSMF. Note that a density dependence of $f_{\rm th}$ may also have a physical interpretation. For example, higher mean densities on $10^2-10^3\,\pc$ scales may result in more clustered star formation, which may reduce thermal losses. However, we stress that our primary motivation was to counteract the excessive thermal losses in the high-density ISM that can be attributed to our limited resolution. 

We use the asymptotic values $f_{\rm th,max}=3$ and $f_{\rm th,min}=0.3$, where the high asymptote $f_{\rm th,max}$ is reached at low metallicity and high density, and vice versa for the low asymptote. As discussed in \citet{Crain2014EagleModels}, where we present variations on the reference model, the choice of the high asymptote is the more important one. Using a value of $f_{\rm th,max}$ greater than unity enables us to reproduce the GSMF down to lower masses. 

Values of $f_{\rm th}$ greater than unity can be motivated on physical grounds by appealing to other sources of energy than supernovae, e.g.\ stellar winds, radiation pressure, or cosmic rays, or if supernovae yield more energy per unit mass than assumed here (e.g.\ in case of a top-heavy IMF). However, we believe that a more appropriate motivation is again the need to compensate for the finite numerical resolution. Galaxies containing few star particles tend to have too high stellar fractions \citep[e.g.][]{Haas2013OwlsI}, which can be understood as follows. The first generations of stars can only form once the halo is resolved with a sufficient number of particles to sample the high-density gas that is eligible to form stars. We do not have sufficient resolution to resolve the smallest galaxies that are expected to form in the real Universe. Hence, the progenitors of the galaxies in the simulations started forming stars, and hence driving winds, too late. As a consequence, our galaxies start with too high gas fractions and initially form stars too efficiently. As the galaxies grow substantially larger than our resolution limit, this initial error becomes progressively less important. Using a higher value of $f_{\rm th,max}$ counteracts this sampling effect as it makes the feedback from the first generations of stars that form more efficient. 

The mean and median values of $f_{\rm th}$ that were used for the feedback from the stars present at $z=0.1$ in Ref-L100N1504 are 1.06 and 0.70, respectively. For Recal-L025N0752 these values are 1.07 and 0.93. Hence, averaged over the entire simulation, the total amount of energy is similar to that expected from supernovae alone. A more detailed discussion of the effects of changing the functional form of $f_{\rm th}$ is presented in \citet{Crain2014EagleModels}. In that work we also present models in which $f_{\rm th}$ is constant or depends on halo mass or dark matter velocity dispersion.

\subsection{Black holes and feedback from AGN}
\label{sec:BHs}

In our simulations feedback from accreting, supermassive black holes (BHs) quenches star formation in massive galaxies, shapes the gas profiles in the inner parts of their host haloes, and regulates the growth of the BHs. 

Models often make a distinction between ``quasar-'' and ``radio-mode'' BH feedback \citep[e.g.][]{Croton2006SA,Bower2006SA,Sijacki2007AGN}, where the former occurs when the BH is accreting efficiently and comes in the form of a hot, nuclear wind, while the radio mode operates when the accretion rate is low compared to the Eddington rate and the energy is injected in the form of relativistic jets. Because cosmological simulations lack the resolution to properly distinguish these two feedback modes and because we want to limit the number of feedback channels to the minimum required to match the observations of interest, we choose to implement only a single mode of AGN feedback with a fixed efficiency. The energy is injected thermally at the location of the BH at a rate that is proportional to the gas accretion rate. Our implementation may therefore be closest to the process referred to as quasar-mode feedback. For \owls\ we found that this method led to excellent agreement with both optical and detailed X-ray observations of groups and clusters \citep{McCarthy2010AGN,McCarthy2011AGN,LeBrun2014CosmoOWLS}. 

Our implementation consists of two parts: i) prescriptions for seeding low-mass galaxies with central BHs and for their growth via gas accretion and merging (we neglect any growth by accretion of stars and dark matter); ii) a prescription for the injection of feedback energy. Our method for the growth of BHs is based on the one introduced by \citet{Springel2005AGN} and modified by \citet{Booth2009AGN} and \citet{Rosas2013BHs}, while our method for AGN feedback is close to the one described in \citet{Booth2009AGN}. Below we summarize the main ingredients and discuss the changes to the methods that we made for \eagle.

\subsubsection{BH seeds}

The BHs ending up in galactic centres may have originated from the direct collapse of (the inner parts of) metal-free dwarf galaxies, from the remnants of very massive, metal-free stars, or from runaway collisions of stars and/or stellar mass BHs (see e.g.\ \citealt{Kocsis2013BHReview} for a recent review). As none of these processes can be resolved in our simulations, we follow \citet{Springel2005AGN} and place BH seeds at the centre of every halo with total mass greater than $10^{10}\,\Msun/h$ that does not already contain a BH. For this purpose, we regularly run the friends-of-friends (FoF) finder with linking length 0.2 on the dark matter distribution. This is done at times spaced logarithmically in the expansion factor $a$ such that $\Delta a = 0.005a$.
The gas particle with the highest density is converted into a collisionless BH particle with subgrid BH mass $m_{\rm BH} = 10^5\,\Msun/h$. The use of a subgrid BH mass is necessary because the seed BH mass is small compared with the particle mass, at least for our default resolution. Calculations of BH properties such as its accretion rate are functions of $m_{\rm BH}$, whereas gravitational interactions are computed using the BH particle mass. When the subgrid BH mass exceeds the particle mass, it is allowed to stochastically accrete neighbouring SPH particles such that BH particle and subgrid masses grow in step.

Since the simulations cannot model the dynamical friction acting on BHs with masses $\la m_{\rm g}$, we force BHs with mass $< 100 m_{\rm g}$ 
to migrate towards the position of the minimum of the gravitational potential in the halo. At each time step the BH is moved to the location of the particle that has the lowest gravitational potential of all the neighbouring particles whose velocity relative to the BH is smaller than $0.25 c_{\rm s}$, where $c_{\rm s}$ is the speed of sound, and whose distance is smaller than three gravitational softening lengths. These two conditions prevent BHs in gas poor haloes from jumping to nearby satellites.

\subsubsection{Gas accretion}
\label{sec:bh_accretion}

The rate at which BHs accrete gas depends on the mass of the BH, the local density and temperature, the velocity of the BH relative to the ambient gas, and the angular momentum of the gas with respect to the BH. Specifically, the gas accretion rate, $\dot{m}_{\rm accr}$, is given by the minimum of the Eddington rate,
\begin{equation}
\dot{m}_{\rm Edd} = \frac{4\pi G m_{\rm BH} m_{\rm p}}{\epsilon_{\rm r} \sigma_{\rm T} c},
\end{equation}
and
\begin{equation}
\dot{m}_{\rm accr} = \dot{m}_{\rm Bondi} \times \min\left (C_{\rm visc}^{-1}(c_{\rm s}/V_\phi)^3,1 \right ), 
\label{eq:mdotaccr}
\end{equation}
where $\dot{m}_{\rm Bondi}$ is the Bondi-Hoyle (\citeyear{Bondi1944Accr}) rate for spherically symmetric accretion, 
\begin{equation}
\dot{m}_{\rm Bondi} = \frac{4\pi G^2 m_{\rm BH}^2 \rho}{(c_{\rm s}^2 + v^2)^{3/2}}.
\end{equation}
Here $m_{\rm p}$ is the proton mass, $\sigma_{\rm T}$ the Thomson cross section, $c$ the speed of light, $\epsilon_{\rm r}=0.1$ the radiative efficiency of the accretion disc, and $v$ the relative velocity of the BH and the gas. Finally, $V_\phi$ is the rotation speed of the gas around the BH computed using equation (16) of \citet{Rosas2013BHs} and $C_{\rm visc}$ is a free parameter related to the viscosity of the (subgrid) accretion disc.
The mass growth rate of the BH is given by \begin{equation} \dot{m}_{\rm BH} = (1-\epsilon_{\rm r}) \dot{m}_{\rm accr}. \end{equation}

The factor $(c_{\rm s}/V_\phi)^3/C_{\rm visc}$ by which the Bondi rate is multiplied in equation~(\ref{eq:mdotaccr}) is equivalent to the ratio of the Bondi and the viscous time scales (see \citealt[][]{Rosas2013BHs}). We set $C_{\rm visc} = 2\pi$ for Ref-L100N1504, but increase the value of $C_{\rm visc}$ by a factor $10^3$ for the recalibrated high-resolution model, Recal-L025N0752, and by a factor $10^2$ for AGNdT9-L050N0752 (see Table~\ref{tbl:subgridpars}). Since the critical ratio of $V_\phi/c_{\rm s}$ above which angular momentum is assumed to reduce the accretion rate scales with $C_{\rm visc}^{-1/3}$, angular momentum is relatively more important in the recalibrated simulations, delaying the onset of quenching by AGN to larger BH masses.
As demonstrated by \citet{Rosas2013BHs}, the results are only weakly dependent on $C_{\rm visc}$ because the ratio of $V_\phi/c_{\rm s}$ above which the accretion rate is suppressed, which scales as $C_{\rm visc}^{-1/3}$, is more important than the actual suppression factor, which scales as $C_{\rm visc}$. 

Our prescription for gas accretion differs from previous work in two respects. First, the Bondi rate is not multiplied by a large, ad-hoc factor, $\alpha$. \citet{Springel2005AGN} used $\alpha=100$ while \owls\ and \citealt{Rosas2013BHs} used a density dependent factor that asymptoted to unity below the star formation threshold. Although the use of $\alpha$ can be justified if the simulations underestimate the gas density or overestimate the temperature near the Bondi radius, the correct value cannot be predicted by the simulations. We found that at the resolution of \eagle, we do not need to boost the Bondi-Hoyle rate for the BH growth to become self-regulated. Hence, we were able to reduce the number of free parameters by eliminating $\alpha$. 
Second, we use the heuristic correction of \citet{Rosas2013BHs} to account for the fact that the accretion rate will be lower for gas with more angular momentum (because the accretion is generally not spherically symmetric as assumed in the Bondi model, but proceeds through an accretion disc). 

\subsubsection{BH mergers}

BHs are merged if they are separated by a distance that is smaller than both the smoothing kernel of the BH, $h_{\rm BH}$, and three gravitational softening lengths, and if their relative velocity is smaller than the circular velocity at the distance $h_{\rm BH}$, $v_{\rm rel} < \sqrt{G m_{\rm BH}/h_{\rm BH}}$, where $h_{\rm BH}$ and $m_{\rm BH}$ are, respectively, the smoothing length and subgrid mass of the most massive BH in the pair. The limit on the allowed relative velocity prevents BHs from merging during the initial stages of galaxy mergers.

\subsubsection{AGN feedback}
\label{sec:AGNfeedback}

AGN feedback is implemented thermally and stochastically, in a manner analogous to energy feedback from star formation. The energy injection rate is $\epsilon_{\rm f} \epsilon_{\rm r} \dot{m}_{\rm accr} c^2$, where $\epsilon_{\rm f} = 0.15$ is the fraction of the radiated energy that is coupled to the ISM. As was the case for the stellar feedback efficiency, $f_{\rm th}$, the value of $\epsilon_{\rm f}$ must be chosen by calibrating to observations, in this case the normalisation of the relation between BH mass and stellar mass. As demonstrated and explained by \citet[][see also \citealt{Booth2009AGN}]{Booth2010DMHaloesBHs}, the value of $\epsilon_{\rm f}$ \emph{only} affects the BH masses, which are inversely proportional to $\epsilon_{\rm f}$. In particular, the outflow rate generated by the AGN and hence also the factor by which the star formation is reduced, are highly insensitive to $\epsilon_{\rm f}$ provided it is nonzero. This can be explained by self-regulation: the BH accretion rate adjusts until the rate at which energy is injected is sufficient for outflows to balance inflows. 

We use the same value for the AGN efficiency as in \owls, $\epsilon_{\rm f}=0.15$ and $\epsilon_{\rm r}=0.1$, which implies that a fraction $\epsilon_{\rm f}\epsilon_{\rm r}=0.015$ of the accreted rest mass energy is returned to the local ISM. As was the case for stellar feedback, the required value will depend on the radiative losses in the ISM, which may depend on the resolution and the precise manner in which the energy is injected. We do not implement a dependence on metallicity, because metals are not expected to dominate the radiative losses at the high temperatures associated with AGN feedback. As shown in Figure~\ref{fig:bh}, a constant value of $\epsilon_{\rm f}=0.15$ yields broad agreement with observations of the relation between BH mass and stellar mass. 

Each BH carries a ``reservoir'' of feedback energy, $E_{\rm BH}$. After each time step $\Delta t$, we add $\epsilon_{\rm f} \epsilon_{\rm r} \dot{m}_{\rm accr} c^2 \Delta t$ to this reservoir. If the BH has stored sufficient energy to heat at least $n_{\rm heat}$ particles of mass $m_{\rm g}$, then the BH is allowed to stochastically heat each of its SPH neighbours by increasing their temperature by $\Delta T_{\rm AGN}$. For each neighbour the heating probability is
\begin{equation}  
P = \frac{E_{\rm BH} }{ \Delta\epsilon_{\rm AGN} N_{\rm ngb} \left <m_{\rm g}\right > },
\end{equation} 
where $\Delta\epsilon_{\rm AGN}$ is the change in internal energy per unit mass corresponding to the temperature increase, $\Delta T_{\rm AGN}$ (we convert the parameter $\Delta T_{\rm AGN}$ into $\Delta\epsilon_{\rm AGN}$ assuming a fully ionised gas with primordial composition), $N_{\rm ngb}$ is the number of gas neighbours of the BH and $\left <m_{\rm g}\right >$ is their mean mass. We then reduce $E_{\rm BH}$ by the expectation value for the injected energy. We use $n_{\rm heat}=1$ and limit the time step of the BHs such that we expect\footnote{Because the expected probability is based on the accretion rate in the previous time step, limiting the BH time step does not guarantee that $P<0.3$. If the probability exceeds 0.3, then we limit it to 0.3 and store the unused energy in $E_{\rm BH}$.} $P<0.3$ (see \S\ref{sec:injection}).

The most important parameter for the AGN feedback is the temperature increase $\Delta T_{\rm AGN}$. Larger values will make individual feedback events more energetic, generally resulting in smaller radiative losses in the ISM. However, larger values will also make the feedback more intermittent. We set $\Delta T_{\rm AGN} = 10^{8.5}\,\K$ in the L100N1504 reference model, but use $10^9\,\K$ for our recalibrated high-resolution model Recal-L025N0752 and model AGNdT9-L050N0752 (see Table~\ref{tbl:subgridpars}). These temperatures exceed the value of $10^8\,\K$ used in \owls\ and the $\Delta T = 10^{7.5}\,\K$ that we use for stellar feedback. As can be seen from equation~(\ref{eq:nhtc}), the critical density above which the feedback energy is expected to be radiated away increases with the value of $\Delta T$. Because the density of the ambient gas around the BH tends to increase with resolution, we found that we need to increase $\Delta T$ when increasing the resolution. Similarly, because the gas density around the BH often reaches values that are much higher than is typical for star-forming gas, we require higher temperature jumps for AGN feedback than for stellar feedback. 

\section{Comparison with observables considered during the calibration of the feedback}
\label{sec:cal_obs}

In this section we will compare the main \eagle\ simulations to $z\sim 0$ observations of the GSMF, the related stellar mass - halo mass relation, galaxy sizes, and the relation between BH mass and stellar mass. Since these observables were considered during the calibration of the subgrid models for feedback, we cannot consider the \eagle\ results reported in this section to be ``predictions''. However, note that we had no control over the slope of the $M_{\rm BH}-M_\ast$ relation and that galaxy sizes were only used to rule out strongly discrepant models (i.e.\ models without a density dependence of the energy feedback from star formation). 

\subsection{The galaxy stellar mass function}
\label{sec:gsmf}

\begin{figure} 
\resizebox{\colwidth}{!}{\includegraphics{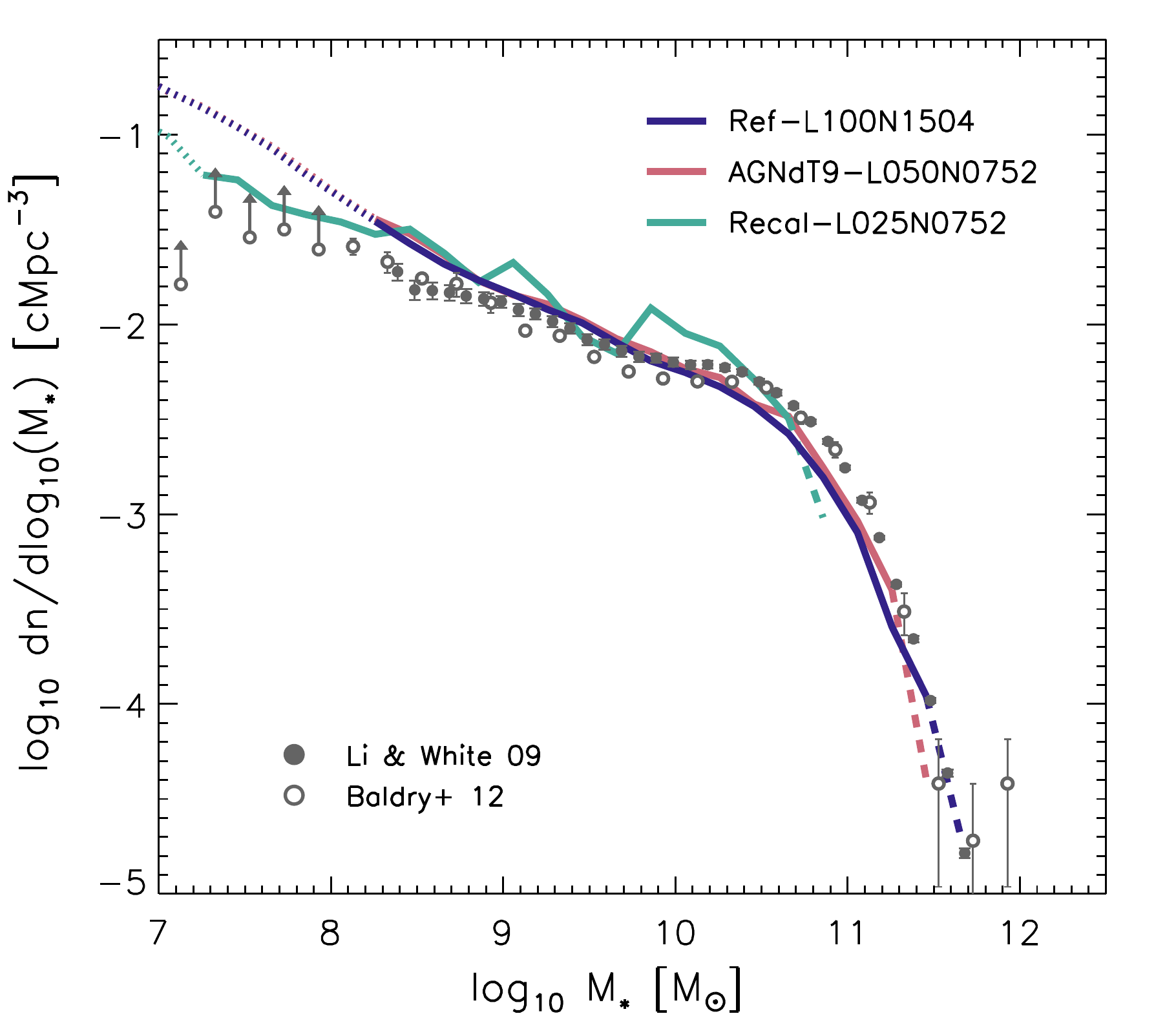}} 
\caption{The galaxy stellar mass function at $z=0.1$ for the \eagle\ simulations Ref-L100N1504 (blue), AGNdT9-L050N0752 (red), and Recal-L025N0752 (green-blue). The curves switch from solid to dashed at the high-mass end when there are fewer than 10 objects per (0.2~dex) stellar mass bin. At the low-mass end the curves become dotted when the stellar mass falls below that corresponding to 100 baryonic particles. Data points show measurements with $1\sigma$ error bars from the GAMA survey (open circles; $z<0.06$; \citealt{Baldry2012GSMF}) and from SDSS (filled circles; $z\sim 0.07$;  \citealt{Li2009GSMF}). The high-resolution model Recal-L025N0752 is noisier because of its small box size. The intermediate-resolution models slightly underestimate the galaxy number density at the knee of the mass function and slightly overestimate the abundance at $M_\ast \sim 10^{8.5}\,\Msun$. The galaxy number density agrees with the data to $\la 0.2$ dex.} 
\label{fig:gsmf} 
\end{figure}

Figure~\ref{fig:gsmf} shows the $z=0.1$ galaxy stellar mass function (GSMF) from \eagle. The dark blue curve shows Ref-L100N1504, the green curve shows the high-resolution simulation Recal-L025N0752, and the red curve shows AGNdT9-L050N0752. Recall that AGNdT9-L050N0752 employs a higher heating temperature for AGN feedback than the reference model, which makes the feedback more efficient. While this is unimportant for the GSMF, we will see in \S\ref{sec:groups} that it offers a significant improvement for the intracluster medium. At the high-mass end the curves switch from a solid to a dashed line style where there are fewer than 10 objects per (0.2~dex) stellar mass bin.
At the low-mass end the curves become dotted when the stellar mass falls below that corresponding to 100 baryonic particles, where sampling effects associated with the limited resolution become important, as can be seen by comparing the intermediate- and high-resolution simulations. 

The GSMF of the high-resolution simulation Recal-L025N0752 is noisier because the box size is too small to provide a representative sample. Note that the main problem is not Poisson noise due to the small number of objects per bin, but the small number of large-scale modes that modulate the local number density of galaxies of various masses. 
Indeed, Fig.~\ref{fig:gsmf_conv} shows that the GSMF of Recal-L025N0752 has the same wiggles as that of Ref-L025N0376, which uses the same box size and, apart from the change in resolution, the same initial conditions. The wiggles that are present for Ref-L025N0376 are absent for model Ref-L100N1504, even though these two simulations use identical resolutions and (subgrid) parameter values. This confirms that the wiggles in the GSMF of Recal-L025N0752 are caused by the small size of its simulation volume. We will therefore focus on the larger volume simulations when comparing the simulated and observed GSMFs.

The simulation results are compared with observations from the Galaxy And Mass Assembly (GAMA) survey (\citealt{Baldry2012GSMF}; open circles) and from SDSS (\citealt{Li2009GSMF}; filled circles). 
For the intermediate-resolution simulations the galaxy number densities agree with the observations to $\la 0.2$~dex over the full mass range for which the resolution and box size are adequate, i.e.\ from $2\times 10^8\,\Msun$ to over $10^{11}\,\Msun$ (slightly below $10^{11}\,\Msun$ for Recal-L025N0752). The observed shape of the GSMF is thus reproduced well.

At fixed number density, the differences in stellar mass between the simulations and observations are smaller than 0.3~dex for Ref-L100N1504 and AGNdT9-L050N0752. Given that even for a fixed IMF, uncertainties in the stellar evolution models used to infer stellar masses are $\sim 0.3$~dex \citep[e.g.][]{Conroy2009SPSSUncertainty,Behroozi2010Uncertainties,Pforr2012SPSSUncertainty,Mitchell2013MassUncertainty}, there is perhaps little point in trying to improve the agreement between the models and the data further.

The subgrid models for energy feedback from star formation and for BH accretion have been calibrated to make the simulated GSMF fit the observed one, so the excellent agreement with the data cannot be considered a successful prediction. However, success was by no means guaranteed given that the computational expense of hydrodynamical simulations severely limits the number of test runs that can be performed and, more importantly, because the freedom built into the model is rather limited. For example, while the mass scale above which AGN feedback becomes dominant is sensitive to the parameter $C_{\rm visc}$ of the subgrid model for BH accretion (see equation \ref{eq:mdotaccr} in \S\ref{sec:bh_accretion}), the efficiency of the AGN feedback was calibrated to the observed relation between BH mass and stellar mass and does not affect the shape of the GSMF \citep{Booth2009AGN,Booth2010DMHaloesBHs}.

\begin{figure*} 
\resizebox{\colwidth}{!}{\includegraphics{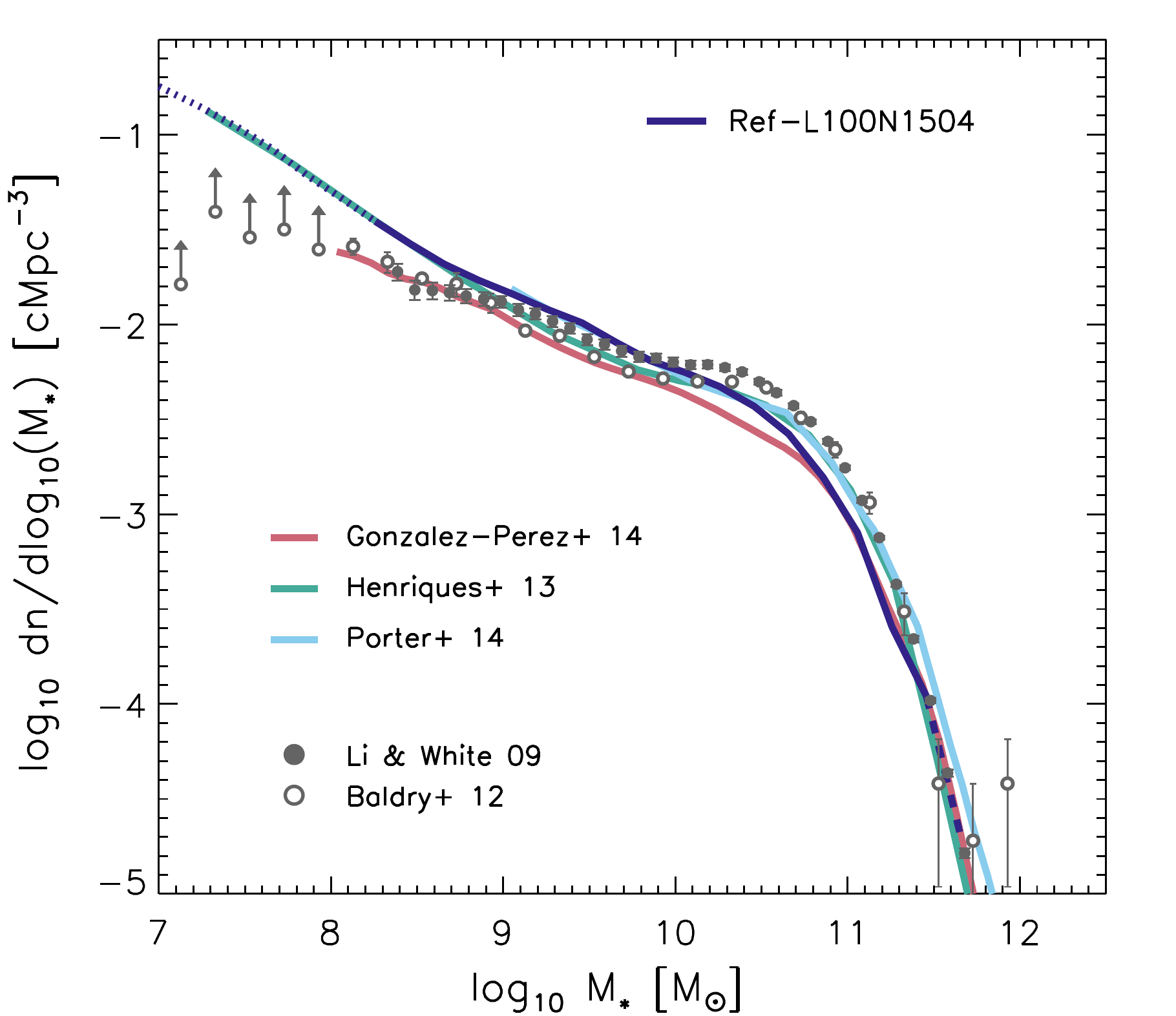}} 
\resizebox{\colwidth}{!}{\includegraphics{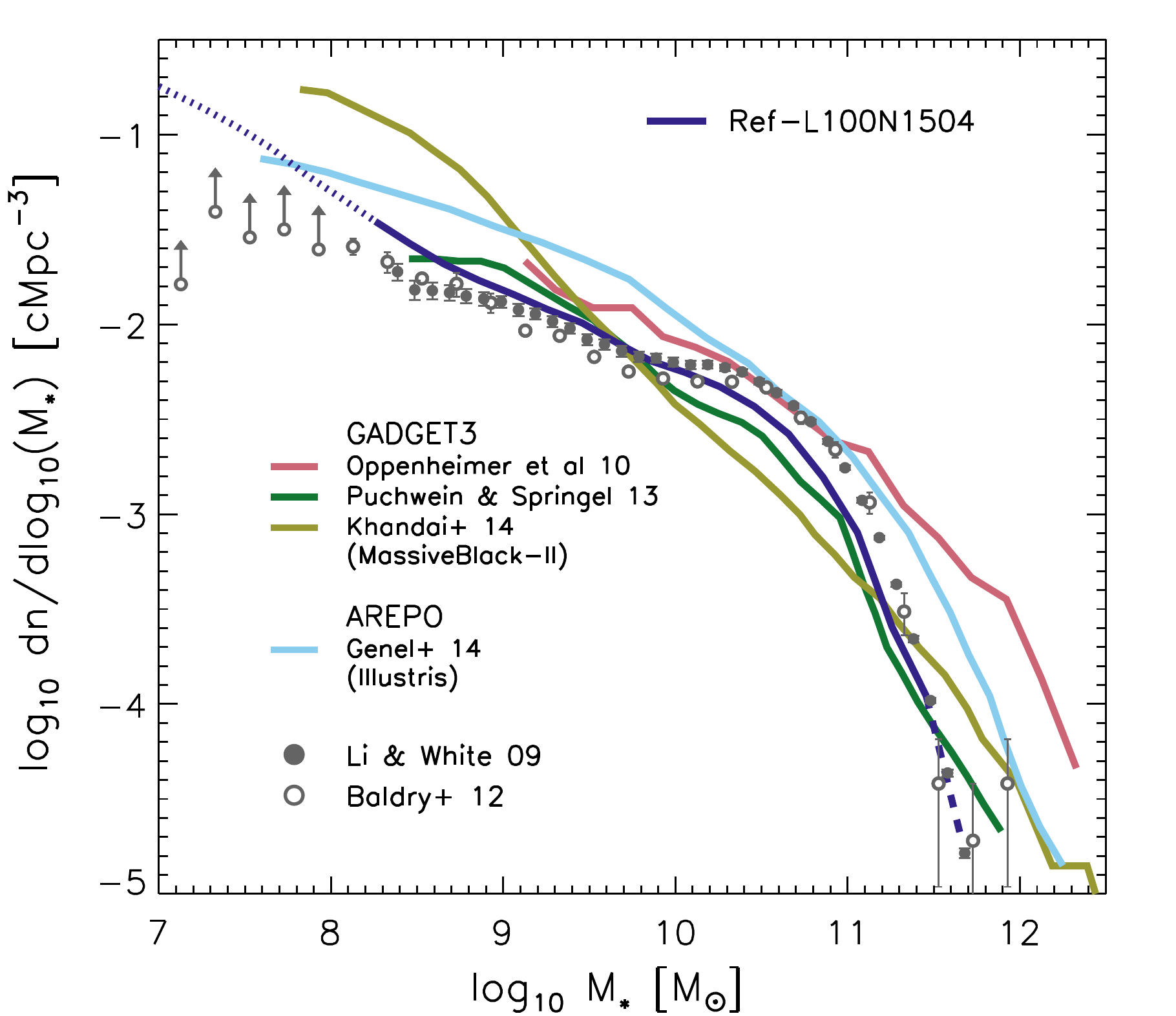}} 
\caption{Comparisons of the GSMF from \eagle's Ref-L100N1504 with the semi-analytic models of \citet{Gonzalez-Perez2014Galform}, \citet{Henriques2013SAM}, and \citet{Porter2014SAM} (left panel) and with the large hydrodynamical simulations of \citet{Oppenheimer2010GSMF}, \citet{Puchwein2013GSMF}, the Illustris simulation \citep[][data taken from \citealt{Genel2014IllustrisEvolution}]{Vogelsberger2014Illustris}, and the MassiveBlack-II simulation \citep{Khandai2014MassiveBlack} (right panel). All models are for a Chabrier IMF (\citealt{Gonzalez-Perez2014Galform} and \citealt{Khandai2014MassiveBlack} have been converted from Kennicutt and Salpeter IMFs, respectively). The \eagle\ curve is dotted when galaxies contain fewer than 100 stellar particles and dashed when there are fewer than 10 galaxies per stellar mass bin.
Except for \citet{Oppenheimer2010GSMF}, all simulations include AGN feedback. Apart from MassiveBlack-II, all models were calibrated to the data (the Galform semi-analytic model of \citealt{Gonzalez-Perez2014Galform} was calibrated to fit the K-band galaxy luminosity function). The agreement with the data is relatively good for both \eagle\ and the semi-analytic models, but \eagle\ fits the data substantially better than the other hydrodynamical simulations do.}
\label{fig:gsmf_other} 
\end{figure*}

Figure~\ref{fig:gsmf_other} shows that the level of correspondence between the data and \eagle\ is close to that attained for semi-analytic models (left panel) and is unprecedented for large, hydrodynamical simulations (right panel). As can be seen from the right panel, even though \citet{Oppenheimer2010GSMF}, \citet{Puchwein2013GSMF}, and Illustris \citep{Vogelsberger2014IllustrisNature,Genel2014IllustrisEvolution} all adjusted their subgrid feedback models to try to match the data, the fits to the data are substantially less good than for \eagle. In particular, their models all produce mass functions that are too steep below the ``knee'' of the Schechter function and too shallow for larger masses. It is worth noting that each of these three groups implemented the feedback from star formation kinetically, scaled the wind velocity with the velocity dispersion of the dark matter, determined the dependence of the wind mass loading on the dark matter velocity dispersion by assuming a constant wind energy, and temporarily turned off the hydrodynamical forces on wind particles to allow them to escape the galaxies. This contrasts with \eagle, where the feedback was implemented thermally rather than kinetically, the feedback energy varied with local gas properties, and the hydrodynamical forces were never turned off. 

Hence, contrary to the other models shown, \eagle's subgrid model does not impose any particular wind velocity or mass loading or any dependence on dark matter or halo properties. The injected energy does depend on the local metallicity and gas density, but the relation between the outflow properties and the energy injected at the star formation site is an outcome of the simulation. \citet{Crain2014EagleModels} will show that while varying the feedback energy with local gas properties is necessary to obtain reasonable galaxy sizes, the $z\sim 0$ GSMF is actually also reproduced by the \eagle\ model that injects a constant energy per unit stellar mass (equal to the energy from supernovae) without any calibration.

While the excellent fit to the low-$z$ GSMF is encouraging, the success of the model can only be judged by comparing to a wide range of observables and redshifts, particularly those that were not considered during the calibration. We will consider a diverse selection of observables in \S\ref{sec:otherobs} and will investigate their evolution in \citet{Furlong2014EagleEvolution} and other future papers.

\subsubsection{Effect of the choice of aperture}
\label{sec:aperture}

\begin{figure} 
\resizebox{\colwidth}{!}{\includegraphics{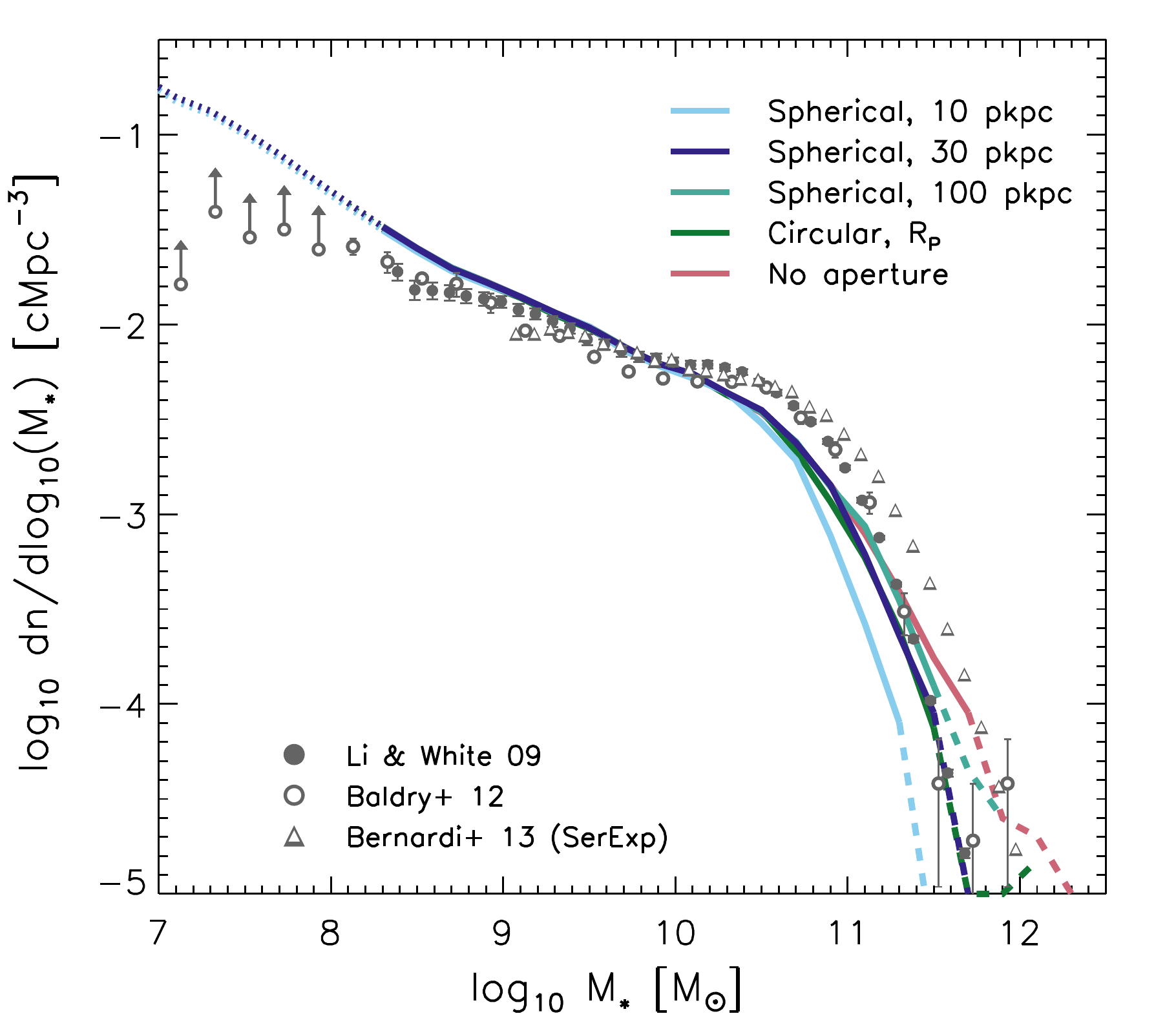}} 
\caption{The effect of the choice of aperture on the GSMF. Curves show the $z=0.1$ GSMF from Ref-L100N1504 for different 3-D apertures: radii of 30, 50, and 100 proper kiloparsec, a 2-D Petrosian aperture, and no aperture at all. In all cases only stellar mass bound to a subhalo is considered. The simulation curves are dotted where galaxies contain fewer than 100 stellar particles and dashed where there are fewer than 10 galaxies per stellar mass bin. Data points indicate observations. The \citet{Li2009GSMF} and \citet{Bernardi2013GSMF} data points are both for SDSS, but use Petrosian magnitudes and integrals of S\'ersic plus exponential fits, respectively. The \citet{Baldry2012GSMF} data points are for the GAMA survey and use integrals of single S\'ersic fits. The choice of aperture is important for $M_\ast > 10^{11}\,\Msun$, both for the simulation and the observations.} 
\label{fig:gsmf_aperture} 
\end{figure}

For the simulations we chose to define a galaxy's stellar mass as the sum of the mass of the stars that are part of a gravitationally bound subhalo and that are contained within a 3-D aperture of radius 30 proper kiloparsec (see \S\ref{sec:simulations}). Figure~\ref{fig:gsmf_aperture} shows the effect of the choice of aperture for Ref-L100N1504. For $M_\ast < 10^{11}\,\Msun$ the results are insensitive to the aperture, provided it is $\ga 30$ pkpc. However, for $M_\ast > 10^{11}\,\Msun$ the aperture does become important, with larger apertures giving larger masses.

The same is true for the observations, as can be seen by comparing the data from \citet{Li2009GSMF} with the re-analysis of SDSS data 
by \citet{Bernardi2013GSMF} (open triangles in Fig.~\ref{fig:gsmf_aperture}). \citet{Baldry2012GSMF} and \citet{Li2009GSMF} are in good agreement, but \citet{Bernardi2013GSMF} find a much shallower bright-end slope than previous analyses. For $M_\ast > 10^{11}\,\Msun$ \citet{Bernardi2013GSMF} attribute substantially  more mass to galaxies than \citet{Li2009GSMF} and \citet{Baldry2012GSMF}.
Part of the difference is due to the assumed mass-to-light ratios (even though all studies assume a Chabrier IMF) and the way in which the background is subtracted (see e.g.\ \citealt{Bernardi2013GSMF} and \citealt{Kravtsov2014GalaxyHalo} for discussion). Most of the difference between \citet{Li2009GSMF} and \citet{Bernardi2013GSMF}
can probably be attributed to the way in which a galaxy's light is measured. \citet{Li2009GSMF} integrate the light within a 2-D aperture of size twice the Petrosian radius, defined to be the radius at which the mean local surface brightness is 0.2 times the mean internal surface brightness. \citet{Bernardi2013GSMF} on the other hand, estimate the total amount of light by integrating S\'ersic plus exponential profile fits. Hence, the \citet{Bernardi2013GSMF} mass function potentially includes intracluster light and the discrepancy between different authors is related to the fact that it is unclear where cD galaxies end. \citet{Baldry2012GSMF} integrate single S\'ersic fits to the light profiles, which we would expect includes less intracluster light than the S\'ersic plus exponential fits of \citet{Bernardi2013GSMF} but more than the Petrosian apertures of \citet{Li2009GSMF}. However, \citet{Bernardi2013GSMF} find that the high-mass end of the  \citet{Baldry2012GSMF} mass function is affected by their redshift cut ($z<0.06$). 

We believe the  \citet{Baldry2012GSMF} and \citet{Li2009GSMF} data to be the most suitable for comparison to our results, since our definition of a galaxy excludes intracluster light. For \citet{Li2009GSMF} this is confirmed by our finding that a 3-D aperture of 30 pkpc gives nearly identical results to a 2-D Petrosian cut, as can be seen from Figure~\ref{fig:gsmf_aperture}. 

Thus, for masses $> 10^{11}\,\Msun$ comparisons of the GSMF with observations would benefit from mimicking the particular way in which the mass is estimated for real data. This would, however, have to be done separately for each survey. For our present purposes this is unnecessary, also because our simulation volume is in any case too small to study the GSMF at masses $\gg 10^{11}\,\Msun$.

\subsubsection{Numerical convergence}
\label{sec:convergence}

\begin{figure*} 
\resizebox{\colwidth}{!}{\includegraphics{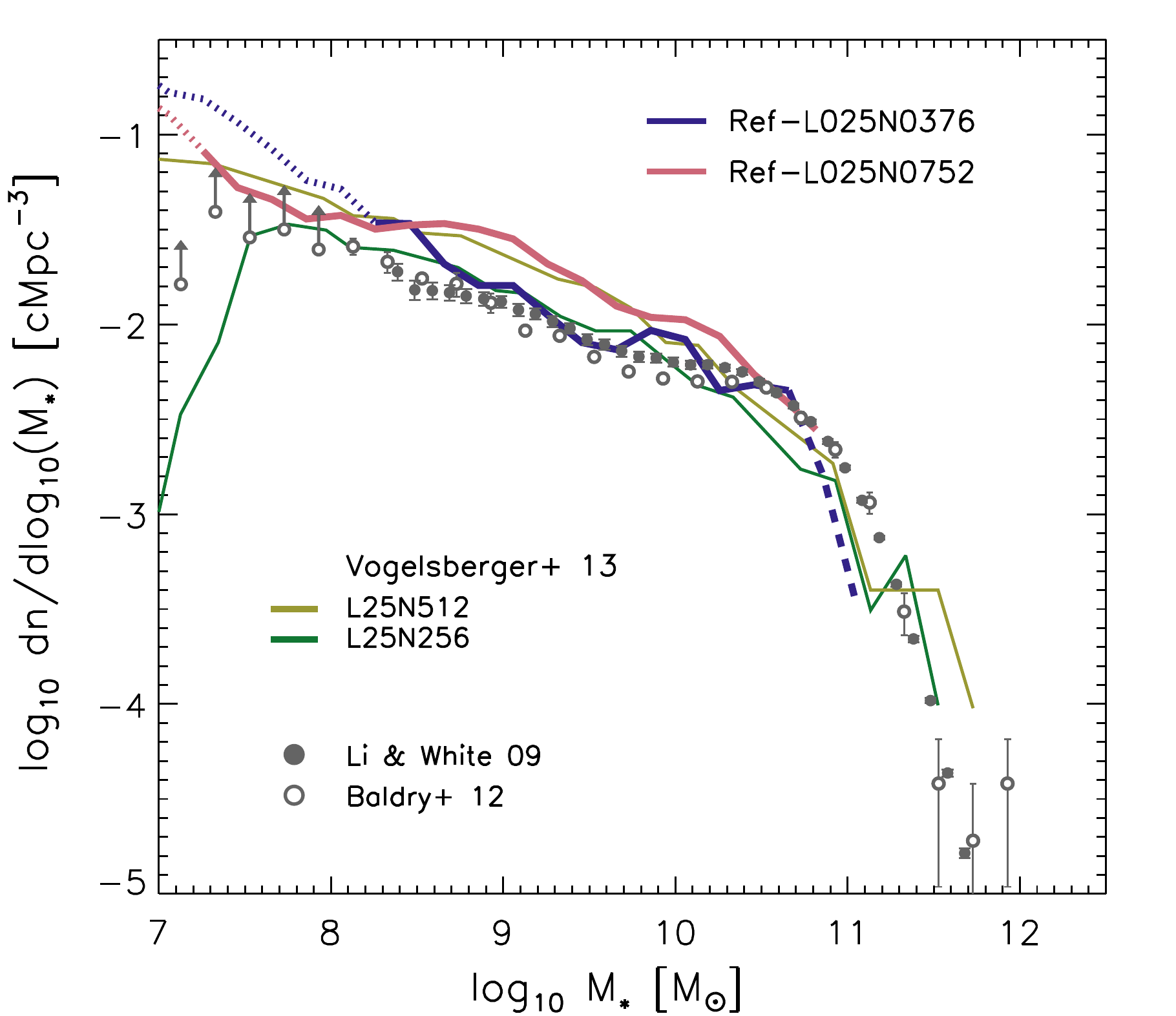}} 
\resizebox{\colwidth}{!}{\includegraphics{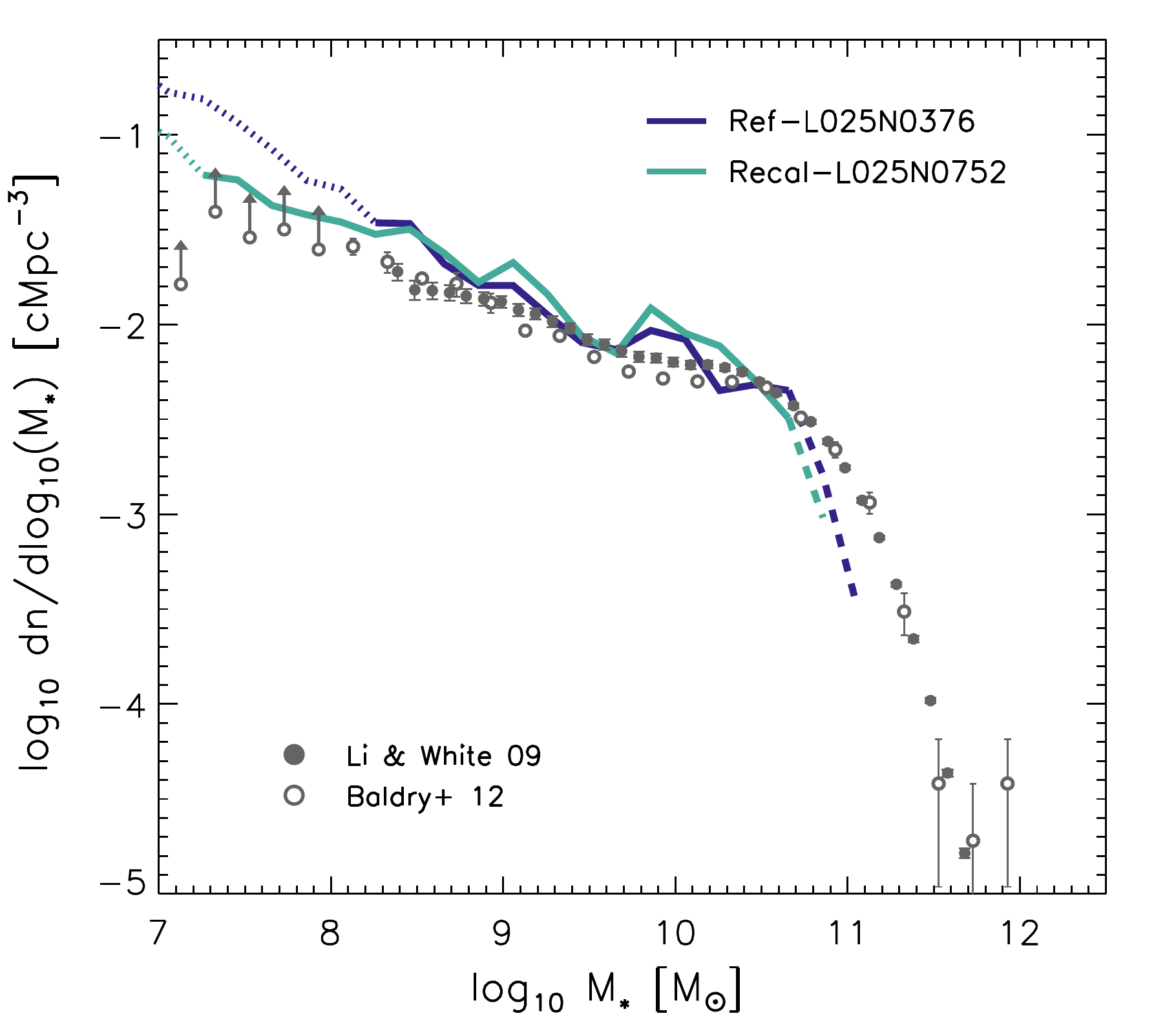}} 
\caption{Strong (left panel) and weak (right panel) tests of the convergence of the GSMF with numerical resolution. Models L025N0752 have a better mass and spatial resolution than L025N0376 by factors of 8 and 2, respectively. The strong convergence test compares models with identical subgrid parameter values, while the weak convergence test compares the original, intermediate-resolution model Ref-L025N0376 with a high-resolution model Recal-L025N0752 for which the parameters of the subgrid models for feedback from star formation and for gas accretion onto BHs were recalibrated in order to reproduce the observed GSMF. For comparison, the thin curves in the left panel show the strong convergence test for the galaxy formation model used for the Illustris simulation as reported by \citet{Vogelsberger2013IllustrisModel}. The \eagle\ curves are dotted where galaxies contain fewer than 100 stellar particles and dashed where there are fewer than 10 galaxies per stellar mass bin.} 
\label{fig:gsmf_conv} 
\end{figure*}

The left panel of Figure~\ref{fig:gsmf_conv} compares the GSMFs for model Ref-L025N0376, which has the same resolution as the largest \eagle\ volume Ref-L100N1504, and the higher-resolution model Ref-L025N0752. The two Ref-L025 simulations use identical subgrid parameters, but the mass and spatial resolution differ by factors of 8 and 2, respectively. In \S\ref{sec:conv_discussion} we termed a comparison between models with identical parameters a ``strong convergence test''. Below $10^9\,\Msun$ the mass function is substantially flatter in the high-resolution model. However, at $M_\ast \sim 10^9\,\Msun$ its GSMF is up to 0.4~dex higher than for the fiducial resolution, leading to disagreement with the data. The largest discrepancy is the stellar mass corresponding to a number density of $\sim 2\times 10^{-2}\,{\rm cMpc}^{-3}$, which is about an order of magnitude higher than observed. 

The thin curves in Figure~\ref{fig:gsmf_conv} show the strong convergence test of \citet{Vogelsberger2013IllustrisModel} using the galaxy formation model that was also used for Illustris. Clearly, the strong convergence is similarly poor. This is somewhat surprising, since Illustris uses a subgrid model for feedback from star formation that was designed to give good strong convergence. In particular, the parameters of the subgrid wind model vary with the velocity dispersion of the dark matter rather than with the properties of the gas and hydrodynamical interactions between the wind and the ISM are not modelled.

That the strong convergence is not particularly good for \eagle\ is unsurprising for the reasons discussed in \S\ref{sec:conv_discussion} and \S\ref{sec:snii}. For $M_\ast < 2\times 10^8\,\Msun$ galaxies in Ref-L025N0376 contain fewer than 100 star particles, which is insufficient to properly sample the feedback from star formation in the context of \eagle's subgrid model. Because the feedback can be modelled down to lower masses in Ref-L025N0752, galaxies with $M_\ast \sim 10^9\,\Msun$ have had systematically different histories than galaxies of a similar mass in Ref-L025N0376. In addition, higher resolution enables the gas density distribution to be populated by particles up to higher densities, where our fiducial implementation of thermal feedback becomes inefficient (equation~\ref{eq:nhtc} in \S\ref{sec:snii}). 

In \S\ref{sec:conv_discussion} we argued that hydrodynamical simulations such as \eagle\ should recalibrate the efficiency of the subgrid feedback when the resolution is changed substantially. In general, keeping the subgrid parameters fixed does not imply that the physical model remains unchanged, since the energy, mass or intermittency associated with the feedback events changes. Moreover, the efficiency of the feedback cannot, in any case, be predicted from first principles, even if the convergence were perfect.

Recal-L025N0752 is our recalibrated high-resolution simulation. As detailed in \S\ref{sec:calibration} and Table~\ref{tbl:subgridpars}, the dependence of the feedback energy per unit stellar mass on the gas density is somewhat different between the different resolutions. However, the mean values of $f_{\rm th}$, which is equal to the expectation value of the amount of injected energy in units of the energy available from core collapse supernovae, are nearly identical: 1.06 at intermediate resolution (for stars formed at $z>0.1$ in Ref-L100N1504) and 1.07 at high resolution (for stars formed at $z>0.1$ in Recal-L025N0752). The asymptotic maximum of $f_{\rm th}$, reached at low metallicity and low gas density, is 3 in both cases.  As detailed in \S\ref{sec:bh_accretion} and Table~\ref{tbl:subgridpars}, Recal-L025N0752 also uses a different value for the parameter that controls the importance of angular momentum in suppressing accretion onto BHs, making the accretion rate more sensitive to the angular momentum of the accreting gas. Without this change, AGN feedback would become important at too low masses. Finally, the high-resolution model uses a higher AGN feedback temperature, $\Delta T_{\rm AGN} = 10^9\,\K$ rather than $10^{8.5}\,\K$, which helps to suppress the increase in the cooling losses that would otherwise occur due to the higher gas densities that are resolved in the higher resolution model. Without this change the AGN feedback would be insufficiently effective.

The right panel of Figure~\ref{fig:gsmf_conv} shows a ``weak convergence test'', i.e.\ a comparison of the GSMFs of the calibrated intermediate resolution model Ref-L025N0376 and the recalibrated high-resolution model Recal-L025N0752. The two curves show some of the same bumps and wiggles, because the initial conditions used for the two simulations share the same large-scale modes. In the mass range for which galaxies in the intermediate-resolution model are resolved with more than 100 star particles ($M_\ast > 2\times 10^8\,\Msun$) the difference in the galaxy number density is smaller than 0.2~dex. We conclude that the weak convergence is good.

\subsection{The relation between stellar mass and halo mass}
\label{sec:eta}

\begin{figure*}
\resizebox{\colwidth}{!}{\includegraphics{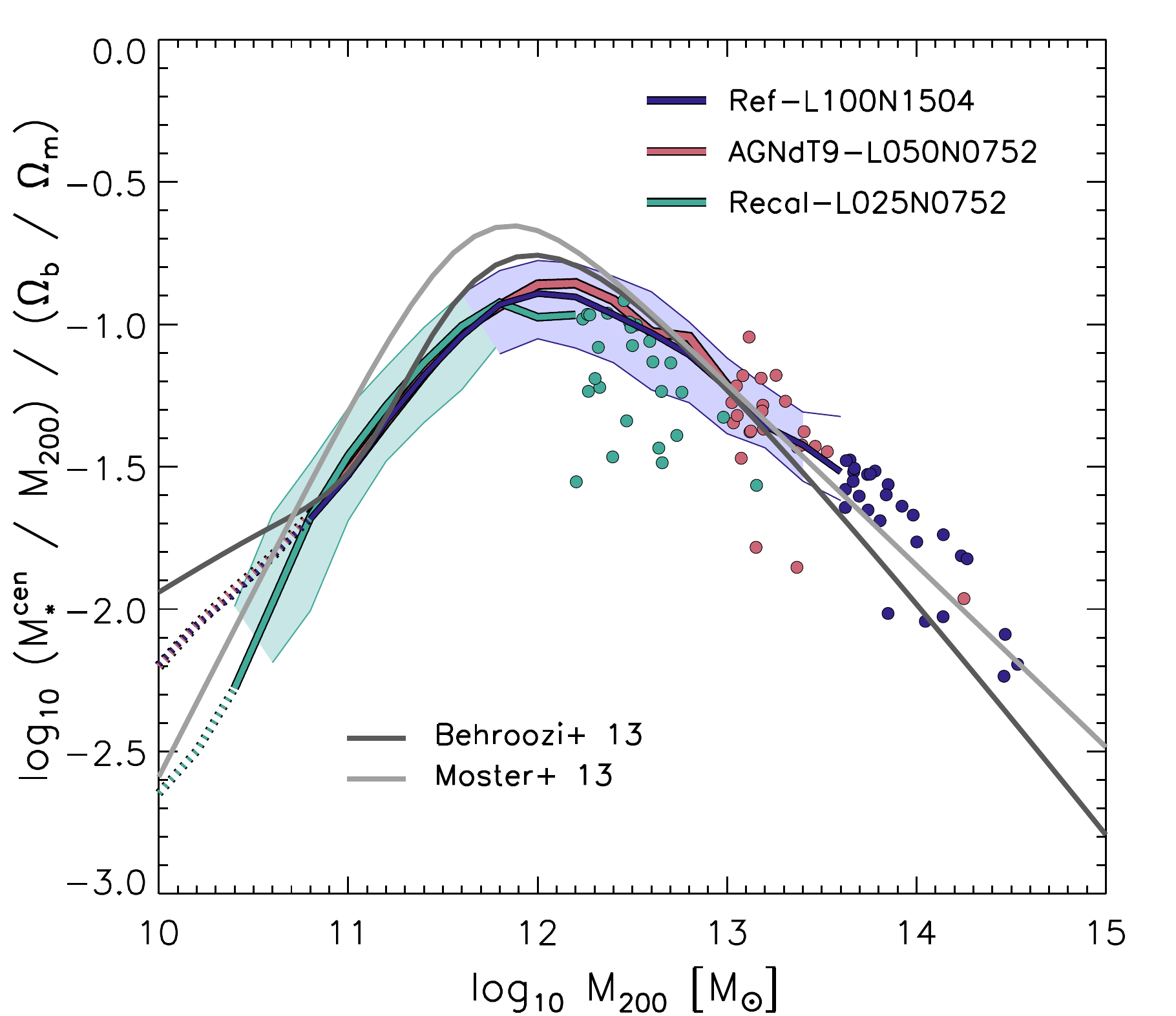}} 
\resizebox{\colwidth}{!}{\includegraphics{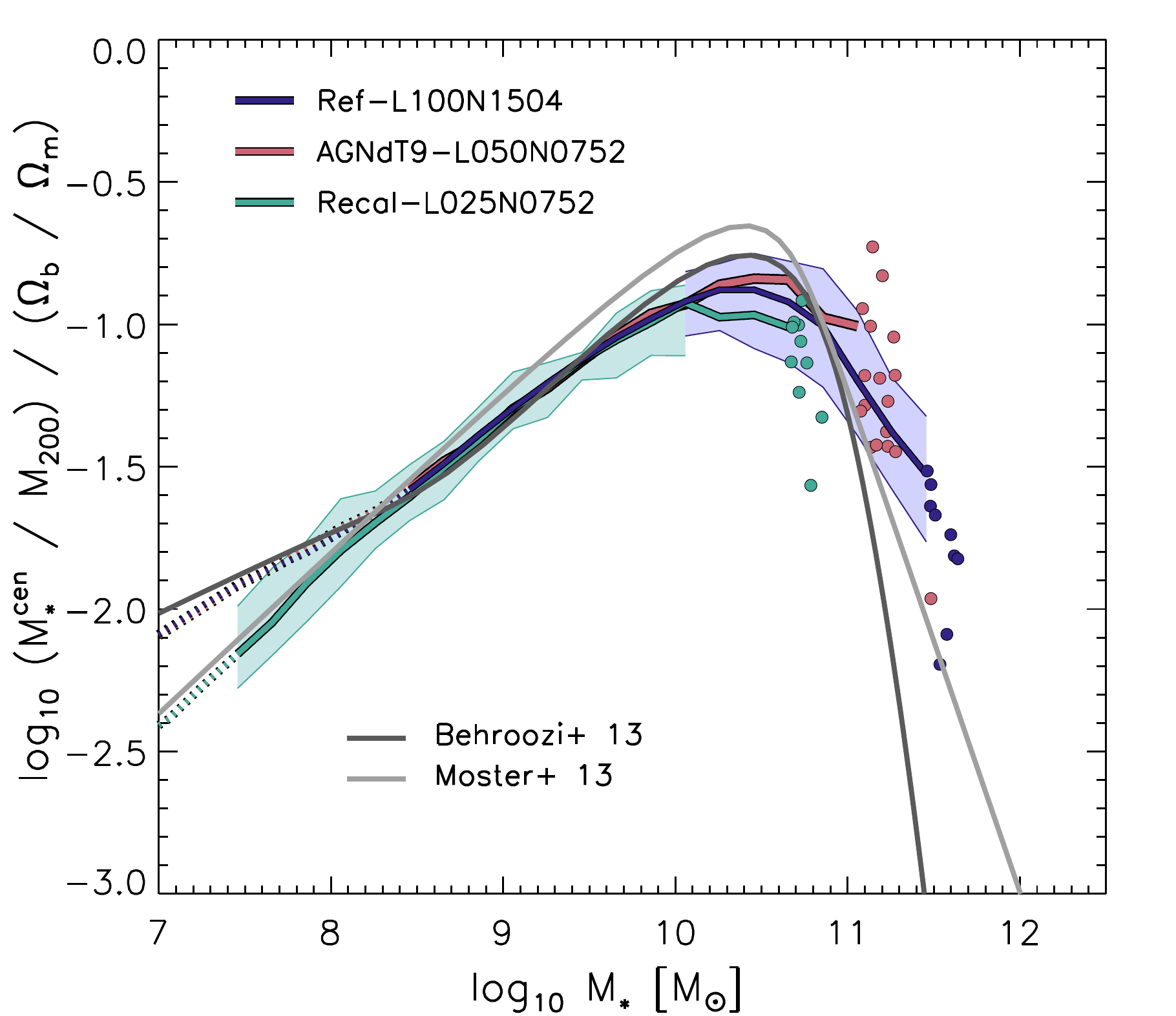}} 
\caption{The ratio of the stellar to halo mass, relative to the universal baryon fraction, as a function of halo mass (left panel) and stellar mass (right panel) for central galaxies. The simulation curves are dotted where there are fewer than 100 stellar particles per galaxy.  Filled circles show individual objects where there are fewer than 10 objects per bin. The shaded regions show the $1\sigma$ scatter in the simulations. For clarity we only show the scatter in Recal-L025N0752 for $M_\ast < 10^{10}\,\Msun$ and in Ref-L100N1504 for $M_\ast > 10^{10}\,\Msun$. The \eagle\ results agree with results inferred from observations through the technique of abundance matching (grey, solid curves; \citealt{Moster2013AbundMatching,Behroozi2013AbundMatching}). The small difference between \eagle\ and the abundance matching in the location and height of the peak are consistent with \eagle's small underestimate of the GSMF around the knee (see Fig.~\ref{fig:gsmf}).} 
\label{fig:eta} 
\end{figure*}

The GSMF can be thought of as a convolution between the mass function of dark matter haloes and a function describing the galaxy content of the haloes as a function of their mass. 
The halo mass function can be predicted accurately when the cosmology is known, but the galaxy content of haloes is very sensitive to the baryonic processes involved in the formation of galaxies. As modelling galaxy formation is \eagle's primary goal, it is of interest to compare the relation between stellar mass and halo mass in the simulations to the relation inferred from observations. Because the subgrid model for feedback was calibrated to fit the $z\sim 0$ GSMF, the relation between stellar and halo mass can hardly be considered a prediction. We therefore discuss this relation in this section, even though we did not calibrate the simulations to fit the relation inferred from observations.

Figure~\ref{fig:eta} shows the ``galaxy formation efficiency'', $(M_\ast / M_{200}) / (\Omega_{\rm b}/\Omega_{\rm m})$, for central galaxies as a function of either the mass of their host halo (left panel) or their stellar mass (right panel). Here the halo mass, $M_{200}$, is defined as the total mass contained within the virial radius $R_{200}$, defined to be the radius within which the mean internal density is 200 times the critical density, $3H^2/8\pi G$, centred on the dark matter particle of the corresponding FoF halo with the minimum gravitational potential (see \S\ref{sec:simulations}). If the baryon fraction in the halo were equal to the cosmic average of $\Omega_{\rm b}/\Omega_{\rm m} \approx 0.16$, then an efficiency of unity would indicate that the stellar mass accounts for all the halo's share of baryons.  
We focus on central galaxies because the strong tidal stripping to which satellite haloes are subject obscures the underlying relation between galaxy formation efficiency and halo mass.

The simulation clearly shows that galaxy formation is most efficient in haloes with mass $\sim 10^{12}\,\Msun$, as has been found by many others. In fact, it would be more appropriate to say that this is the mass where galaxy formation is ``least inefficient'' as the efficiency is only $\sim 10$\% at the peak. The efficiency is sharply peaked at a stellar mass of $\sim 10^{10.4}\,\Msun$, which corresponds to the onset of the knee in the GSMF (Fig.~{\ref{fig:gsmf}). As is the case for most models of galaxy formation, in \eagle\ the sharp reduction at lower masses is mostly due to stellar feedback, while the drop off at higher masses can in part be attributed to inefficient cooling, but is mostly caused by AGN feedback. 

Although halo masses can be measured observationally, e.g.\ from gravitational lensing or satellite kinematics, the errors are still relatively large and it is difficult to disentangle central and satellite galaxies. In Figure~\ref{fig:eta} we therefore compare with results obtained through the abundance matching technique. In its most basic form abundance matching relates central galaxies to haloes by matching the observed GSMF to the halo mass function predicted from a collisionless simulation, assuming that the stellar masses of galaxies increase monotonically with the masses of their host haloes \citep[e.g.][]{Vale2004AbundanceMatching}. Modern versions allow for scatter and evolution, and assume that the masses of satellite galaxies are set at the last time they were centrals. 

Figure~\ref{fig:eta} compares \eagle\ to the abundance matching results of \citet{Behroozi2013AbundMatching} and \citet{Moster2013AbundMatching}. Note that the abundance matching studies assumed the WMAP7 cosmology, whereas we assume the Planck cosmology. For \eagle\ we use the total mass of the halo in the hydrodynamical simulation, whereas abundance matching studies use collisionless simulations. Because feedback processes reduce halo masses, we expect $M_{\rm 200}$ to be biased high by $\sim 10$\% for the abundance matching results \citep[e.g.][]{Sawala2013HaloMass,Cui2014HaloMass,Velliscig2014HaloMass,Cusworth2014ClusterMass,Martizzi2014ClusterMass,Sawala2014EagleZooms,Vogelsberger2014Illustris}, but this effect is small compared to the dynamic range shown\footnote{For $M_{200} \ll 10^{10}\,\Msun$ the systematic errors in the abundance matching results are likely to be much greater because only a small fraction of such low-mass haloes may host galaxies \citep{Sawala2013HaloMass,Sawala2014EagleZooms}.}. Beyond the peak the results become increasingly sensitive to the aperture used to measure the galaxy's light. For example, \citet{Kravtsov2014GalaxyHalo} show that using the \citet{Bernardi2013GSMF} GSMF as input increases the efficiency by $\sim 0.5$ dex at $M_{\rm 200}  = 10^{14}\,\Msun$ relative to the values of \citet{Behroozi2013AbundMatching} and \citet{Moster2013AbundMatching}. However, as discussed in \S\ref{sec:aperture}, our use of a fixed 30~pkpc aperture means that comparison to \citet{Bernardi2013GSMF} is inappropriate at the high-mass end. In \S\ref{sec:groups} we will show that a more robust comparison with observations of the total stellar content of massive galaxies reveals good agreement with \eagle.

The convergence with resolution is good and the galaxy formation efficiency in \eagle\ is very close to that inferred from abundance matching. This was of course to be expected, given the good convergence and the good agreement with the observations for the GSMF. 
The peak efficiency is 0.1--0.2 dex lower in \eagle\ and is reached at a slightly ($\sim 0.2$~dex) higher stellar mass, which is consistent with the fact that \eagle\ slightly undershoots the observed GSMF at the knee (see Fig.~\ref{fig:gsmf}). 

\subsection{Galaxy sizes}
\label{sec:sizes}

\begin{figure}
\resizebox{\colwidth}{!}{\includegraphics{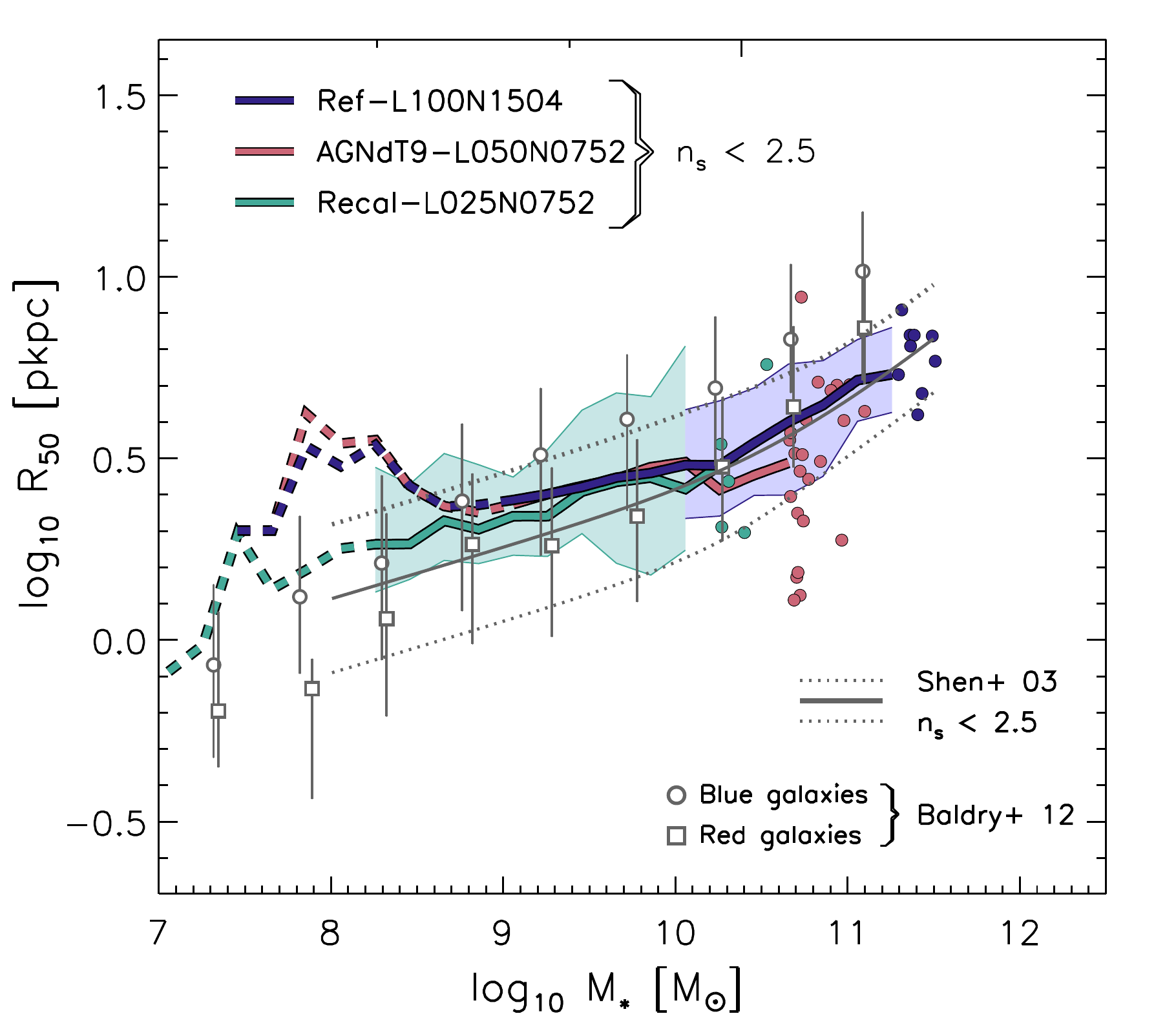}} 
\caption{Galaxy size as a function of stellar mass for galaxies at $z=0.1$ in proper kiloparsec. The coloured curves show the median, projected half-mass radii for the simulations and the shaded regions show the $1\sigma$ scatter. For clarity we only show the scatter in Recal-L025N0752 for $M_\ast < 10^{10}\,\Msun$ and in Ref-L100N1504 for $M_\ast > 10^{10}\,\Msun$. The simulation curves are dotted below the resolution limit of 600 stellar particles. Where there are fewer than 10 galaxies per bin, individual objects are shown as filled circles. The models are compared with S\'ersic half-light radii from SDSS (\citealt{Shen2003Sizes}; the grey, solid line shows the median and the grey dotted lines indicate 1$\sigma$ scatter) and GAMA (\citealt{Baldry2012GSMF}; data points with error bars indicate the $1\sigma$ scatter, shown separately for blue and red galaxies). The simulations and \citet{Shen2003Sizes} only include late-type galaxies, i.e.\ a S\'ersic index $n_{\rm s} < 2.5$.}
\label{fig:sizes}
\end{figure}

The parameters of the subgrid model for feedback from star formation and AGN were calibrated to observations of the $z\sim 0$ GSMF. The parameter that controls the importance of the angular momentum of the gas in suppressing BH accretion was set to a value for which AGN feedback causes the GSMF to turn over at a mass similar to what is observed. 
As will be shown in \citet{Crain2014EagleModels}, we found that for \eagle, calibration of the stellar feedback is actually unnecessary to reproduce the GSMF. Fixing the amount of energy injected per unit stellar mass to that available in the form of core collapse supernovae, i.e.\ $f_{\rm th}=1$, works well, as does the physically motivated dependence on the gas metallicity that we use (eq.~\ref{eq:f(Z)}). However, such models produce galaxies that are far too compact because of excessive radiative losses at high gas densities, and we can show analytically that these spurious cooling losses are caused by our limited numerical resolution (see \S\ref{sec:snii}). 

We consider it reassuring that the breakdown of the subgrid model for feedback from star formation at high density is understood and leads to a clear conflict with observations. On the other hand, the fact that such an unrealistic model has no trouble matching the observed GSMF emphasizes the importance of comparing to a wide range of observables. 

To counteract the numerical radiative losses occurring at high gas densities, we introduced a dependence of the feedback energy from star formation on the gas density, while keeping both the maximum and mean amounts of energy reasonable (see \S\ref{sec:calibration}).  
Although we could not afford the computational expense of calibrating the models to fit both the $z\sim 0$ GSMF and the size distribution in detail, we did reject models that produced galaxies that were far too small. As a consequence of this strategy, the $z\sim 0$ galaxy sizes cannot be regarded as true predictions.

Figure~\ref{fig:sizes} plots the median value of the half-mass radius, $R_{50}$, i.e.\ the radius that encloses 50 per cent of the stellar mass in projection, as a function of galaxy stellar mass. 
 The half-mass radii were determined by fitting S\'ersic laws to the projected, azimuthally averaged surface density profiles, as in \citet{McCarthy2012RotSize}. Following \citet{Shen2003Sizes}, only galaxies with S\'ersic index $n_{\rm s}<2.5$ are included. For Ref-L1001504, 94\% of the galaxies with more than 600 star particles have $n_{\rm s}<2.5$. 

The high-resolution Recal-L025N0752 agrees very well with the intermediate-resolution models for $M_\ast > 10^{9}\,\Msun$, which corresponds to about 600 star particles for the intermediate-resolution runs. For this mass the median $R_{50}$ is about three and a half times the maximum gravitational softening length (see Table~\ref{tbl:sims}). Hence, we take the stellar mass $600 m_{\rm g}$ as the minimum value for which we can measure half-mass radii. We thus require six times more stellar particles to measure sizes than we need to measure mass. 

The simulations are compared to data from SDSS \citep{Shen2003Sizes} and GAMA \citep{Baldry2012GSMF}. Note that the observations fit surface brightness profiles and provide half-light radii rather than half-mass radii, so the comparison with the models is only fair if the stellar mass-to-light ratio does not vary strongly with radius. As mentioned above, \citet{Shen2003Sizes} select galaxies with $n_{\rm s}<2.5$, as we have done here. \citet{Baldry2012GSMF} on the other hand present results separately for red and blue galaxies, finding that the latter are $\sim 0.2$ dex more extended at fixed stellar mass.  \citet{Shen2003Sizes} use Petrosian apertures, which we expect to yield results similar to the 3-D apertures of 30 pkpc that we use for the simulations (see \S\ref{sec:aperture}). 

For $M_\ast \gg 10^8\,\Msun$ \citet{Shen2003Sizes} agree better with the \citet{Baldry2012GSMF} results for red galaxies, even though $n_{\rm s} < 2.5$ should pick out more disky and hence bluer galaxies. The differences between the two data sets are indicative of the level of correspondence between independent measurements of observed galaxy sizes. 

For $10^9 < M_\ast/\Msun < 10^{10}$ the simulation results fall in between those of \citet{Baldry2012GSMF} for red and blue galaxies. For $M_\ast< 10^9\,\Msun$ and $M_\ast> 10^{10}\,\Msun$ the simulations agree very well with the sizes of blue and red galaxies, respectively. At $10^{11}\,\Msun$ the red sample of \citet{Baldry2012GSMF} gives sizes that are about 0.1--0.2 dex larger than found for both the simulations and the data from \citet{Shen2003Sizes}. This difference may be due to the fact that \citet{Shen2003Sizes} use Petrosian sizes, whereas \citet{Baldry2012GSMF} do not. Indeed, if we do not impose any 3-D aperture, then the simulation curve follows the results of the red sample nearly exactly for $M_\ast \ga 10^{11}\,\Msun$, while the sizes of lower-mass galaxies remain unchanged (not shown). The agreement with \citet{Shen2003Sizes} is excellent: the difference with the simulations is $\le 0.1$ dex for all models and for the full range of stellar mass. 

For $M_\ast > 10^{10}\,\Msun$ the scatter in the sizes of the simulated galaxies is similar to the observed dispersion, but at lower masses it appears to be smaller. This could be due to a lack of resolution or some other deficiency in the simulations or halo finder, but it could also be due to observational errors or to the fact that we have ignored variations in the stellar mass-to-light ratio and dust extinction.

\subsection{The relation between BH mass and stellar mass}
\label{sec:magorrian}

Figure~\ref{fig:bh} shows the mass of the central supermassive BH as a function of the galaxy's stellar mass. The simulation results are compared with the compilation of observations from \citet{McConnell2013BH}. The observed stellar mass was obtained by extrapolating a fit to the mass profile of the bulge inferred from kinematic data. Because the observed galaxies were selected to be early-type, the bulge likely dominates the stellar mass, at least for the massive systems. 

The three \eagle\ simulations give nearly identical results, indicating good convergence. For $M_\ast \ll 10^{10}\,\Msun$ the BH mass asymptotes to $10^5\,\Msun/h$, which is the mass of the seed BHs that are inserted into FoF haloes with mass $>10^{10}\,\Msun/h$ that do not already contain BHs. As can be seen from Fig.~\ref{fig:eta}, a halo mass of $10^{10}\,\Msun$ corresponds to $M_\ast\sim 10^8\,\Msun$. Above $M_\ast \sim 10^{10}\,\Msun$ the relation between BH mass and stellar mass steepens, but it quickly flattens off to a relation that agrees very well with the observations for $M_\ast \ga 10^{11}\,\Msun$. The rapid growth of the BHs between $M_\ast = 10^{10}$ and $10^{11}\,\Msun$ coincides with the steepening of the GSMF (compare Fig.~\ref{fig:gsmf}) and the sharp increase in the fraction of galaxies that are passive (right panel of Fig.~\ref{fig:ssfr}). This is understandable, as the AGN feedback associated with the rapid BH growth quenches star formation.
 
The agreement with the observations is good, although the observed scatter is larger. In terms of the normalisation of the $M_{\rm BH}$-$M_\ast$ relation the good agreement is perhaps not a surprise. The normalisation is determined by the assumed efficiency of the AGN feedback, $\epsilon_{\rm f}\epsilon_{\rm r}$, i.e.\ the amount of energy that is injected per unit of accreted mass \citep[e.g.][]{Booth2009AGN,Booth2010DMHaloesBHs}}. We used the same value ($\epsilon_{\rm f}\epsilon_{\rm r} = 0.015$) as was used for \owls\ and \cosmo, which \citet{Booth2009AGN} and \citet{LeBrun2014CosmoOWLS} found to give agreement with the observed $M_{\rm BH}$-$M_\ast$ relation. Fig.~\ref{fig:bh} shows that this efficiency also works for \eagle, even though the mass resolution of \eagle\ is nearly two orders of magnitude better than for \owls\ and about 3 orders of magnitude better than for \cosmo. Note, however, that we used higher AGN heating temperatures than the $\Delta T_{\rm AGN} = 10^8\,\K$ that was used in \owls\ (see Table~\ref{tbl:subgridpars}).

It would clearly be desirable to extend the comparison to observations to lower masses, but in this regime a more careful analysis is required. This is because of the importance of systematic and selection effects for the observations \citep[e.g.][]{Lauer2007BHBias,Schulze2011BHBias} and because a bulge-to-disc decomposition would be necessary for the simulations since most low-mass galaxies are disky. The same issues likely also affect the comparison of the scatter.

\begin{figure} 
\resizebox{\colwidth}{!}{\includegraphics{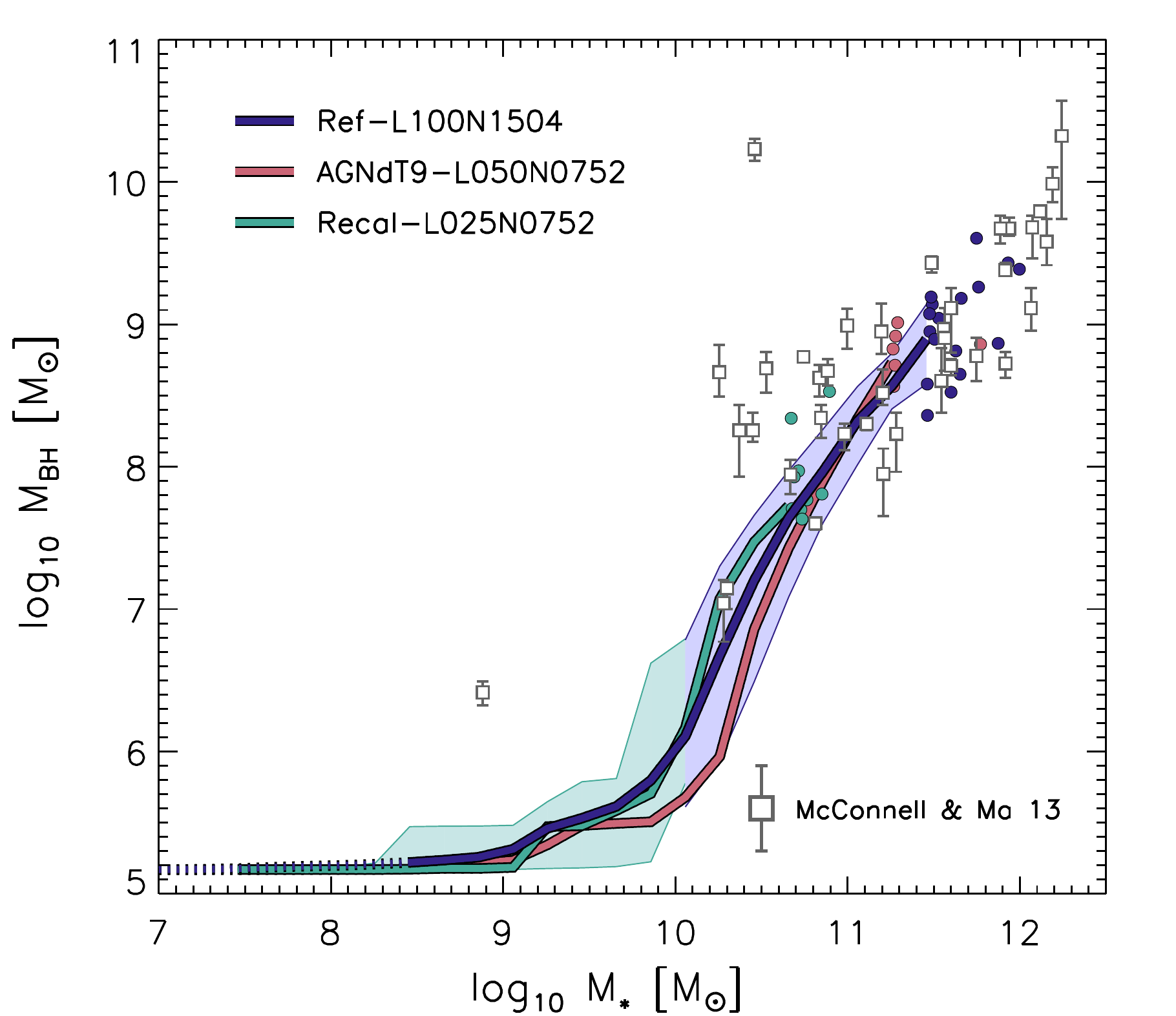}} 
\caption{The relation between the mass of the central supermassive black hole and the stellar mass of galaxies. The coloured curves show the median relations for the simulations and the shaded regions show the $1\sigma$ scatter. For clarity we only show the scatter in Recal-L025N0752 for $M_\ast < 10^{10}\,\Msun$ and in Ref-L100N1504 for $M_\ast > 10^{10}\,\Msun$. Where there are fewer than 10 objects per bin, individual objects are shown as filled circles.
Data points with $1\sigma$ error bars show the compilation of observations from \citet{McConnell2013BH}. The simulations show the total stellar mass (within a 3-D aperture of 30~pkpc), while observations show bulge masses. However, the observed galaxies were selected to be early-type. The simulations agree with the observations, although the observed scatter is larger.}
\label{fig:bh} 
\end{figure}

\section{Comparison with other observations}
\label{sec:otherobs}

In this section we will compare the results of \eagle\ to a diverse set of low-$z$ observations of galaxies, galaxy clusters, and the IGM. The results reported in this section were not used to calibrate the subgrid models for feedback and can therefore be considered predictions that can be used as independent consistency checks. During the testing phase, we did look at earlier, more basic versions of some of the plots shown here, so most of the predictions cannot be considered blind. However, we have not adjusted any model parameters to improve the results shown in this section. 

There are two exceptions to the above statements. First, we plotted the metal column density distributions (\S\ref{sec:cddf}) for the first time after the simulations had finished, so this was a truly blind prediction. Second, the discrepancy between the gas fraction in clusters predicted by Ref-L100N1504 and inferred from X-ray observations that will be discussed in \S\ref{sec:groups} was the motivation for running model AGNdT9-L050N0752. This model represents an educated guess in terms of the modifications to the subgrid AGN feedback, because we could only afford to calibrate models using volumes of 25 cMpc on a side, which are too small to contain clusters of galaxies. 

The observables presented in this section were not selected because the models reproduce them accurately. They were selected because they give a broad overview of the $z\sim 0$ \eagle\ universe, because we had the tools to compute them, and because we are currently not preparing separate papers on them. Future papers will present more observables as well as results for higher redshifts. 

\subsection{Specific star formation rates and passive fractions}
\label{sec:ssfr}

\begin{figure*}
\resizebox{\colwidth}{!}{\includegraphics{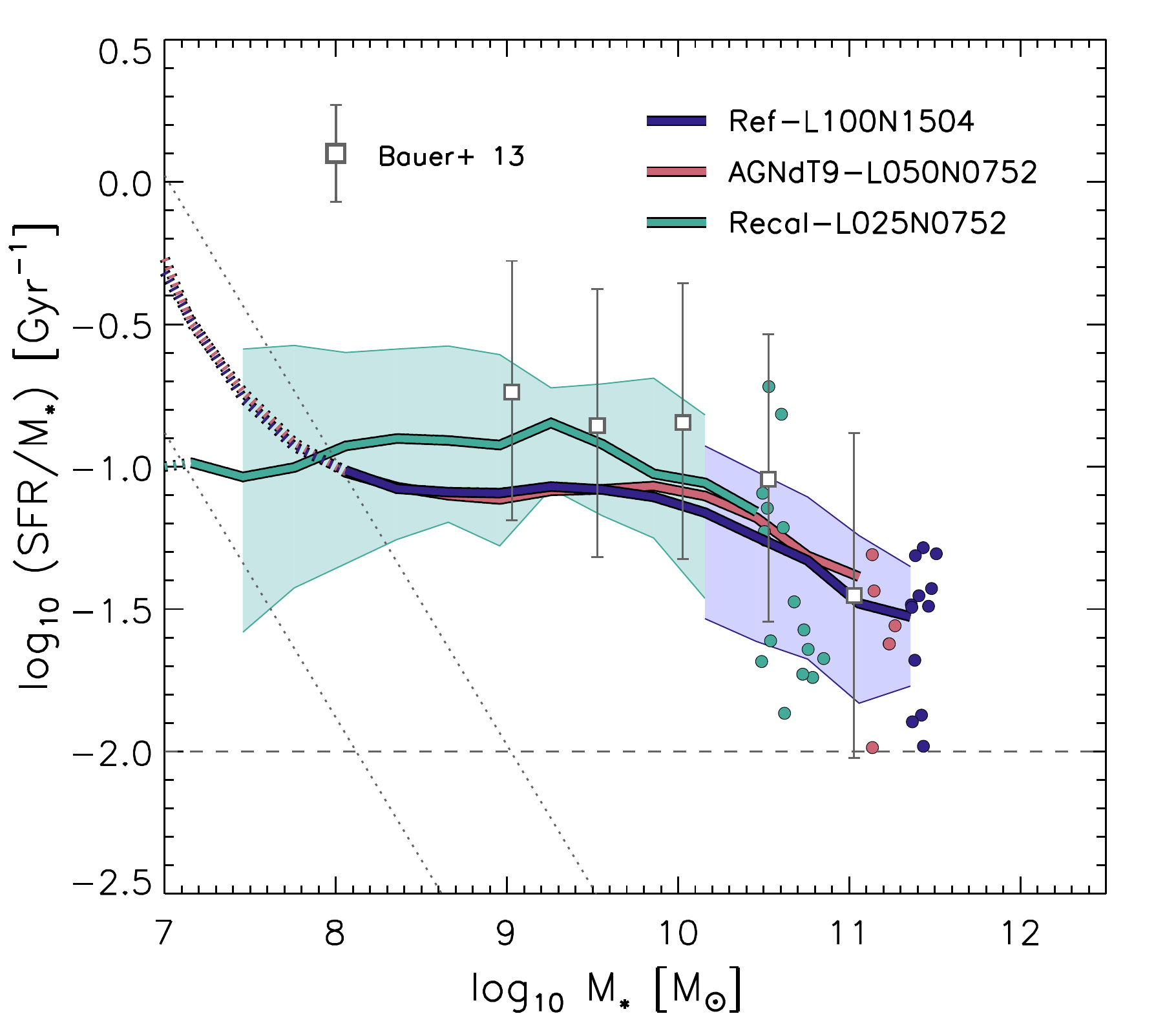}} 
\resizebox{\colwidth}{!}{\includegraphics{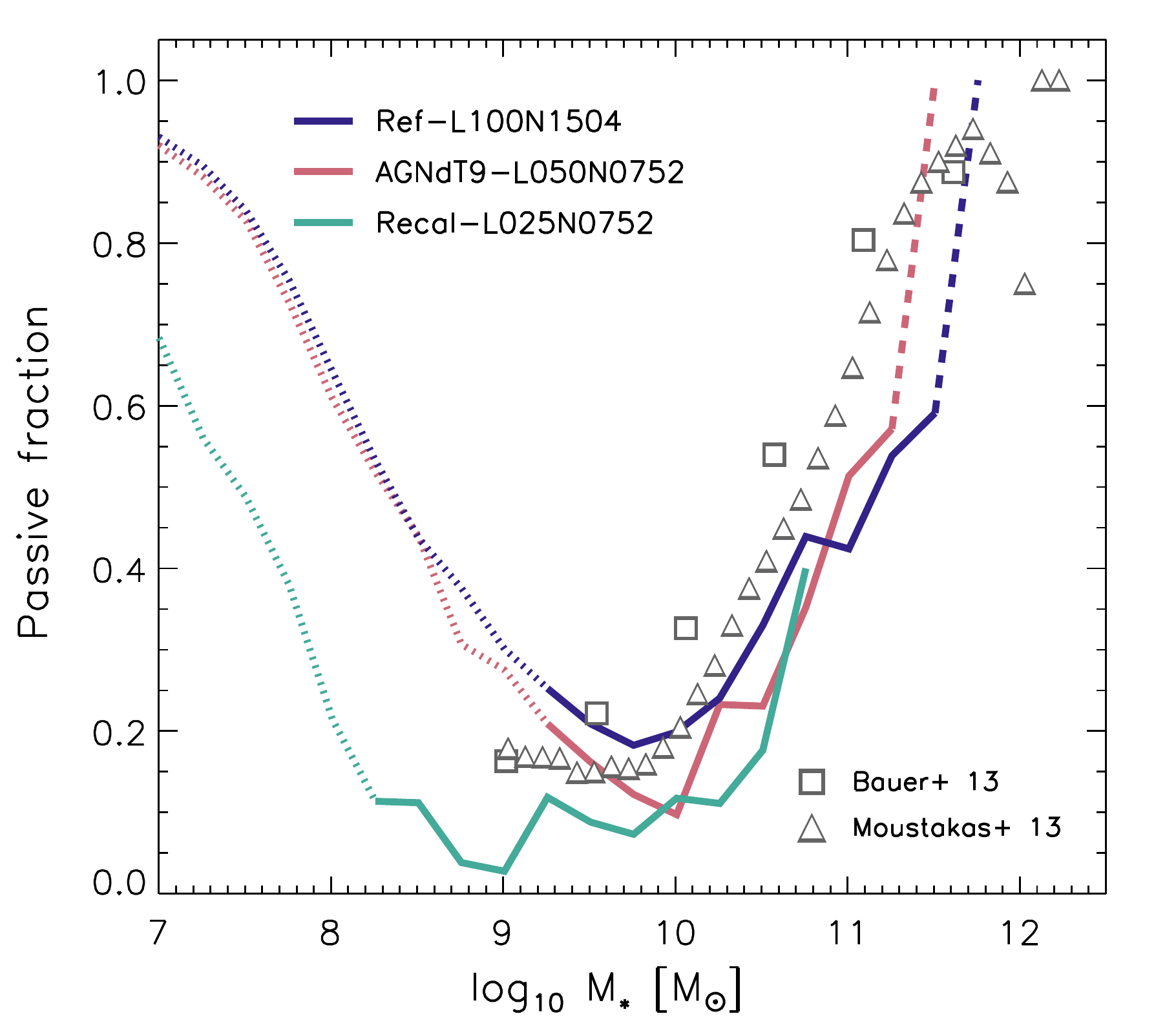}} 
\caption{\emph{Left panel:} Specific star formation rate, $\dot{M}_\ast/M_\ast$, for actively star-forming galaxies as a function of stellar mass at $z=0.1$. Galaxies are classified as star-forming if their SSFR $> 10^{-2}\,\Gyr^{-1}$, indicated by the horizontal, dashed line. The coloured curves show simulation medians and the shaded regions show the $1\sigma$ scatter. For clarity we only show the scatter in Recal-L025N0752 for $M_\ast < 10^{10}\,\Msun$ and in Ref-L100N1504 for $M_\ast > 10^{10}\,\Msun$, all at $z=0.1$. The higher and lower diagonal lines correspond to 10 star-forming gas particles (assuming $n_{\rm H} = 0.1~\cm^{-3}$) at intermediate and high resolution, respectively. To the left of these lines the curves are dotted to indicate that the results are unreliable due to sampling effects. In particular, the sharp upturns at the lowest masses trace lines of fixed numbers of star-forming gas particles. 
The data points show observations from GAMA ($0.05 < z < 0.32$; \citealt{Bauer2013ssfr}) with the error bars indicating the $1\sigma$ scatter. 
\emph{Right panel:} Fraction of passive galaxies, i.e.\ galaxies with SSFR $< 10^{-2}\,\Gyr^{-1}$, as a function of stellar mass at $z=0.1$. In both panels the simulation curves are dotted where they are unreliable due to poor resolution ($< 10$ star-forming gas particles) and dashed were there are $<10$ objects per bin. Data points show observations from
\citet{Bauer2013ssfr} and \citet{Moustakas2013GSMF}.}
\label{fig:ssfr}
\end{figure*}

The left panel of Figure~\ref{fig:ssfr} shows the specific star formation rate (SSFR), $\dot{M}_\ast/M_\ast$, of actively star-forming galaxies as a function of stellar mass. Here, galaxies are classified to be star-forming if the ${\rm SSFR} > 0.01~\Gyr^{-1}$, which is indicated by the horizontal, dashed line in the left panel.
The higher and lower diagonal lines in the left panel indicate the SSFR corresponding to 10 star-forming gas particles (assuming a gas density of $n_{\rm H}=10^{-1}\,\cm^{-3}$, the star formation threshold that we impose at the metallicity $Z=0.002$) at intermediate and high resolution, respectively. To the left of these curves resolution effects become important, which we indicate by using dotted lines. In particular, the increase in the SSFR at low stellar mass that is clearly visible for the intermediate resolution simulations is a numerical effect: the curves trace lines of constant numbers of star-forming particles. Compared with the intermediate-resolution models, the high-resolution simulation Recal-L025N0752 predicts slightly higher SSFRs. The difference is 0.2 dex at $M_\ast = 10^9\,\Msun$ and less than 0.1 dex above $10^{10}\,\Msun$. 

The models are compared with observations from \citet{Bauer2013ssfr}, who measured the SSFRs of $\sim 73,000$ galaxies from the GAMA survey using spectroscopic H$\alpha$ measurements and dust corrections based on Balmer decrements. The intermediate-resolution simulations agree with the data at the high-mass end, but underpredict the SSFR at low masses, reaching a maximum discrepancy of 0.3 -- 0.4 dex at $10^9\,\Msun$. The high-resolution model also underpredicts the SSFR, but the discrepancy is less than 0.2 dex. These differences are comparable to the systematic uncertainty in the data. For example, even for a fixed IMF the systematic uncertainty in the stellar mass, which shifts the data parallel to the diagonal lines, is $\sim 0.3$~dex \citep{Conroy2009SPSSUncertainty,Behroozi2010Uncertainties,Pforr2012SPSSUncertainty,Mitchell2013MassUncertainty} and the systematic error in the star formation rate, which shifts the data vertically, is likely to be at least as large \citep[e.g.][]{Moustakas2006SfrIndicators}. The scatter in the simulations is $\sim 50$\% smaller than observed, but the observed scatter includes measurement and systematic uncertainties. 

The right panel of Figure~\ref{fig:ssfr} shows the fraction of galaxies that are passive as a function of stellar mass. For the simulations we classify galaxies as passive if they have ${\rm SSFR} < 0.01~\Gyr^{-1}$, but the observational papers use somewhat different and varying criteria. We leave a more precise comparison for future work, e.g.\ using colours and accounting for dust extinction for the simulated galaxies. At low stellar masses the curves become dashed where there are, on average, fewer than 10 star-forming gas particles in a galaxy with ${\rm SSFR} = 0.01~\Gyr^{-1}$. These parts of the curves are unreliable and the upturn of the passive fraction at low mass is thus due to the limited resolution of the simulations. This interpretation is confirmed by the fact that the upturn shifts to eight times lower masses if the particle mass is decreased by a factor of eight, switching from the intermediate resolution Ref-L100N1504 to the high-resolution Recal-L025N0752. 

For $M_\ast \gg 10^9\,\Msun$, where the simulations are close to converged, both the simulations and the observations show a strong increase of the passive fraction with mass, from $\sim 10$ per cent at $10^9\,\Msun$ to $\sim 90$ per cent at $10^{11.5}\,\Msun$. Relative to the data, the simulation curves are shifted towards higher stellar masses by about 0.3 dex. This difference is similar to the systematic uncertainty in the observed stellar masses. We also find shifts of similar magnitudes if we vary the critical SSFR below which simulated galaxies are classified as passive by a factor of two. 

We conclude that in the regime where the simulations can be trusted, the predicted SSFRs and passive fractions are slightly lower than the observations but agree with them to within the expected (systematic) errors.

\subsection{Tully-Fisher relation}
\label{sec:tf}

\begin{figure}
\resizebox{\colwidth}{!}{\includegraphics{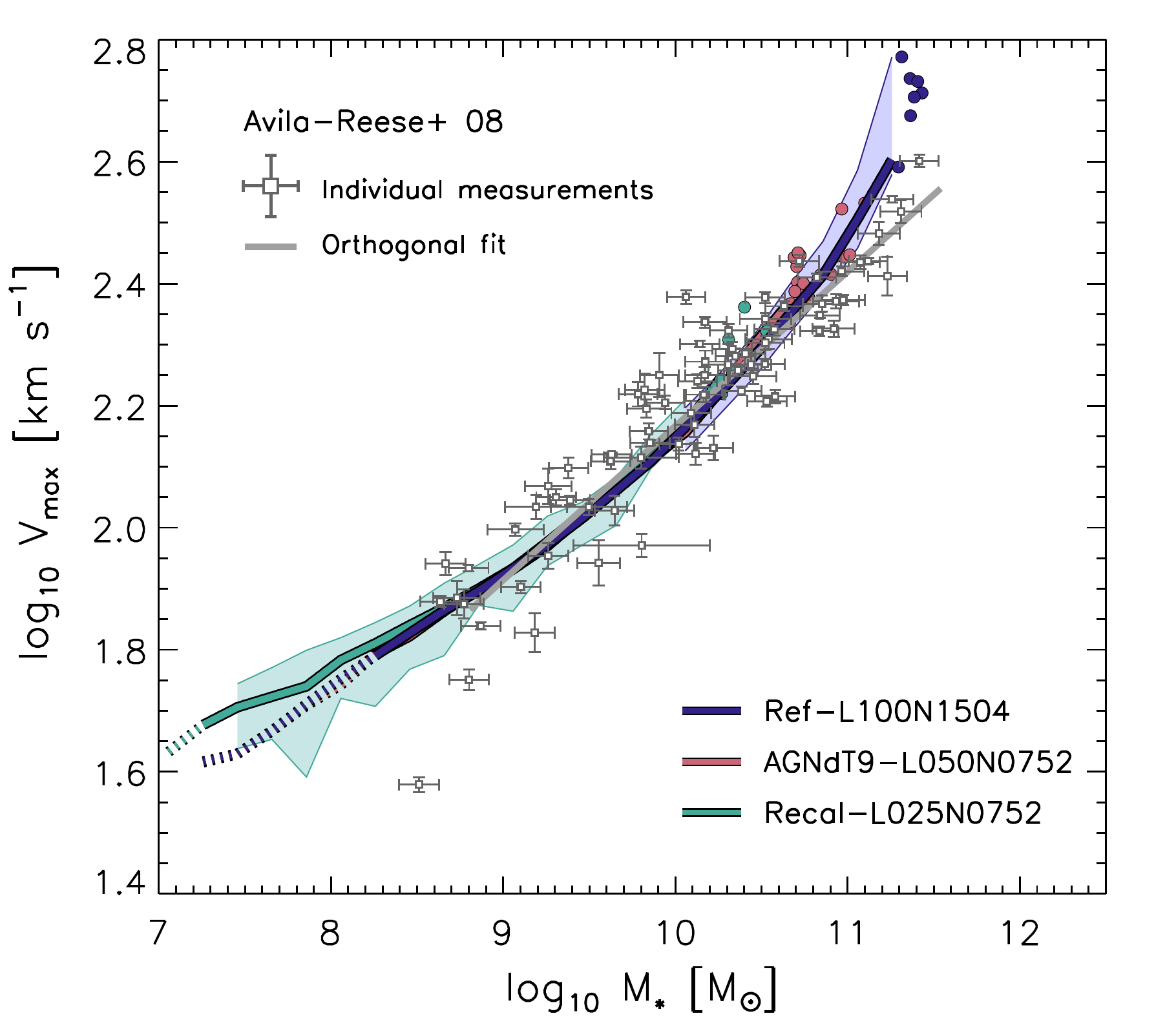}} 
\caption{The relation between the maximum of the rotation curve and stellar mass, i.e.\ an analogue of the Tully-Fisher relation, for late-type galaxies at $z=0.1$. The coloured curves show the medians for the simulations. The curves are dotted below the resolution limit of 100 stellar particles. Where there are fewer than 10 galaxies per bin, individual objects are shown as filled circles. The shaded regions show the $1\sigma$ scatter in the simulations. For clarity we only show the scatter in Recal-L025N0752 for $M_\ast < 10^{10}\,\Msun$ and in Ref-L100N1504 for $M_\ast > 10^{10}\,\Msun$. The simulation results only include galaxies with S\'ersic index $n_{\rm s} < 2.5$ and are based on maximum circular velocities.  The data points with $1\sigma$ error bars correspond to the set of homogenised observations of disc galaxies compiled by \citet{Avila-Reese2008TF} and the grey line indicates the median. The model predictions are in remarkable agreement with the data.}
\label{fig:tf}
\end{figure}

Figure~\ref{fig:tf} shows the relation between the maximum of the rotation curve and stellar mass for disc galaxies, i.e.\ a close relative of the Tully-Fisher relation \citep{Tully1977TF}. 
For the simulations we classify galaxies with S\'ersic index $n_{\rm s}<2.5$ as late-type, as we did when considering galaxy sizes (\S\ref{sec:sizes}). We use circular velocities ($v_{\rm c}=\sqrt{GM(<r)/r}$) rather than trying to estimate rotation velocities, since the latter become noisy for galaxies that are not resolved with many particles. 

The data points with $1\sigma$ error bars correspond to the set of homogenised observations of disc galaxies compiled by \citet{Avila-Reese2008TF} and the grey line indicates the median.
The stellar masses have been reduced by 0.15 dex, which is necessary to convert to a Chabrier IMF (Avila-Reese, private communication). In addition, following \citet{McCarthy2012RotSize} and \citet{Dutton2011TF}, we applied a small correction to the stellar masses using the expression given in the appendix of \citet{Li2009GSMF} to improve the consistency with those derived from more accurate five-band SDSS data. 

All simulations track each other very closely, implying excellent numerical convergence. The simulations are in excellent agreement with the data. Over the mass range $10^9 \la M_\ast/\Msun <10^{11}$ the difference in velocity between the models and the data compiled by \citet{Avila-Reese2008TF} is less than 0.03~dex, which is smaller than the 0.1~dex $1\sigma$ error on the fit to the observations. At higher masses, which are only probed by Ref-L100N1504, the difference with the observations increases, reaching 0.12 dex at $M_\ast = 10^{11.3}\,\Msun$. However, most of these very massive galaxies do not look disky and would probably not be selected by \citet{Avila-Reese2008TF}. 

Note that we have not attempted to analyse the simulations and the data in the same manner, because this would go beyond the scope of the current study. As mentioned above, we use maximum circular velocities, whereas the observations are based on maximum gas rotation velocities, which may show more scatter if the orbits are not all circular. In addition, the observations probe only the inner parts of the halo, whereas we consider the entire halo. \citet{McCarthy2012RotSize} found that for the \gimic\ simulations the maximum circular velocities are nearly always reached within two effective radii for $M_\ast \ga 10^{9.5}\,\Msun$, and should therefore be easily accessible to the observations, but it is possible that for smaller masses the observations underestimate the maximum rotation velocity. 

\subsection{Mass-metallicity relations}
\label{sec:Zm}

\begin{figure*}
\resizebox{\colwidth}{!}{\includegraphics{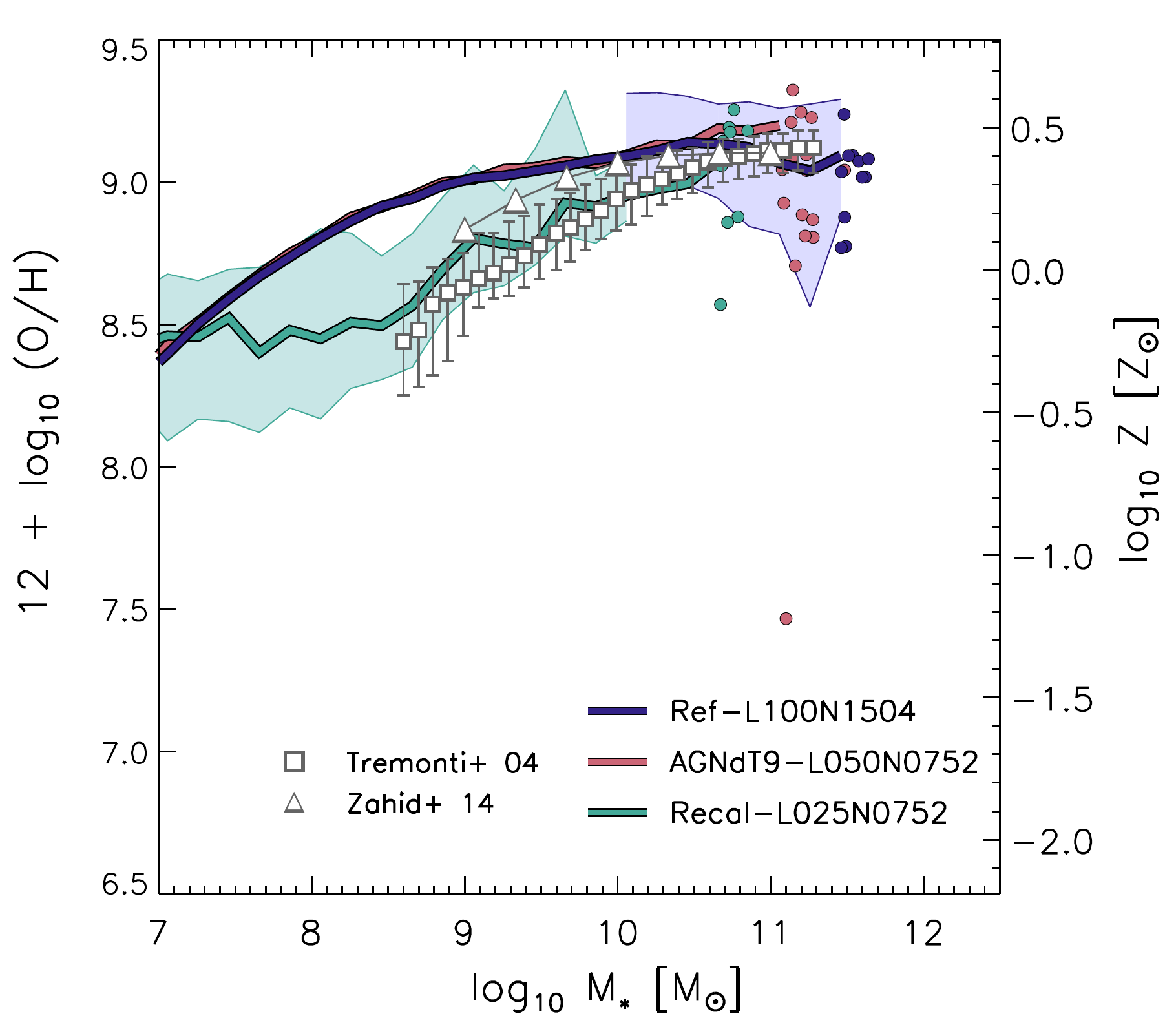}} 
\resizebox{\colwidth}{!}{\includegraphics{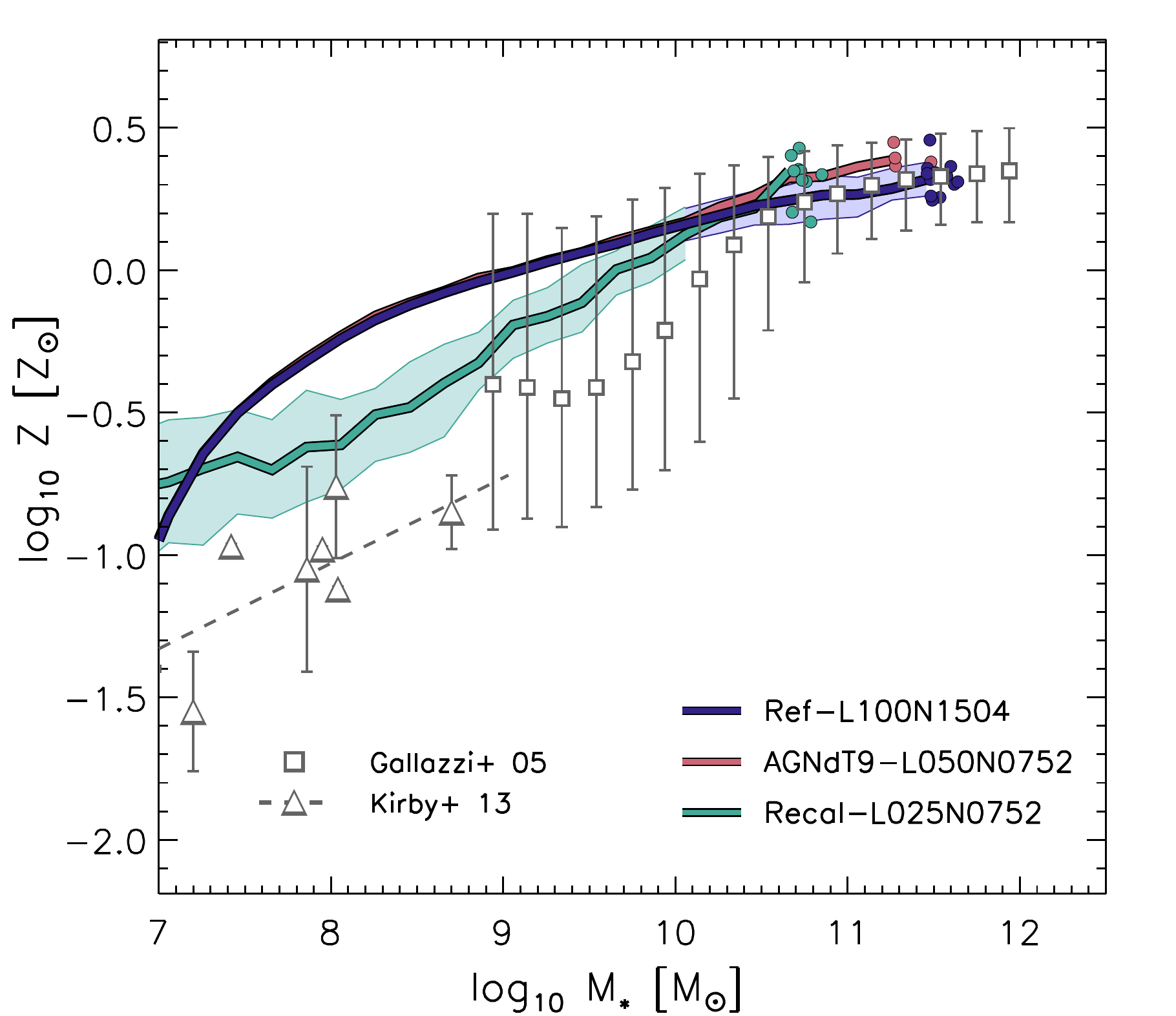}} 
\caption{The metallicity of the ISM (left panel) and of stars (right panel) as a function of stellar mass. The conversion from the absolute oxygen abundances shown along the left $y$-axis in the left panel to the metallicities relative to solar shown along the right $y$-axis  assumes $12 + \log_{10}({\rm O}/{\rm H})_\odot = 8.69$ \citep{AllendePrieto2001Solar}. Note that the two panels show the same range in metallicity. Curves show the median relations for the simulations at $z=0.1$, where we take ISM to be all star-forming gas, and the shaded regions show the $1\sigma$ scatter. For clarity we only show the scatter in Recal-L025N0752 for $M_\ast < 10^{10}\,\Msun$ and in Ref-L100N1504 for $M_\ast > 10^{10}\,\Msun$. Where there are fewer than 10 galaxies per bin, individual objects are shown as filled circles. The high-mass galaxies with very low gas metallicities correspond to objects that are nearly devoid of gas, leading to sampling problems in the simulations. The data points show observations reported by \citet{Zahid2014Zgas} and \citet{Tremonti2004Zgas} for gas, and by \citet{Gallazzi2005Zstars} and \citet{Kirby2013Zstars} for stars (converted to solar abundances assuming $\Zsun = 0.0127$ and $12 + \log_{10}({\rm Fe}/{\rm H})_\odot = 7.52$, respectively). The dashed line in the right panel shows the best-fit relation given by \citet{Kirby2013Zstars}, which includes also lower-mass galaxies than shown here.}
\label{fig:mz}
\end{figure*}

The left panel of Figure~\ref{fig:mz} shows the metallicity of the ISM, which we take to be star-forming gas for the simulations, as a function of stellar mass. For both the intermediate- and the high-resolution models the gas metallicity increases with stellar mass and flattens off for $M_\ast > 10^{10}\,\Msun$. However, the high-resolution simulation, Recal-L025N0752, predicts systematically lower metallicities. For $M_\ast \ga 10^{10}\,\Msun$ the difference is less than 0.15 dex, but it increases with decreasing mass, reaching a maximum of 0.4 dex at $M_\ast \sim 10^{8.5}\,\Msun$. Because there is no clear mass below which the two resolutions diverge, it is unclear where to put the resolution limit and we therefore have not dotted any part of the curves.

Interestingly, model Ref-L025N0752 (not shown) yields a mass-metallicity relation that agrees better with Ref-L100N1504 than the prediction of Recal-L025N0752 does, particularly for $M_\ast < 10^9\,\Msun$. The high-resolution run again predicts lower metallicities than the intermediate-resolution version, but the maximum difference is smaller than 0.2~dex. For $M_\ast < 10^{7.5}\,\Msun$ the metallicity is actually lower at intermediate resolution than at high resolution. Hence, for the mass-metallicity relation the strong convergence is considerably better than one might infer from the comparison of Ref-L025N0752 and Recal-L025N0752. Recall that the latter was recalibrated to fit the GSMF, which meant the efficiency of feedback had to be increased relative to the reference model, particularly at $M_\ast \sim 10^9\,\Msun$ (see Fig.~\ref{fig:gsmf_conv}). Apparently, the stronger outflows in Recal-L025N0752 reduce the metallicity of the ISM. 
Thus, the ``strong convergence'' is better than the ``weak convergence''. This is possible because in this case the weak convergence test compares simulations that were each calibrated to fit the GSMF, not the mass-metallicity relation. 

The two sets of observations that are shown in the left panel of Figure~\ref{fig:mz} are both derived from SDSS data. 
\citet{Tremonti2004Zgas} estimated the metallicity statistically based
on theoretical model fits to various strong emission lines, while \citet{Zahid2014Zgas} derived metallicities using the R23 strong line method as calibrated by \citet{Kobulnicky2004Zgas}. The two studies do not agree with each other. In particular, while \citet{Tremonti2004Zgas} and \citet{Zahid2014Zgas} agree at $M_\ast \sim 10^{11}\,\Msun$, the former find a steeper relation than the latter, resulting in metallicities that are about 0.2 dex lower for $10^9 - 10^{10}\,\Msun$. The difference is due to the uncertain calibration of the emission-line diagnostics. In fact, as shown by \citet{Kewley2008ZCalibration}, the systematic uncertainty is even larger than suggested by this plot. For example, the empirical calibration of \citet{Pilyugin2005ZCalibration} yields a metallicity that is 0.75~dex lower than that of \citet{Tremonti2004Zgas} at $10^{11}\,\Msun$ and an almost flat relation with stellar mass, dropping by only 0.2 dex when the stellar mass decreases to $10^9\,\Msun$. Besides the calibration issues, the gas phase abundance likely underestimates the total metallicity of the ISM because a non-negligible fraction of the metals may condense onto dust-grains \citep[e.g.][]{Dwek1998Dust,Mattsson2012Dust}.
Finally, the systematic uncertainty in the stellar mass, for a fixed IMF, is about 0.3~dex \citep[e.g.][]{Conroy2009SPSSUncertainty}. 

The metallicities predicted by the simulations are also subject to significant systematic uncertainties unrelated to the galaxy formation physics. Even for a fixed IMF, the nucleosynthetic yields are uncertain at the factor of two level \cite[e.g.][]{Wiersma2009Chemo}. However, we choose not to simply re-scale the simulation metallicities  within this uncertainty because that would make them inconsistent with the radiative cooling rates used during the simulation. 

Given the large systematic uncertainties in both the normalisation and the shape of the observed mass-metallicity relation, and the systematic uncertainties in the yields adopted in the simulations, care needs to be taken when comparing the models and the data. We will nevertheless proceed to make such a comparison. 

The median mass-metallicity relations predicted by the intermediate-resolution simulations agree with \citet{Zahid2014Zgas} to better than 0.2 dex at all masses and to better than 0.1 dex for $M_\ast > 10^{9.5}\,\Msun$, but the observed relation is steeper at  lower masses. The predicted scatter is larger than observed by \citet{Tremonti2004Zgas}, particularly for the highest masses. The scatter in the gas metallicity of these massive objects is large in the simulations because they typically contain very few star-forming gas particles. This causes strong sampling effects and large variations in time following AGN outbursts.

The median metallicity predicted by the high-resolution model Recal-L025N0752 matches \citet{Tremonti2004Zgas} to better than 0.2 dex over the full mass range covered by both the simulation and the observations ($10^{8.5} < M_\ast/\Msun < 10^{11}$) and to better than 0.1 dex for $M_\ast > 10^{9.2}\,\Msun$. Apparently, the increase in the efficiency of energy feedback from star formation that is required to make the GSMF fit the observations (and which was implemented by changing the density dependence of the efficiency, see \S\ref{sec:calibration}), simultaneously decreases the metallicity of the ISM of low-mass galaxies to the values observed by \citet{Tremonti2004Zgas}. 

The predicted relations between stellar metallicity and mass are shown in the right panel of Figure~\ref{fig:mz} and compared with observations from SDSS from \citet{Gallazzi2005Zstars} and for dwarf galaxies from \citet{Kirby2013Zstars}. The trends and differences largely parallel those seen for the gas-phase abundances in the left panel. For $M_\ast \ga 10^9\,\Msun$ simulation Recal-L025N0752 is relatively close to the data, but at lower masses all models predict higher metallicities than observed by \citet{Kirby2013Zstars}. As was the case for the gas metallicity, the (strong) convergence is actually much better than suggested by this figure. For $M_\ast > 10^{7.5}\,\Msun$ simulation Ref-L025N0752 (not shown) predicts a stellar metallicity that is lower, but within 0.1 dex of the metallicity predicted by Ref-L100N1504. Model AGNdT9-L050N0752 predicts slightly higher metallicities than Ref-L100N1504 for $M\gg 10^{10}\,\Msun$, which agrees better with the data.

The main difference between the conclusions that can be drawn from the gas and stellar metallicities concerns the scatter. While the scatter in the gas phase abundances was overestimated in the simulations, the scatter in the stellar abundances appears to be strongly underestimated. However, it would be surprising for the scatter in the observed stellar metallicity to be so much larger than the observed scatter in the gas phase metallicity, which suggests that the scatter in the observed stellar metallicities may be dominated by errors. Indeed, while the mean relation from the CALIFA integral field survey is close to that of \citet{Gallazzi2005Zstars}, the scatter is about a factor of two smaller \citep{Gonzalez2014CalifaZM}.

\subsection{X-ray observations of the intracluster medium}
\label{sec:groups}

In this section we will consider some parameters that are commonly measured from X-ray observations of the intragroup and intracluster gas. The comparison to observations is more like-for-like than in previous sections, because all simulation results are derived by applying observational analysis techniques to virtual X-ray observations of the simulations. Simulation Recal-L025N0752 is not considered here because the simulation box is too small to produce clusters of galaxies.  

The methods used to generate the plots are identical to those employed for \cosmo\ in \citet{LeBrun2014CosmoOWLS} and we refer the reader to \S2.2 of that paper for details. Briefly, gas density, temperature and metallicity profiles are determined by fitting single temperature, single metallicity ``Astrophysical Plasma Emission Code'' (APEC) \citep{Smith2001APEC} models to synthetic \emph{Chandra} X-ray spectra in three-dimensional radial bins centred on the minimum of the gravitational potential in the halo. Mass profiles are obtained by fitting the functions proposed by \citet{Vikhlinin2006} to the density and temperature profiles and assuming hydrostatic equilibrium. We then determine the radius within which the mean internal density equals 500 times the critical density, $R_{500, {\rm hse}}$ , and the corresponding spherical overdensity mass,$M_{500, {\rm hse}}$. We will use the subscript ``hse'' to indicate that the quantity has been inferred from virtual observations under the assumption of hydrostatic equilibrium (which holds only approximately, see \citealt{LeBrun2014CosmoOWLS} and references therein). Mean X-ray temperatures and elemental abundances within $R_{500, {\rm hse}}$ are determined by fitting APEC models to a single radial bin. We include all $z=0$ haloes with FoF mass $ > 10^{12.5}\,\Msun$ but plot only results for haloes with $M_{500, {\rm hse}} > 10^{13}\,\Msun$ for which the correspondence between $M_{\rm 500}$ and $M_{500, {\rm hse}}$ is good for most objects, except that $M_{500, {\rm hse}}$ is systematically biased low by $\sim 20$ per cent \citep[see Fig.~B1 of][]{LeBrun2014CosmoOWLS}. 

\begin{figure}
\resizebox{\colwidth}{!}{\includegraphics{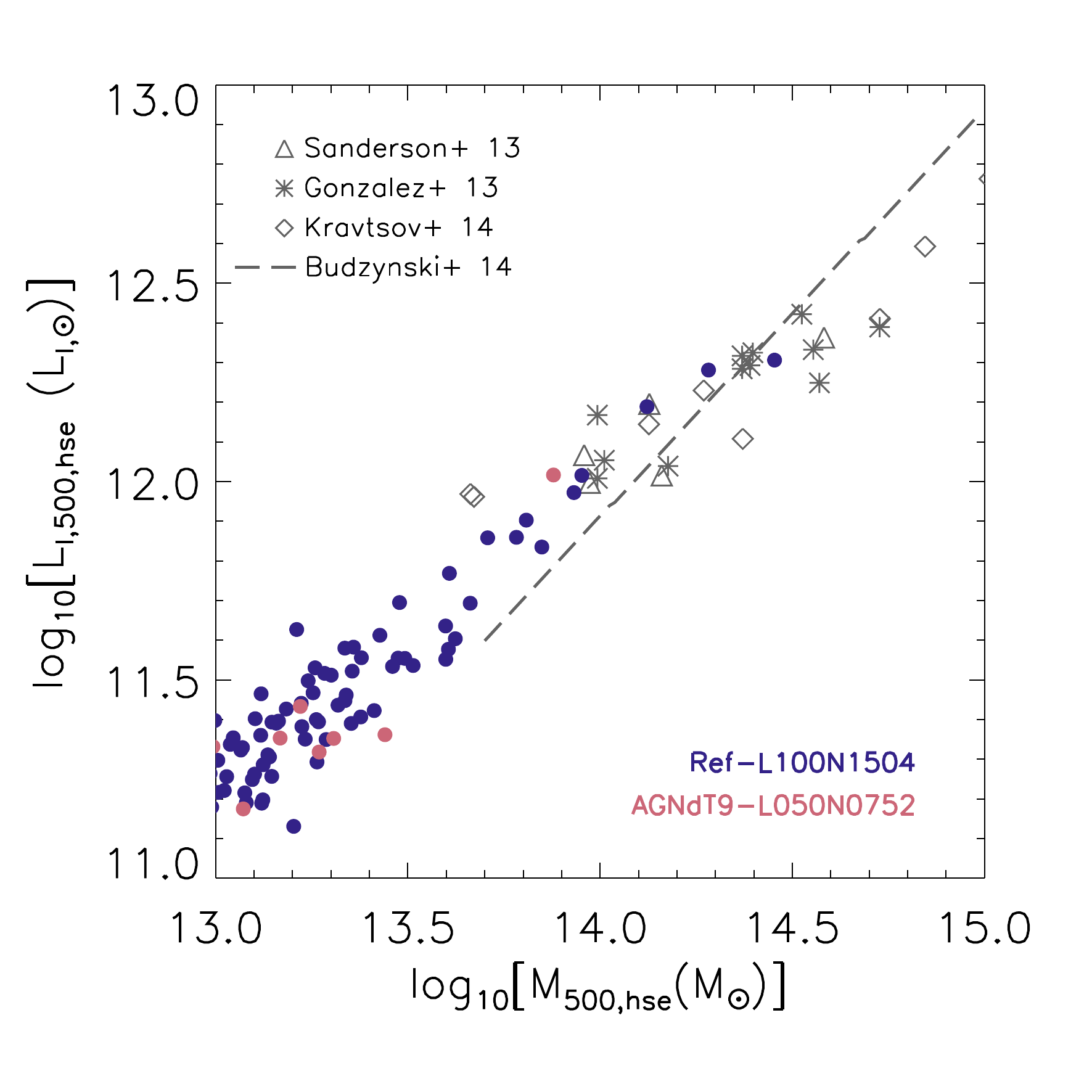}} 
\caption{\textit{I}-Band luminosity within $R_{500,{\rm hse}}$ as a function of $M_{500,{\rm hse}}$ at $z=0$. Black data points represent observations of \citet{Sanderson2013}, \citet{Gonzalez2013}, and \citet{Kravtsov2014GalaxyHalo}, and the dashed black line represents the SDSS image stacking results of \citet{Budzynski2014}. Where necessary, observations were converted to the $I$-band following \citet{LeBrun2014CosmoOWLS}. The observational studies and the simulations both include contributions from satellites and diffuse intracluster light (ICL). The simulations agree well with the data.}
\label{fig:Lopt_groups}
\end{figure}

Figure~\ref{fig:Lopt_groups} shows the (Cousins) $I$-band luminosity within $R_{500,{\rm hse}}$ as a function of $M_{500,{\rm hse}}$. Each point corresponds to a single simulated or observed object.
The predicted luminosity-mass relation matches the observations very well. As the $I$-band luminosity is a proxy for stellar mass and the simulations were calibrated to the observed GSMF, this may at first sight not be surprising. However, the high-mass tail of the GSMF was not calibrated to any observations, because the test simulations were too small to contain such rare objects. Moreover, here we plot the total luminosity within $R_{500}$, a radius that exceeds the aperture used for the GSMF by more than an order of magnitude. Hence, the results shown here include contributions from satellites and the intracluster light, both for the observations and simulations. 

\begin{figure}
\resizebox{\colwidth}{!}{\includegraphics{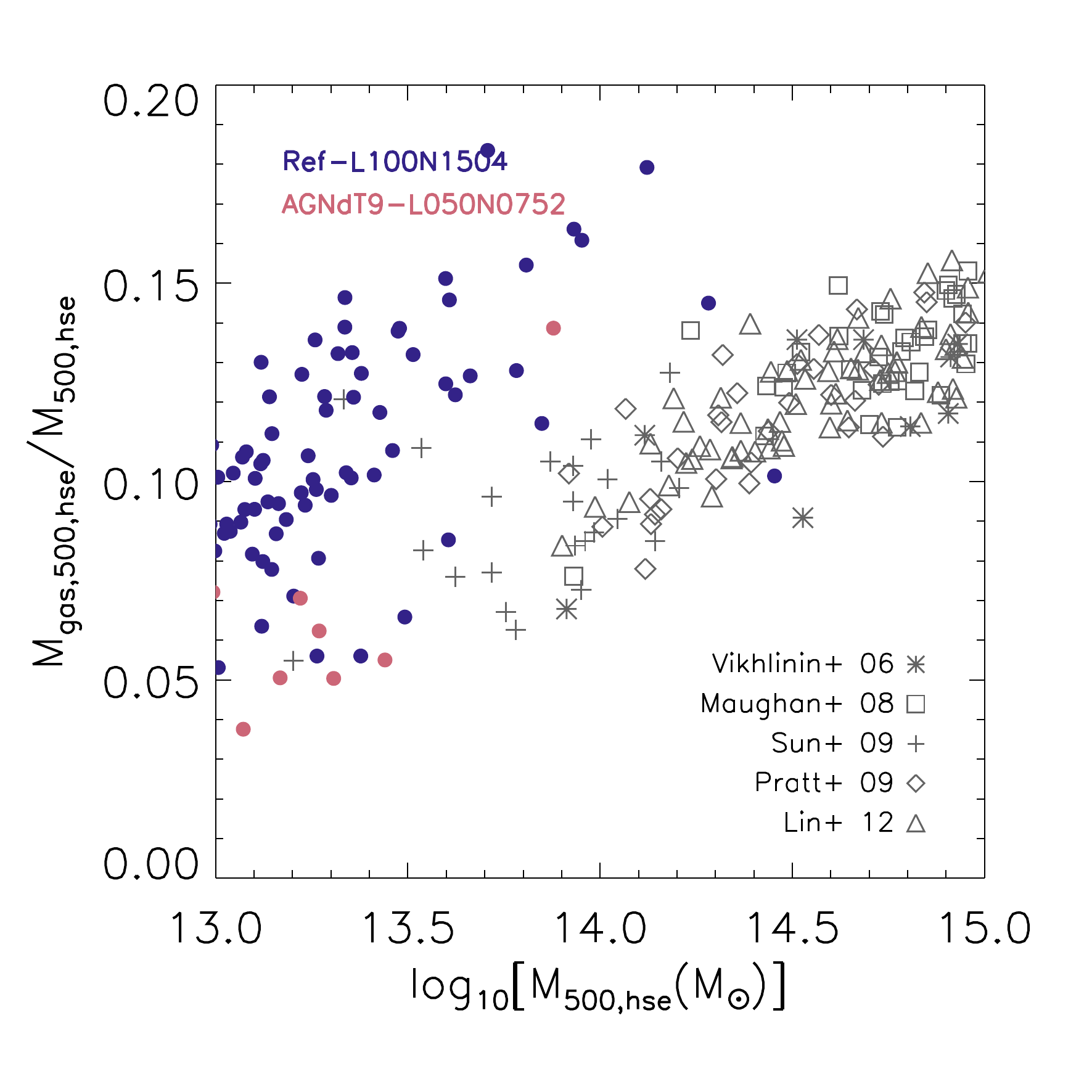}} 
\caption{The $z=0$ gas mass fraction within $R_{500,{\rm hse}}$ as a function of $M_{500,{\rm hse}}$. All quantities are inferred from (virtual) X-ray observations. Black data points represent observations of \citet{Vikhlinin2006}, \citet{Maughan2008}, \citet{Allen2008}, \citet{Pratt2009}, \citet{Sun2009}, and \citet{Lin2012}. The reference model overpredicts the gas fractions. Model AGNdT9-L050N0752, which employs a higher heating temperature for AGN feedback, performs well for groups of galaxies, but may also overpredict the gas fraction in higher mass ($\ga 10^{14}\,\Msun$) clusters.}
\label{fig:fgas_groups}
\end{figure}

Figure~\ref{fig:fgas_groups} shows the gas mass fraction, $M_{\rm gas,500,hse}/M_{\rm 500,hse}$ as a function of mass $M_{\rm 500,hse}$. Because the gas mass is derived from the (virtual) X-ray data, it only correctly accounts for gas that has a temperature similar to that of the gas that dominates the X-ray emission. For the reference model the gas mass inferred from X-ray observations, under the assumption of hydrostatic equilibrium, is about 0.2~dex higher than observed, except perhaps for the two most massive objects. 

\citet{LeBrun2014CosmoOWLS} have shown that the gas fraction is particularly sensitive to the temperature to which the AGN heat the surrounding gas in our subgrid prescription for AGN feedback. In particular, higher heating temperatures, which correspond to more energetic but less frequent bursts, eject the gas more effectively, yielding lower gas fractions. This was the motivation for running model AGNdT9-L050N0752, which uses a heating temperature $\Delta T_{\rm AGN}$ of $10^9\,\K$, compared with $10^{8.5}\,\K$ for the reference model. Before running this model, we used a 25~cMpc version to (approximately) recalibrate the BH accretion model so as to maintain the good match with the GSMF, in particular the location of the knee. We could, however, not afford to run multiple 50~cMpc models and could therefore not calibrate to observations of groups of galaxies.  

As can be seen from Figure~\ref{fig:fgas_groups}, contrary to model Ref-L100N1504, model AGNdT9-L050N0752 does appear to reproduce the observations of group gas fractions. That is, for $M_{500,{\rm hse}} < 10^{13.5}\,\Msun$ the simulation points agree with an extrapolation of the observations for high-mass systems. 
There is a strong hint that the gas fraction may again become too high for more massive clusters, although with only 1 object with $M_{500,{\rm hse}}> 10^{13.5}\,\Msun$ it is hard to judge the significance of this deviation.

\citet{LeBrun2014CosmoOWLS} found that the \cosmo\ simulations, which use $2\times 1024^3$ particles in $400~\hcMpc$ volumes, reproduce these and many other observations of groups and clusters over the full mass range of $10^{13}-10^{15}\,\Msun$ for $\Delta T_{\rm AGN} = 10^8\,\K$. This may seem surprising given that \eagle\ requires higher values of $\Delta T_{\rm AGN}$. Note, however, that because the particle mass in \cosmo\ is more than 3 orders of magnitudes larger than for \eagle, the energy in individual AGN feedback events in \cosmo\ is still much larger than that in AGNdT9-L050N0752. 

\begin{figure}
\resizebox{\colwidth}{!}{\includegraphics{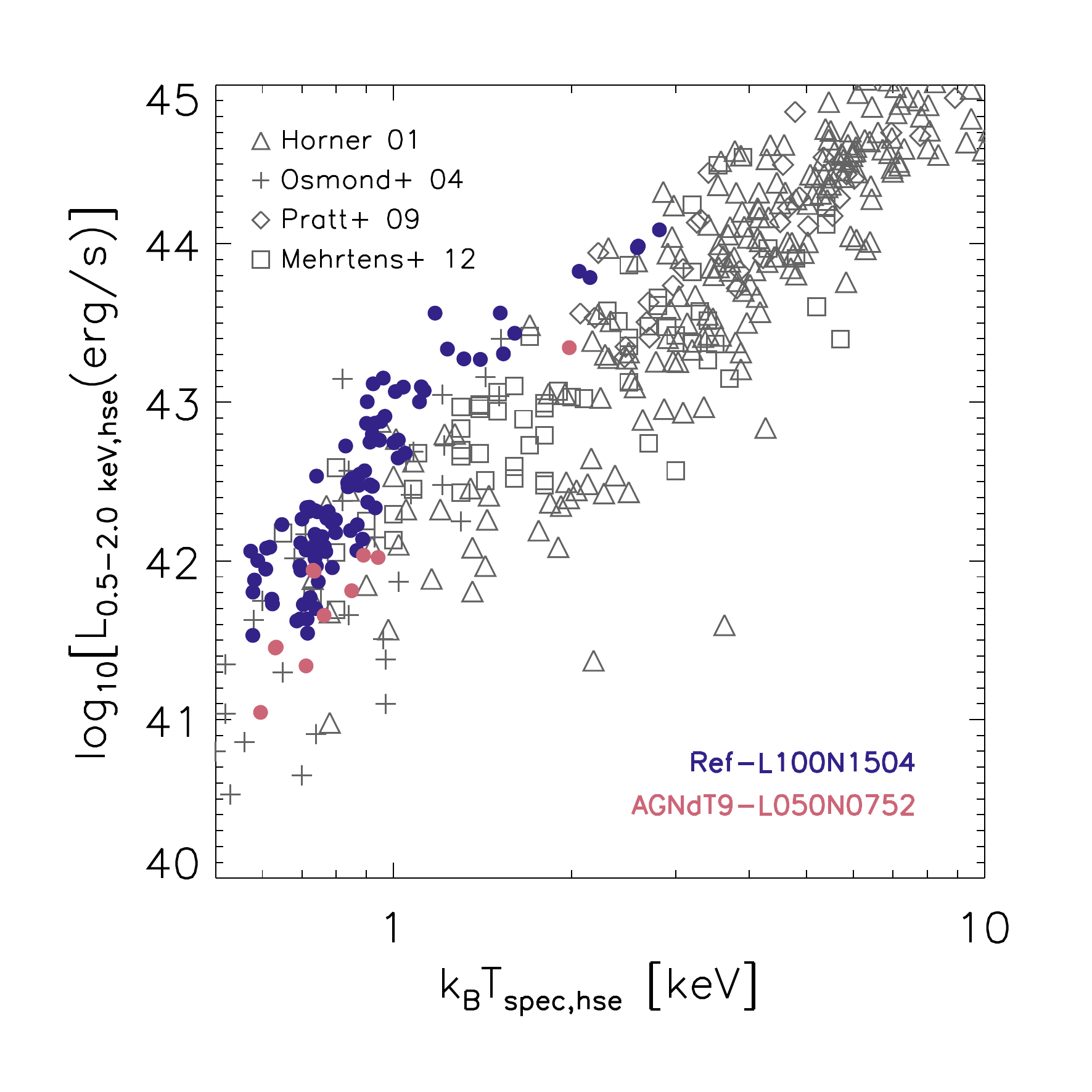}} 
\caption{The soft (0.5--2.0 keV) X-ray luminosity as a function of the X-ray temperature at $z=0$. Only points for which $M_{500,{\rm hse}}> 10^{13}\,\Msun$ are shown. Black data points represent observations of \citet{Horner2001}, \citet{Osmond2004}, \citet{Pratt2009}, and \citet{Mehrtens2012}. The reference model predicts too high X-ray luminosities for clusters above 1 keV, but simulation AGNdT9-L050N0752 is consistent with the data.}
\label{fig:Lx_groups}
\end{figure}

Figure~\ref{fig:Lx_groups} shows the X-ray luminosity in the 0.5--2.0 keV band as a function of the temperature measured from the (virtual) X-ray data. For the reference model the agreement with the observations is reasonably good at low temperatures (the lack of simulated points with $L \ll 10^{42}\,\erg\,\s^{-1}$ is due to the fact that we only selected systems with $M_{500,{\rm hse}}> 10^{13}\,\Msun$), but the predicted luminosity is about a factor of three too high above 1 keV. Model AGNdT9-L050N0752 appears to match the data well, but more objects with $k_{\rm B}T> 1~{\rm keV}$ are needed to better assess the degree of correspondence. 

\subsection{Column density distributions of intergalactic metals}
\label{sec:cddf}

The galactic outflows that we invoke to reproduce observations of galaxies also disperse heavy elements into the IGM. Furthermore, the winds shock-heat the gas, which may, in turn, change its ionisation balance. Hence, it is interesting to compare the predicted distribution of intergalactic metal ions to the observations. This is a strong test for the model, since the subgrid feedback was only calibrated to match the stellar properties of galaxies.

\begin{figure*}
\resizebox{\colwidth}{!}{\includegraphics{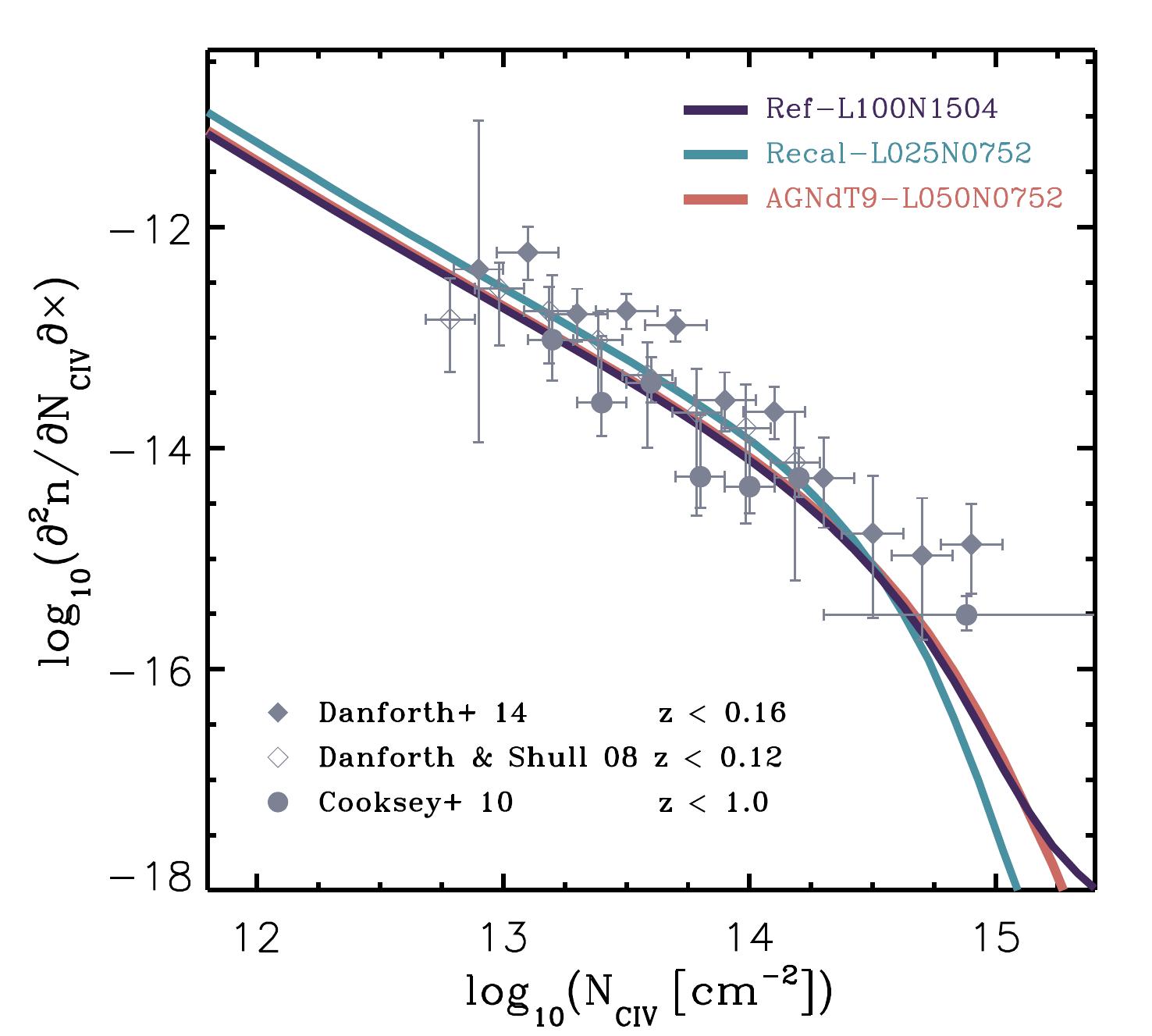}} 
\resizebox{\colwidth}{!}{\includegraphics{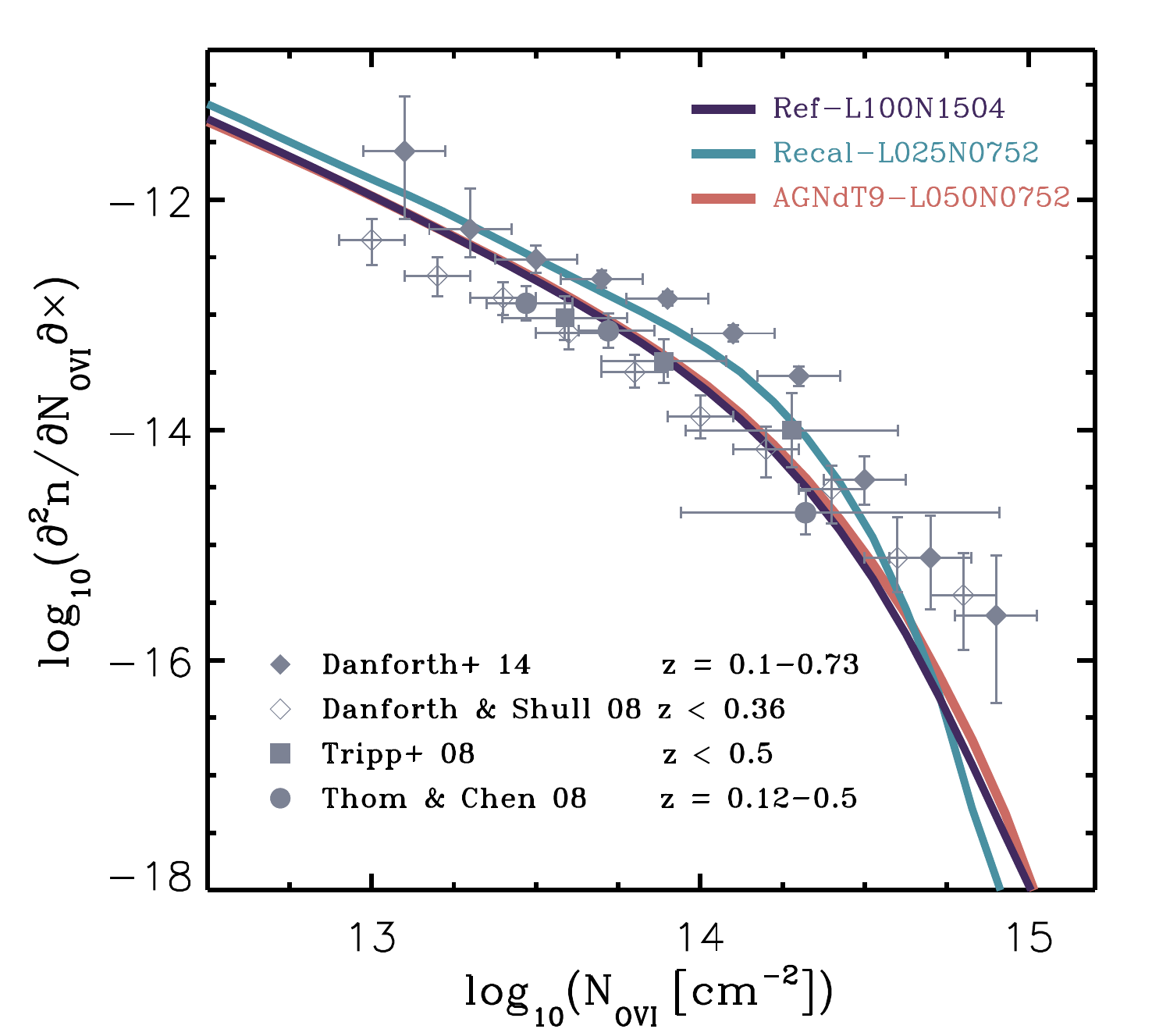}} 
\caption{The column density distribution functions of \CIV\ (left panel) and \OVI\ (right panel). The coloured curves show the simulation predictions and the data points with $1\sigma$ error bars indicate observations taken with STIS/FUSE \citep{Danforth2008LowzAbs}, COS \citep{Danforth2014LowzAbs}, STIS/FUSE/GHRS \citep{Cooksey2010CIV}, and STIS \citep{Thom2008OVI, Tripp2008OVI}. The redshift ranges of the observations vary and are indicated in the legend. For clarity we only show the simulation results for $z=0.27$. The predictions are consistent with the data.}
\label{fig:cddf}
\end{figure*}

Figure~\ref{fig:cddf} compares the predicted column density distribution functions (CDDFs) of \CIV\ (left panel) and \OVI\ (right panel) with measurements derived from quasar absorption line observations, mainly from the Hubble Space Telescope (HST). Note that this prediction was completely blind. 

The CDDF is conventionally defined as the number of absorbers per unit column density, $N$, and per unit absorption distance, $dX$. The number of absorbers per unit absorption distance is obtained from the quantity that is actually observed, the number of absorbers per unit redshift, via $dX = dz (H_0/H(z))(1+z)^2$. The redshift ranges of the observations vary and are indicated in the legend. All observations are for $z<1$ and most for much lower redshift. For clarity we only show the simulation results for our $z=0.27$ snapshots. However, limiting the comparison to $z=0.27$ does not affect our conclusions because the evolution is weak. 

For the simulations we
compute ion fractions for each gas particle using \textsc{cloudy} photoionisation models, assuming the gas is in ionisation equilibrium and exposed to the \citet{Haardt2001UVB} model for the UV/X-ray background from galaxies and quasars. We then obtain the CDDF by projecting the simulation cube onto a 2-D grid and applying SPH interpolation to compute the ion column density in each cell. We use a grid cell size of 10~ckpc, which is sufficiently small to obtain convergence, and have verified that projection effects are negligible by comparing results obtained from simulations using different box sizes. 

Observationally, the CDDF is obtained by decomposing the identified absorption features into Voigt profiles and grouping those into systems using criteria that differ between observers and that are not always well defined. We intend to mimic the observational procedures more closely in future work. From Figure~\ref{fig:cddf} it is clear that the differences between different sets of observations exceed the reported statistical uncertainties, suggesting the presence of significant systematic errors. Particularly for \OVI, the analysis of COS spectra by \citet{Danforth2014LowzAbs} yields systematically more absorbers than the earlier analyses of STIS/FUSE/GHRS data by \citet{Danforth2008LowzAbs}, \citet{Thom2008OVI}, and \citet{Tripp2008OVI}. 

As discussed in \S\ref{sec:Zm}, even for a fixed IMF the nucleosynthetic yields are uncertain at the factor of two level \cite[e.g.][]{Wiersma2009Chemo}. This suggests that we are free to rescale the metal column densities, i.e.\ to shift the curves in Figure~\ref{fig:cddf} horizontally by up to 0.3~dex. However, doing so would break the self-consistency of the simulations as the metal abundances determine the cooling rates. 

The simulation predictions generally agree well with the data, falling in between the different sets of observations, both for \CIV\ and \OVI. The simulations appear to produce too few ultra-strong absorbers, i.e.\ systems with column densities $\sim 10^{15}\,\cm^{-2}$. However, the frequency of these extremely rare systems is particularly sensitive to systematics and hence requires a more careful comparison. 

For \OVI\ the difference between Ref-L100N1504 and Recal-L025N0752 is substantial for $N_{\ionsubscript{O}{VI}} \sim 10^{14}\,\cm^{-2}$ with the high-resolution model yielding up to a factor of 3 more absorbers. However, this does not lead to any disagreement with the data as all simulations fall in between the different sets of observations. Recall that in low-mass galaxies feedback from star formation is more effective in the recalibrated, high-resolution model Recal-L025N0752 than in the reference model. It is interesting that while this boost in the feedback efficiency decreases the metallicity of the ISM (Fig.~\ref{fig:mz}), it boosts the abundance of metal ions in the IGM. It is tempting to conclude that the more effective feedback transports more metals from galaxies into the IGM. However, whether this is the case is not clear from the results presented here due to the importance of ionisation corrections. 

In future work we will compare with high-redshift data and with absorption line observations of the gas around galaxies of known mass. For now, we are encouraged by the fact that a model that was calibrated to the GSMF and galaxy sizes, also yields good agreement with observations of intergalactic metals. 

\section{Summary and discussion}
\label{sec:summary}

We have introduced the \eagle\ project, where \eagle\ stands for ``Evolution and Assembly of GaLaxies and their Environments''. \eagle\ consists of a suite of large, hydrodynamical cosmological simulations. In this introductory paper we have focused on a set of simulations for which the subgrid parameters for feedback were calibrated to match the observed $z\sim 0$ galaxy stellar mass function (GSMF), subject to the constraint that the galaxy sizes must also be reasonable. \citet{Crain2014EagleModels} will present models in which the subgrid physics is varied.

The largest \eagle\ simulation, Ref-L100N1504, uses nearly 7 billion ($2\times 1504^3$) particles in a 100 cMpc box. This corresponds to an initial baryonic particle mass of $1.8\times 10^6\,\Msun$ and a force resolution of 0.7~proper kpc (smaller at high redshift), which we refer to as ``intermediate resolution''. The resolution was chosen to marginally resolve the Jeans scales in the warm ($T\sim 10^4\,\K$) ISM. The high-resolution model, Recal-L025N0752, has eight times better mass resolution and two times better spatial resolution, thus resolving a galaxy like the Milky Way with $\sim 10^6$ particles. 

The simulations were run with the code \textsc{gadget}~3, but with a modified implementation of SPH, the time stepping, and the subgrid models. The simulations include subgrid prescriptions for (element-by-element) radiative cooling, star formation, stellar evolution and mass loss, energy feedback from star formation, the growth of supermassive BHs, and AGN feedback. The prescription for star formation accounts for the observation that stars form from molecular clouds and that the \HI-H$_2$ transition depends on metallicity. The subgrid model for accretion onto BHs accounts for the fact that angular momentum suppresses the accretion rate.

The most critical parts of the model are the implementations of energy feedback from star formation and AGN. We argued that present-day simulations of representative volumes cannot predict the efficiency of the feedback processes from first principles because of their reliance on subgrid models, because of spurious radiative losses due to the limited resolution, and because they lack the resolution and do not include all the physics necessary to model the structure of the ISM.

We discussed some of the implications of the inability to predict the efficiency of the feedback from first principles. We argued that current cosmological simulations can predict neither BH nor stellar masses, which implies that the subgrid models for feedback need to be calibrated to observations. Another consequence is that it is difficult to distinguish different physical feedback mechanisms that operate nearly simultaneously, such as winds driven by supernovae and radiation pressure. Furthermore, unless one can demonstrate that the model does not suffer from overcooling due to limited numerical resolution, one cannot conclude that there is a need for a new, physical feedback process just because the implemented feedback is insufficiently effective.

Because the spurious radiative losses depend on the resolution, one may have to recalibrate when the resolution is changed. We termed this ``weak convergence'' as opposed to the ``strong convergence'' that corresponds to the same physical model giving consistent results at different resolutions. However, we argued that most subgrid models for feedback effectively change with resolution even if the subgrid parameters are kept constant. 

The quest for strong convergence of simulations that lack the resolution to model the ISM has led to significant sacrifices, which generally involve disabling aspects of the hydrodynamics during feedback. Examples include temporarily turning off radiative cooling, temporarily turning off hydrodynamical forces, and making the feedback efficiency dependent on dark matter velocity dispersion rather than on local properties of the gas. However, until the cooling losses can be predicted, even fully converged simulations will be unable to predict stellar and BH masses from first principles. We therefore prefer to minimize the sacrifices and to opt for weak convergence. Nevertheless, we demonstrated that the strong convergence of our model is reasonably good (Fig.~\ref{fig:gsmf_conv}).

Motivated by the above considerations, we chose to keep our subgrid models for feedback as simple as possible. We employ only one type of stellar feedback and hence we do not distinguish between stellar winds, radiation pressure, and core collapse supernovae. Similarly, we include only one type of AGN feedback and therefore do not implement separate ``quasar'' and ``radio modes''. We find that a more complex approach is not required to match observational data.

We implement both feedback from star formation and AGN thermally using the stochastic prescription of \citet{DallaVecchia2012Winds}. By injecting the energy stochastically rather than at every time step, we can specify both the temperature jump of the heated gas and the expectation value for the amount of energy that is injected. This enables us to better mimic the physical conditions associated with observed feedback processes, in particular the high heating temperatures that suppress the initial radiative losses, than would otherwise be possible given the limited resolution of the simulations. The velocities and mass loading factors of galactic winds are thus not imposed, but are an outcome of the simulation.

The temperature jump associated with feedback events is chosen to balance the need to minimize both the initial, radiative losses (which are largely numerical) and the time between feedback events (to allow for self-regulation). The probability of heating events then needs to be calibrated by comparing the simulation results for some observable to real data.
The subgrid efficiency of the AGN feedback, i.e.\ the expectation value for the amount of energy that is injected into the ISM per unit of accreted gas mass, is constant and was chosen to match the normalisation of the observed relation between the masses of galaxies and their central supermassive BHs. This parameter is, however, unimportant for observables other than the masses of BHs.
The subgrid efficiency of the feedback from star formation, $f_{\rm th}$, i.e.\ the expectation value for the amount of energy that is injected into the ISM in units of the energy available from core collapse supernovae, was chosen to reproduce the observed GSMF for $M_\ast < 10^{10.5}\,\Msun$, i.e.\ below the knee of the Schechter function. Finally, the value of the parameter that controls the sensitivity of the BH accretion rate to the angular momentum of the surrounding gas was adjusted to make the mass function turn over at the onset of the exponential drop of the observed GSMF. 

We made $f_{\rm th}$ a function of both metallicity and density.
We use a physically motivated metallicity dependence with $f_{\rm th}$ dropping when the metallicity is increased from values $\ll 0.1Z_\odot$ to $\gg 0.1 Z_\odot$. This reduction in the efficiency is meant to capture the increase in radiative losses that is expected when metal-line cooling becomes important, which happens for $Z > 0.1 Z_\odot$ at the temperatures relevant for gas shock-heated in galactic winds \citep[e.g.][]{Wiersma2009Cooling}. 

While a constant value of $f_{\rm th}=1$, or a pure metallicity dependence, each give an excellent fit to the GSMF, they result in galaxies that are far too compact \citep{Crain2014EagleModels}. This happens because, at the resolution of \eagle, the stochastic implementation for stellar feedback is still subject to numerical radiative losses at high gas densities, as we demonstrated analytically. To compensate for these spurious losses, we increase $f_{\rm th}$ at high gas densities. However, $f_{\rm th}$ never exceeds 3 and the mean value is smaller than 1.1. 

We compared \eagle\ to a diverse set of observations of the low-redshift Universe, carefully distinguishing between observations that were considered during the calibration (the GSMF and thus also the directly related $M_\ast - M_{200}$ relation, galaxy sizes, and the $M_{\rm BH} - M_\ast$ relation) and those that were not. We came to the following conclusions:
\begin{itemize}
\item The observed GSMF is reproduced over the range $10^8 < M_\ast/\Msun \la 10^{11}$. At fixed mass, the difference in number density relative to the data is $\la 0.2$ dex. At fixed number density, the difference in mass is smaller than 0.3 dex (Fig.~\ref{fig:gsmf}). Even for a fixed IMF, this discrepancy is comparable to the systematic uncertainty in the observed masses due to stellar evolution alone. This level of agreement with the data is close to that obtained by semi-analytic models and is unprecedented for hydrodynamical simulations (Fig.~\ref{fig:gsmf_other}).
\item Three-dimensional apertures of 30 proper kpc, which we used throughout the paper, give results close to the Petrosian masses that are often used for observations, e.g.\ by SDSS. For $M_\ast > 10^{11}\,\Msun$ larger apertures yield higher masses (Fig.~\ref{fig:gsmf_aperture}).
\item The stellar mass - halo mass relation for central galaxies is close to that inferred from abundance matching. The efficiency of galaxy formation, $M_\ast/M_{200}$, peaks at the halo mass $M_{200} \sim 10^{12}\,\Msun$ and at the stellar mass $M_\ast \sim 10^{10.4}\,\Msun$ (Fig.~\ref{fig:eta}). 
\item Disc galaxy sizes are well matched to the observations. Over the full range of stellar mass, $10^8 < M_\ast/\Msun < 10^{11.5}$, the median stellar half-mass radii of late-type galaxies agree with the observed half-light radii to within 0.1 dex (Fig.~\ref{fig:sizes}).
\item The median relation between BH mass and stellar mass agrees with the observations, but the scatter in the model is smaller than observed. The simulations predict that galaxies with total stellar masses of $10^9-10^{10}\,\Msun$ typically host BHs with masses that fall below the extrapolation of the high-mass power-law relation (Fig.~\ref{fig:bh}).
\item The predicted relation between the median specific star formation rate ($\dot{M}_\ast/M_\ast$; SSFR) and stellar mass for star-forming galaxies, i.e.\ the ``main sequence of star formation'', agrees with the observations to within 0.2 dex over the observed range of $10^9 < M_\ast/\Msun <10^{11}$ at high-resolution and to within 0.35 dex at intermediate resolution (Fig.~\ref{fig:ssfr}, left panel). 
\item The predicted fraction of galaxies that are passive, which we define as SSFR $< 10^{-2}\,\Gyr^{-1}$ for the simulations, increases sharply with stellar mass between $10^{10}$ and $10^{11.5}\,\Msun$, in agreement with the observations (Fig.~\ref{fig:ssfr}, right panel). 
\item The predicted median relation between the maximum of the rotation curve and stellar mass of late-type galaxies, i.e.\ a close analogue of the Tully-Fisher relation, agrees with the observations to better than 0.03 dex over the observed mass range of $10^9 \la M_\ast/\Msun <10^{11}$ (Fig.~\ref{fig:tf}). 
\item The relations between ISM metallicity and stellar mass and between stellar metallicity and stellar mass are predicted to flatten with stellar mass. For the gas the predicted median metallicities agree with the observed values to within 0.1 dex for $M_\ast > 10^{9.5}\,\Msun$ at intermediate resolution and down to the lowest observed mass, $M_\ast \sim 10^{8.5}\,\Msun$, at high resolution. At lower masses the predicted relations are less steep than extrapolations of the observed trends. For the stellar metallicities the discrepancies are larger. For $M_\ast > 10^{10}\,\Msun$ all simulations agree with the data to better than 0.2~dex, but the difference increases with decreasing mass. At $M_\ast \sim 10^8\,\Msun$ the stellar metallicities in the intermediate- and high-resolution simulations are higher than observed by about 0.7 and 0.3~dex, respectively. 
\item For the mass-metallicity relations the strong convergence is significantly better than the weak convergence, i.e.\ simulations that keep the subgrid parameters fixed converge better with numerical resolution than simulations for which the feedback is (re)calibrated to the $z\sim 0$ GSMF at each resolution. Hence, the increase in the efficiency of the feedback from star formation that was applied at high resolution in order to match the observed GSMF, simultaneously steepens the $Z(M_\ast)$ relations, improving the agreement with the data.  
\item A comparison to observations of groups and clusters of galaxies with $M_{500,{\rm hse}} > 10^{13}\,\Msun$, where the subscript ``hse'' indicates that the quantity was estimated from virtual observations under the assumption of hydrostatic equilibrium, revealed that:
\begin{itemize}
\item The predicted relation between the total $I$-band light within $R_{500,{\rm hse}}$ and $M_{500,{\rm hse}}$ agrees with the data. Note that this includes contributions from satellites and intracluster light (Fig.~\ref{fig:Lopt_groups}).
\item The gas mass fractions, $M_{\rm gas,500,hse}/M_{\rm 500,hse}$, are overestimated by about 0.2~dex in the reference model. For $M_{\rm 500,hse} < 10^{13.5}\,\Msun$ this can be remedied by increasing the subgrid AGN heating temperature, as implemented in model AGNdT9-L050N0752. At higher masses this change may be insufficient, although larger simulation volumes are needed to confirm this (Fig.~\ref{fig:fgas_groups}).
\item The reference model predicts soft X-ray luminosities that are about 0.5~dex higher than observed for clusters with spectroscopic temperatures $\sim 1$~keV. However, model AGNdT9-L050N0752 is consistent with the observations (Fig.~\ref{fig:Lx_groups}).
\end{itemize}
\item The column density distributions of intergalactic \CIV\ and \OVI\ are in good agreement with the data, falling in between the results obtained by different surveys (Fig.~\ref{fig:cddf}).
\end{itemize}

Hence, in the resolved mass range, which spans $10^9 \la M_\ast/\Msun \la 10^{11}$ for some observables and $10^8 \la M_\ast/\Msun \la 10^{11}$ for others, \eagle\ agrees with a diverse set of low-redshift observations of galaxies. At the same time, \eagle\ reproduces some key observations of intergalactic metals. The only discrepancies found in this work that substantially exceed observational uncertainties concern the gas and stellar metallicities of dwarf galaxies, which are too high, and 
the predictions of the reference model for X-ray observations of the intracluster medium. The metallicity problem is only substantial at intermediate resolution, so it is possible that it can be resolved simply by increasing the resolution further. We already demonstrated that the problem with groups of galaxies can be remedied by increasing the heating temperature used in the subgrid model for AGN feedback, as implemented in model AGNdT9-L050N0752, without compromising the successes of the reference model. However, larger volumes are needed to judge whether the increase in the heating temperature that was implemented in this model suffices to obtain agreement with the data for massive ($M_{500} \ga 10^{14}\,\Msun$) clusters of galaxies. 

In future papers we will test many more predictions of \eagle. Although we will undoubtedly uncover problems, so far we have no reason to believe that the results shown here are unrepresentative. We will show that the success of \eagle\ extends to other areas that have in the past proven to be challenging for hydrodynamical simulations, such as the bimodal distribution of galaxies in colour-magnitude diagrams. We will also demonstrate that the relatively good agreement with the data is not limited to low redshift. In addition to further exploring the models that have been presented here, we plan to use the larger suite of physical models presented in \citet{Crain2014EagleModels} to gain insight into the physical processes underlying the observed phenomena. Finally, we have already begun to carry out higher-resolution resimulations of individual structures \citep[e.g.][]{Sawala2014EagleZooms,Sawala2014ChosenFew} with the code used for \eagle. 

Although the relatively good agreement between \eagle\ and the observations, as well as that between other recent, hydrodynamical simulations of representative volumes and the data \citep[e.g.][]{Vogelsberger2014IllustrisNature}, is encouraging, we should keep in mind that we have not attempted to model many of the physical processes that may be important for the formation and evolution of galaxies. For example, \eagle\ does not include a cold interstellar gas phase, radiation transport, magnetohydrodynamics, cosmic rays, conduction, or non-equilibrium chemistry, and \eagle\ does not distinguish between different forms of energy feedback from star formation and between different forms of AGN feedback. We argued that at present there are good reasons for such omissions, but many of those arguments would no longer apply if the numerical resolution were increased by several orders of magnitude. While it will take some time for simulations of representative volumes to attain the resolution that is required to model the cold ISM, simulations of individual objects can already do much better. Ultimately, simulations should be able to predict the efficiencies of the most important feedback processes and hence to predict, rather than calibrate to, the GSMF. 

We hope that \eagle\ will prove to be a useful resource for the community.\footnote{We intend to make the simulation output public in due course, starting with galaxy properties, which we will make available using the SQL interface that was also used for the Millennium simulation \citep{Lemson2006Database,Springel2005Millennium}. Details will be provided on the \eagle\ web site, see \url{http://eagle.strw.leidenuniv.nl/}.} 
The agreement with observations is sufficiently good for the simulations to be used in ways that have so far been reserved for semi-analytic models of galaxy formation. At the same time, because hydrodynamical simulations provide much more detailed 3-D information, make fewer simplifying assumptions, and simultaneously model the galaxies and the IGM, \eagle\ enables one to ask many questions that are difficult to investigate with semi-analytic models.

\section*{Acknowledgments}

We would like to thank Volker Springel for sharing his codes and for his advice and discussions. We gratefully acknowledge discussions with Jarle Brinchmann, Shy Genel, Justin Read, Debora Sijacki. We are also thankful to Martin Bourne and Laura Sales for their contributions to the initial phase of the project, Amandine Le Brun for her help with the X-ray plotting routines, Peter Draper and Lydia Heck for their help with the computing resources in Durham, and to Wojciech Hellwing for help with computing in Poland. We are also grateful to all the people working on the analysis of the \eagle\ simulations and would like to thank the anonymous referee for a constructive report.
This work used the DiRAC Data Centric system at Durham University, operated by the Institute for Computational Cosmology on behalf of the STFC DiRAC HPC Facility (www.dirac.ac.uk). This equipment was funded by BIS National  E-infrastructure capital grant ST/K00042X/1, STFC capital grant ST/H008519/1, and STFC DiRAC Operations grant ST/K003267/1 and Durham University. DiRAC is part of the National E-Infrastructure. We also gratefully acknowledge PRACE for awarding us access to the resource Curie based in France at Tr\`es Grand Centre de Calcul. This work was sponsored by the Dutch National Computing Facilities
Foundation (NCF) for the use of supercomputer facilities, with financial support from the Netherlands Organization for Scientific
Research (NWO) and by the HPC Infrastructure for Grand Challenges of Science and Engineering Project, co-financed by the European Regional Development Fund under the Innovative Economy Operational Programme and conducted at the Institute for Mathematical and Computational Modelling at University of Warsaw. The research was supported in part by the European Research Council under the European Union's Seventh Framework
Programme (FP7/2007-2013) / ERC Grant agreements 278594-GasAroundGalaxies, GA 267291 Cosmiway, and 321334 dustygal, the Interuniversity Attraction Poles Programme initiated by the Belgian Science Policy Office ([AP P7/08 CHARM]), the National Science Foundation under Grant No.\ NSF PHY11-25915, the UK Science and Technology Facilities
Council (grant numbers ST/F001166/1 and ST/I000976/1), Rolling and Consolodating Grants to the ICC, Marie Curie Reintegration Grant PERG06-GA-2009-256573.

\bibliographystyle{mn2e} 
\bibliography{ms}

\appendix 

\section{Hydrodynamics}
\label{app:hydro}

Recently, much effort has been directed at solving a well-known issue with the standard SPH implementation: multi-valued particle pressure and large artificial viscosity causing unphysical surface tension at contact discontinuities \citep[for a detailed description of the problem see, e.g.][]{Agertz:2007fk}. This surface tension impedes the development of hydrodynamical instabilities resulting in poor mixing of gas phases, which could in principle compromise simulations of galaxy formation \citep[e.g.][]{Sijacki2012Hydro,Nelson2013GasAccretion}. Several solutions have been suggested in order to smooth the pressure at contact discontinuities \citep[e.g.][]{Ritchie:2001fk,Price:2008kx,Read:2010uq,Saitoh:2013uq,Hopkins:2013lr}, and to reduce the artificial viscosity away from shocks \citep[e.g.][]{Morris:1997fj,Cullen:2010qy}. 

As described in more detail below, we employ the fully conservative SPH formulation derived by \citet{Hopkins:2013lr}, of which the solutions suggested by \citet{Ritchie:2001fk}, \citet{Read:2010uq} and \citet{Saitoh:2013uq} are special cases. We use the artificial viscosity switch from \citet{Cullen:2010qy} and a switch for artificial conduction similar to that of \citet{Price:2008kx}. We apply the time step limiters of \citet{Durier:2012fj}. 

We adopt the $C^2$ \citet{Wendland:1995} kernel with $N_{\rm ngb} = 58$ neighbours. This kernel inhibits particle pairing \citep{Dehnen:2012uq} and the number of neighbours was chosen to give an effective resolution that is close to that of the cubic spline kernel with 48 neighbours that was used in \owls. 

The methods used here are collectively referred to as ``Anarchy'' and will be described in more detail in Dalla Vecchia (in preparation), who also demonstrates its performance on standard hydrodynamical tests. In \citet{Schaller2014EagleSPH} we compare the results of \eagle\ cosmological simulations with different hydrodynamics and time stepping schemes. Consistent with previous work \citep[e.g.][]{Scannapieco2012Aquila}, we find that our results are generally substantially less sensitive to changes in the hydrodynamical techniques than to reasonable variations in the subgrid physics. 

\subsection{SPH}
\label{sec:sph}

Following \cite{Hopkins:2013lr}, the generalised equation of motion is
\begin{equation}
m_i\frac{{\rm d}\mathbf{v}_i}{{\rm d}t}=-\sum_{j=1}^N x_i x_j\left[
\frac{P_i}{y_i^2}f_{ij}\nabla_i W_{ij}(h_i) +
\frac{P_j}{y_j^2}f_{ji}\nabla_i W_{ij}(h_j) \right],
\label{eqmot}
\end{equation}
where $m_i$, $\mathbf{v}_i$ and $P_i$ are the particle mass, velocity and pressure, respectively; $W_{ij}$ is the SPH kernel;  $h_i$ is the SPH smoothing length; $f_{ij}$ is the correction term for variable smoothing lengths (the so-called ``grad-$h$'' term), given by
\begin{equation}
f_{ij} = 1 - \frac{\tilde{x}_j}{x_j}\left(\frac{h_i}{n_D \tilde{y}_i}\frac{\partial y_i}{\partial h_i}\right)
\left[1 + \frac{h_i}{n_D \tilde{y}_i}\frac{\partial \tilde{y}_i}{\partial h_i}\right]^{-1},
\label{fij}
\end{equation}
where $n_{\rm D}$ is the number of spatial dimensions. In the above equations, $\tilde{x}_i$ and its SPH smoothed value, $\tilde{y}_i=\sum_j \tilde{x}_j W_{ij}(h_i)$, define the particle volume, $\tilde{V}_i=\tilde{x}_i/\tilde{y}_i$. The particle smoothing length is defined by the relation
\begin{equation}
\frac{4\pi}{3} h_i^3=N_{\rm ngb} \tilde{V}_i ,
\end{equation}
where $N_{\rm ngb}$ is the number of neighbouring particles.\footnote{Note that the number of neighbours, $N_{\rm ngb}$, is a parameter and not the actual number of particles within the kernel.} In our implementation we chose $\tilde{x}_i=m_i$ and $\tilde{y}_i\equiv\rho_i=\sum_j m_j W_{ij}(h_i)$, the SPH particle density.

The remaining quantities, $x_i$ and $y_i=\sum_j x_j W_{ij}(h_i)$, define the ``thermodynamical volume'', and can be chosen in order to obtain a smooth representation of the pressure. Since we follow the evolution of the gas pseudo entropy, $A \equiv P/\rho^\gamma$, the natural choice is then $x_i=m_i A_i^{1/\gamma}$ and $y_i\equiv \bar{P}_i^{1/\gamma}=\sum_{j=1}^N m_j A_j^{1/\gamma} W_{ij}(h_i)$ as suggested by \cite{Read:2010uq}. With this definition, the weighted pressure, $\bar{P}_i$, is now single-valued and varies smoothly through contact discontinuities.

In practice, it is convenient to define a weighted density that can be used in the conversion between thermodynamical quantities (entropy, internal energy, temperature) and that can be predicted for inactive particles. We define the weighted density by writing the entropic function, $P=A\rho^{\gamma}$, as follows:
\begin{equation}
\bar{P}_i = A_i \left(\frac{1}{A_i^{1/\gamma}}\sum_{j=1}^N m_j A_j^{1/\gamma} W_{ij}(h_i)\right)^{\gamma} = A_i \bar{\rho}_i^{\gamma}.
\end{equation} 
Note that this definition of the density is the only one that is consistent with the definition of the pressure \citep{Read:2010uq}.

The formulation of the SPH equation in terms of the pressure and entropy thus introduces the notion of a weighted density  $\bar\rho_i$. Despite having the units of a density, this quantity should not be confused with the physical density $\rho_i = \sum_j m_j W_{ij}(h_i)$. The weighted density should be thought of as an intermediate quantity required for the calculation of other thermodynamics quantities and for the SPH equation of motion. As a consequence, both densities must be used in the subgrid recipes. If  the model requires a density (cooling, enrichment), then we use the physical density $\rho_i$. On the other hand, if the quantity of interest is the pressure or the temperature, then we use the weighted density $\bar\rho_i$ for consistency with the SPH equations.

Finally, equation~(\ref{eqmot}) can be written as
\begin{eqnarray}
\frac{{\rm d}\mathbf{v}_i}{{\rm d}t}&=&-\sum_{j=1}^N m_j  \left[
\frac{A_j^{1/\gamma}}{A_i^{1/\gamma}}\frac{\bar{P}_i}{\bar{\rho}_i^2}f_{ij}\nabla_i W_{ij}(h_i) ~ + \right . \nonumber \\
&&
\left . \frac{A_i^{1/\gamma}}{A_j^{1/\gamma}}\frac{\bar{P}_j}{\bar{\rho}_j^2}f_{ji}\nabla_i W_{ij}(h_j) \right] ,
\label{eqmot3}
\end{eqnarray}
where the grad-$h$ terms are (see equation~\ref{fij}):
\begin{equation}
f_{ij} = 1 - \frac{1}{A_j^{1/\gamma}}\left(\frac{h_i}{n_D \rho_i}\frac{\partial\bar{P}_i^{1/\gamma}}{\partial h_i}\right)
\left[1 + \frac{h_i}{n_D \rho_i}\frac{\partial \rho_i}{\partial h_i}\right]^{-1}.
\end{equation}

\subsubsection{Injection of feedback energy}
\label{sec:injection}

When the equations of SPH are formulated using the pressure and entropy as main variables, particles do not carry a   numerical field for their internal energy. This quantity has to be computed as a weighted sum over the particle  neighbours in the same way as the density is computed in other formulations of SPH. Energy from feedback events can  hence not be implemented by simply increasing the internal energy of the particle by some amount $\Delta u$. Furthermore, because the weighted density, $\bar{\rho}_i$, and the entropic function, $A_i$, of a particle are coupled, a na\"ive change of $A_i$ during energy injection would be incorrect as the corresponding weighted density would also change, making the total thermal energy of the gas (across all particles in the simulation volume) change by an amount different from $\Delta u$.  

In Anarchy this problem is partially solved by performing a series of iterations during which $A_i$ and $\bar\rho_i$ are changed until the two quantities have converged: 
\begin{eqnarray}  A_{i,n+1} &=& {(\gamma-1) (u_{\rm old} +\Delta u) \over \bar\rho_{i,n}^{\gamma-1}} , \nonumber\\ \bar\rho_{i,n+1} &=& {\bar\rho_{i,n}\,A_{n}^{1/\gamma}-m_iW(0)A_{i,n}^{1/\gamma}+m_iW(0)A_{i,n+1}^{1/\gamma}\over  A_{i,n+1}^{1/\gamma}}, 
\end{eqnarray} 
where $m_i$ is the mass of particle $i$ and $W$ is the kernel function. This approximation is valid for reasonable values of $\Delta u$ and is crucial for injecting thermal feedback in the gas phase. 

For high thermal jumps with more than one particle being heated, as can for example occur for our AGN feedback scheme, the approximation provided by these iterations is not sufficiently accurate to properly conserve energy. We hence limit the amount of energy that can be injected in the gas phase by AGN in a single event by limiting the heating probability to  $0.3$ (effectively limiting the number of particles being heated at the same time in a given neighbourhood) for which tests show that the correct amount of energy is distributed to the gas. 

\subsection{Artificial viscosity} 
\label{sec:viscosity}

SPH requires artificial viscosity to capture shocks. The artificial viscosity switch has been implemented following \citet{Cullen:2010qy}. Their algorithm enables a precise detection of shocks and avoids excessive viscosity in pure shear flows. As in \citet{Cullen:2010qy}, particles are assigned individual values of the viscosity coefficient, $\alpha_{{\rm v},i}$. This is recomputed at every time step $n$, and if it exceeds the value at the previous step, $\alpha_{{\rm v},i}^n>\alpha_{{\rm v},i}^{n-1}$, the viscosity coefficient is set to $\min{(\alpha_{{\rm v},i}^n, \alpha_{\rm v,max})}$. If $\alpha_{{\rm v},i}^n\leq\alpha_{{\rm v},i}^{n-1}$, the viscosity coefficient decays towards $\alpha_{{\rm v},i}^n$ on a time scale proportional to the particle's sound-crossing time, $\tau_i=h_i/(0.1 c_i)$:
\begin{equation}
\alpha_{{\rm v},i} = \alpha_{{\rm v},i}^n + (\alpha_{{\rm v},i} - \alpha_{{\rm v},i}^n)\,e^{-\Delta t/\tau_i}\,,
\end{equation}
and limiting the minimum allowed value, $\alpha_{{\rm v},i}\geq\alpha_{\rm v,min}\geq 0$.
We adopt $\alpha_{\rm v,min}=0.05$ in order to facilitate particle ordering, and allow the coefficient to range up to $\alpha_{\rm v,max}=2$. We found that if the number of neighbours is sufficiently large ($\sim 10^2$), the calculation of the velocity divergence in \textsc{gadget} is sufficiently accurate for standard hydrodynamical tests. Therefore, we did not implement any expensive matrix calculation of the velocity divergence \citep{Cullen:2010qy,Read:2012qy,Hu2014SPHGal}.

\subsection{Entropy diffusion} 
\label{sec:diffusion}

SPH is by construction non-diffusive. However, some diffusion mechanism is required during mixing of gas phases in order to mimic thermal conduction. We do not attempt to model physical diffusion; the implemented diffusion is purely numerical. We also do not implement diffusion to solve numerical problems at contact discontinuities; these are solved by the adopted SPH scheme.

The thermal energy, $u$, is diffused according to the following equation \citep[e.g.][]{Monaghan:1997mz,Price:2008kx},
\begin{equation}
\frac{{\rm d}u_i}{{\rm d}t}=\sum_{j=1}^N\alpha_{{\rm d},ij} v_{{\rm d},ij} \frac{m_j }{\rho_{ij}}\left(u_i - u_j\right)\nabla_i W_{ij}(h_i,h_j) ,
\end{equation}
where $v_{{\rm d},ij}={\rm max}(c_i + c_j + \mathbf{v}_{ij}\cdot\mathbf{r}_{ij}/r_{ij},0)$, and the diffusion coefficient, $\alpha_{{\rm d},ij}$, density and kernel derivative are averages among particle pairs.
The purely numerical switch, similar to the one of \cite{Price:2008kx}, is triggered by the spatial second derivative of the internal energy
\begin{equation}
\dot{\alpha}_{{\rm d},i}=\beta\frac{h_i \nabla^2_i u_i}{\sqrt{u_i}} ,
\end{equation}
where the growth speed of $\alpha_{{\rm d},i}$ can be tuned through the coefficient $\beta$. We adopt $\beta=0.01$. With this choice, diffusion is mild and there is no need of any further limiter in the presence of gravity. Finally, the diffusion coefficient evolves with time as
\begin{equation}
\alpha_{{\rm d},i}(t+\Delta t) = \alpha_{{\rm d},i}(t) - \left(\frac{\alpha_{{\rm d},i}(t) - \alpha_{{\rm d,min}}}{\tau_i} - \dot{\alpha}_{{\rm d},i}\right)\Delta t ,
\end{equation}
where the decay time scale, $\tau_i$, is the same as employed in the artificial viscosity, and $\alpha_{{\rm d,min}}=0$. We set the maximum allowed value to $\alpha_{{\rm d,max}}=1$, but this is unimportant because $\alpha_{{\rm d},i}\ll 1$ even for large discontinuities in the internal energy.

\subsection{Time stepping}
\label{sec:timestepping}

The accuracy of the time integration is increased by using a time-step limiter \citep[e.g.][]{Saitoh:2013uq}. We adopted the solution of \cite{Durier:2012fj} which ensures that sudden changes in the particle internal energy, e.g.\ caused by feedback, are promptly captured and propagated to neighbouring particles by shortening their time step and by activating them. We set the maximum ratio of neighbouring particles' time steps to four.

\section{Generation of the initial conditions}
\label{app:ics}

We have made two types of initial conditions: dark matter only with
all particles the same mass, and dark matter with gas.  The dark
matter with gas simulations are created starting from a corresponding
dark matter only simulation so we first describe how the dark matter
only initial conditions were made.

\subsection{Building dark matter only initial conditions}

The initial conditions are created in three steps.  Firstly, a particle
load, representing an unperturbed homogeneous periodic universe in a
3-torus is produced. Secondly, a realisation of a Gaussian random
density field with the appropriate linear power spectrum is created
over the 3-torus. Thirdly the displacements and velocities, consistent
with the pure growing mode of gravitational instability, are
calculated from the Gaussian realisation and applied to the particle
load producing the initial conditions.

 The unperturbed particle loads for the dark matter only initial
conditions have a glass-like particle distribution produced by
applying the method first described in \cite{White1994LesHouches}.  This method, available as an option in the \textsc{gadget-2} code \citep{Springel2005Gadget2}, was
applied, with periodic boundary conditions, to make a ``primitive'' cubic glass distribution with $47^3$
particles.  The particle loads required for each of the \eagle\ initial
conditions were built by tiling this primitive cubic glass file $n$
times in each of the three principal coordinate directions across a
larger cubic 3-torus, giving particle loads with a glass distribution
with $(47n)^3$ particles.

 The dark matter only initial conditions were generated using the
\ICgen\ code using the method described in \cite{Jenkins20102lpt} to create second order Lagrangian perturbation theory (2lpt)
resimulation initial conditions. The \ICgen\ code outputs
Zeldovich initial conditions plus a ``2lpt mass'' for each particle.
The \eagle\ version of \textsc{gadget}~3 is then used to solve
a Poisson equation sourced by the 2lpt masses placed at their
unperturbed positions. The solution of this Poisson equation yields
second-order Lagrangian growing mode displacements and velocities for
each particle. Adding these to the Zeldovich displacements and
velocities of all the particles produces the final 2lpt initial
conditions. The 2lpt masses can then be discarded and the usual equations of motion are solved by integrating the initial
conditions forward in time.

\begin{table*}
\caption{The phases for the \eagle\ simulation volumes are taken from
the public multiscale Gaussian white noise field \emph{Panphasia} \citep{Jenkins2013ICs}.  For completeness we publish the phases for all the volumes in
the \eagle\ series, but note that we have not yet carried out baryonic simulations in boxes greater than 100~cMpc. These periodic cubic volumes have side-lengths
given by $6.25\times2^n$~cMpc, where $n$ is an integer, $n=0-10$. }
\begin{tabular}{rl}
\hline
  Box size  &     Phase Descriptor \\
(cMpc)\\
\hline
6.25     &   [Panph1,L19,(40044,38524,52597),S3,CH2062909610,EAGLE\_L0006\_VOL1]  \\
12.5     &   [Panph1,L18,(34546,48586,31987),S3,CH1284484552,EAGLE\_L0012\_VOL1]  \\
25\phantom{00}     &   [Panph1,L17,(22872,9140,6502),S3,CH1193192352,EAGLE\_L0025\_VOL1]  \\
50\phantom{00}     &   [Panph1,L16,(9358,44124,48606),S3,CH1323953302,EAGLE\_L0050\_VOL1]  \\
100\phantom{00}   &   [Panph1,L16,(31250,23438,39063),S12,CH1050187043,EAGLE\_L0100\_VOL1]  \\
200\phantom{00}     &   [Panph1,L16,(27398,55228,10498),S3,CH664747129,EAGLE\_L0200\_VOL1]  \\
400\phantom{00}     &   [Panph1,L16,(11324,24834,60541),S3,CH846509636,EAGLE\_L0400\_VOL1]  \\
800\phantom{00}     &   [Panph1,L16,(65448,27937,42773),S3,CH773405482,EAGLE\_L0800\_VOL1]  \\
1600\phantom{00}     &   [Panph1,L15,(18083,14638,23364),S3,CH1829653368,EAGLE\_L1600\_VOL1]  \\
3200\phantom{00}    &   [Panph1,L14,(2152,5744,757),S3,CH1814785143,EAGLE\_L3200\_VOL1]  \\
6400\phantom{00}     &   [Panph1,L13,(3868,2093,2715),S3,CH1320830929,EAGLE\_L6400\_VOL1]  \\
 \hline                                  
\end{tabular}
\label{tbl:eaglephases}
\end{table*}

\subsection{Choice of phases}
 Generating a Gaussian random field requires choosing a set of random
phases.  For the \eagle\ simulations we take these phases from \emph{Panphasia}
which is a public multiscale Gaussian white noise field
\citep{Jenkins2013ICs,Jenkins2013Panphasia}. Using \emph{Panphasia} provides a simple way to
publish the linear phases that define the \eagle\ volumes.
Table~\ref{tbl:eaglephases} lists the ``phase descriptors'' which define the
location of the phase information of each volume within the much
larger \emph{Panphasia} field \citep{Jenkins2013ICs}. These phase descriptors
define the phases on all scales and uniquely determine the phases not
only for the simulations published here, but for any possible zoom
simulation of any subregion of these volumes, and at any resolution
(down to sub-Earth mass resolution if needed) in the future.  In
principle sufficient information is provided in this paper to
enable anyone to re-run these simulations, or to resimulate objects identified from the \eagle\ database.  The information required is provided
by the combination of the phase descriptors, the cosmological parameters
and the linear matter power spectrum, and for the volumes themselves
the details of how the particle load was constructed.

\subsection{Particle indexing}
 To make it possible to trace particles easily between the initial
conditions and snapshots, each particle in the initial conditions was
given a unique 42-bit integer index.  The index was generated by
assigning each particle a location on a space-filling Peano-Hilbert
curve defined with a resolution of 14 bits per Cartesian coordinate
over the simulation volume.  The location for each particle was determined
from its unperturbed position in the particle load. The particle index
therefore encodes a Lagrangian position for the particle.  Using a
42-bit index allows the Lagrangian position to be determined to a
cubic cell of side length 1/16384 of the box size. This is small
compared to the interparticle separations of particles in the initial
conditions, which means that each particle has a unique index. The primitive $47^3$ glass file and routines to calculate the
Peano-Hilbert indices are available at \url{http://eagle.strw.leidenuniv.nl/}.

\subsection{Making the full initial conditions}
  The initial conditions for the hydrodynamical simulations are
generated from the dark matter only sets of initial conditions.  Each
dark matter particle is replaced with a pair of particles consisting
of a dark matter particle and gas particle with a combined mass equal
to that of the original dark matter particle.  The ratio of the gas and dark matter particles is equal to $\Omega_{\rm baryon}/(\Omega_{\rm
matter}-\Omega_{\rm baryon})$. These particle pairs are
positioned so that the centre of mass of the pair corresponds to the
position of the original particle in the dark matter only initial
conditions.  The particle pairs are aligned with the (1,1,1)
coordinate direction and the gas particle is positioned in the (1,1,1)
direction relative to its corresponding dark matter particle.  The
magnitude of the displacement between the pair is chosen so that an
initial cubic grid with mean density in the dark matter only initial
conditions would transform into a body-centred cubic grid with dark
matter (gas) particles at the centres of cubic cells made of gas (dark
matter) particles.
  
 For the hydrodynamical simulations the index of the dark matter
particles is taken to be exactly twice that of the corresponding index
in the dark matter only initial conditions.  The index of the gas
particle is chosen to be one more than its corresponding dark matter
particle.  Thus, all dark matter particles have even indices, and all
gas particles odd indices. 

\end{document}